\documentclass{SciPost}

\hypersetup{
    colorlinks,
    linkcolor={red!50!black},
    citecolor={blue!50!black},
    urlcolor={blue!80!black}
}

\usepackage[bitstream-charter]{mathdesign}
\usepackage{ulem}
\usepackage{bm}
\usepackage{aas_macros}

\DeclareSymbolFont{usualmathcal}{OMS}{cmsy}{m}{n}
\DeclareSymbolFontAlphabet{\mathcal}{usualmathcal}

\newcommand{\change}[1]{#1}

\renewcommand{\d}[0]{{\rm{fl}}}
\newcommand{\dd}{\ensuremath{\mathrm{d}}}

\newcommand{\del}[0]{\partial }
\newcommand{\sH}[0]{{\mathcal{H}}}
\newcommand{\vol}[2]{\hspace{-0.8mm}\mbox{$\text{d}^{\hspace{-0.0mm}#1}$}\hspace{-0.2mm}#2\hspace{0.8mm}\ }
\renewcommand{\v}[1]{\bm{#1} }
\newcommand{\vv}[0]{\bm{v} }
\newcommand{\vw}[0]{\bm{w} }
\newcommand{\vJ}[0]{\bm{J} }
\newcommand{\vx}[0]{\bm{x} }
\newcommand{\vr}[0]{\bm{r} }
\newcommand{\vs}[0]{\bm{s} }

\newcommand{\vz}[0]{\bm{z} }
\newcommand{\vp}[0]{\bm{p} }
\newcommand{\vq}[0]{\bm{q} }
\newcommand{\vk}[0]{\bm{k} }

\newcommand{\vnabla}[0]{\bm{\nabla} }
\newcommand{\hvz}{\hat{\v{z}}}
\newcommand{\varvol}[2]{\hspace{-0.0mm}\mbox{$\text{d}^{\hspace{-0.0mm}#1}$}\hspace{-0.2mm}#2\hspace{0.8mm}\!}
\renewcommand{\d}{\text{fl}}

\fancypagestyle{SPstyle}{
\fancyhf{}
\lhead{\colorbox{scipostblue}{\bf \color{white} ~SciPost Physics Lecture Notes }}
\rhead{{\bf \color{scipostdeepblue} ~Submission }}

\fancyfoot[C]{\textbf{\thepage}}
}

\begin{document}

\pagestyle{SPstyle}

\begin{center}{\Large \textbf{\color{scipostdeepblue}{
Large-scale structures of the Universe:\\physics, phenomenology, statistics
}}}\end{center}

\begin{center}\textbf{
Cora Uhlemann\textsuperscript{1$\star$},
}\end{center}

\begin{center}
{\bf 1} Fakultät für Physik, Universität Bielefeld, Postfach 100131, 33501 Bielefeld, Germany\\
$\star$ \href{mailto:cuhlemann@physik.uni-bielefeld.de}{\small cora.uhlemann@uni-bielefeld.de}
\end{center}

\section*{\color{scipostdeepblue}{Abstract}}
\textbf{\boldmath{%
In this series of lectures, we seek to describe the evolution of the cosmic large-scale structure. We will discover the cosmic web - the large-scale skeleton of matter traced by galaxies. It arises from the interplay of the gravitational pull of dark matter and the expansion driven by dark energy. Major large-scale galaxy surveys map the distribution of matter and galaxies across most of the sky, spanning over 10 billion years of cosmic history. I will guide you through some of the principles and challenges behind predicting the statistical properties of the matter and galaxy distribution in vast cosmic volumes. In particular we discuss the underlying nonlinear physics and resulting non-Gaussian statistics that need to be predicted to extract fundamental physics from observational data.
}}

\vspace{\baselineskip}

\noindent\textcolor{white!90!black}{%
\fbox{\parbox{0.975\linewidth}{%
\textcolor{white!40!black}{\begin{tabular}{lr}%
  \begin{minipage}{0.6\textwidth}%
    {\small Copyright attribution to authors. \newline
    This work is a submission to SciPost Physics Lecture Notes. \newline
    License information to appear upon publication. \newline
    Publication information to appear upon publication.}
  \end{minipage} & \begin{minipage}{0.4\textwidth}
    {\small Received Date \newline Accepted Date \newline Published Date}%
  \end{minipage}
\end{tabular}}
}}
}


\vspace{10pt}
\noindent\rule{\textwidth}{1pt}
\tableofcontents
\noindent\rule{\textwidth}{1pt}
\vspace{10pt}


\section{Introduction}
\label{sec:intro}

A multitude of cosmological probes (including the Cosmic Microwave Background discussed in parallel) have established that the dominant matter component is of unknown dark origin, exhibiting only very small, if any, non-gravitational interactions. Dark matter is indispensable for our understanding of cosmic structure formation, as it is able to cluster early due to the absence of forces opposing gravity, as well as providing the environment for galaxy formation \cite{Wechsler2018,DesjacquesBiasReview2018}. This property is highly beneficial from a theoretical point of view, as it admits a two-step approach:
\begin{enumerate}
\item solving purely gravitational collisionless dynamics for the dark matter component dominating the large scales, and then
\item tackling the more complicated interplay between gravitational and baryonic effects that lead to the formation of galaxies and mostly affect smaller scales.
\end{enumerate}
Splitting the problem allowed for major advances in the theoretical description \cite{Zeldovich1970,Peebles1980,BernardeauReview2002,Baumann2012} and numerical modelling \cite{Springel2005,HorizunRun4,QuijoteSims,Euclid2025FlagshipSim} of cosmic large scale structure. To take full advantage of observational data from massive galaxy surveys \cite{DESY6_Photoz,KiDS-Legacycosmo,DESIDR1,Euclid_overview_2024,LSST}, such as testing the cosmological standard model and probing fundamental physics, we need to push our predictions to higher precision and smaller scales.

Cosmological simulations have revealed striking universal characteristics of the cosmic web as a whole \cite{Gott1986,Kofman1988,Shandarin1989,vandeWeygaert2009,Neyrinck2012,Codis2018} and individual structures such as the density profiles of bound dark matter halos \cite{NFW1997,Angulo2017} and voids \cite{Hamaus2014}. The halo model has emerged as a useful tool to describe the outcome of dark matter clustering  \cite{PressSchechter1974,Peacock2000,Seljak2000,MoWhite1996,ShethTormen1999,ShethMoTormen2001,CooraySheth2002} by associating all dark matter with halos as bound structures. Then, the statistical properties of the large-scale density can be obtained from a prediction of the number and spatial distribution of the halos, as well as from the distribution of matter within each halo. 

This lecture series progresses in three stages: Section~\ref{sec:DMdynamics} describes theoretical approaches to describe the dark matter dynamics. Section~\ref{sec:SummaryStats} introduces key clustering statistics and showcases how they can be computed using those approaches. Finally, Section~\ref{sec:DMtoObs} describes how predictions for dark matter can be translated to key observables probed by galaxy surveys, in particular galaxy clustering and weak lensing, and Section~\ref{sec:conclusion} provides a short conclusion.

\paragraph{Basic assumptions}
In the bulk of the lectures, we will  assume a `vanilla' flat $\Lambda$CDM cosmology, and thus the validity of General Relativity on cosmological scales. In this model, initial perturbations are adiabatic (same across all species), Gaussian and almost scale-invariant. Dark energy is a cosmological constant with a simple equation of state $w=-1$. We will mostly neglect subtle imprints of massive neutrinos, which cause a slight suppression of structures on nonlinear scales. 

Baryons and cold dark matter can be considered as ``cold'' in the sense that the constituent particles are non-relativistic. We can effectively use the sub-Horizon approximation as most of structure formation happens well within the Hubble horizon. Those two aspects simplify the treatment substantially and mean that we can often use our intuition for Newtonian gravity.

In a cold dark matter universe, there are many aspects that make our treatment easier. In particular, large-scale fluctuations are sufficiently small (and on certain scales close to linear even today). We can perform a perturbative expansion in fluctuations on large scales. Structure formation proceeds hierarchically from small to large scales, which is in contrast to hot dark matter for example. Simulations of large volumes can assume a background cosmology.

\paragraph{Spacetime}

We start from a Friedmann-Lemaitre-Robinson-Walker (FLRW) background spacetime with linear scalar perturbations with the line element
\begin{equation}
ds^2 = -\left(1+2\Psi(\vx,t)\right) dt^2 + a^2(t)(1+2\Phi(\vx,t))d\vx^2\,,
\end{equation}
with scale factor $a(t)$ encoding the overall expansion of space and two scalar potentials, the gravitational potential $\Psi$ and $\Phi$. It is useful to switch to comoving coordinates $\v{r}=a(t)\vx$ and conformal time $\eta$ defined through 
\begin{equation}
\label{eq:conformaltime}
d\eta=\frac{dt}{a(t)}=\frac{da}{a(t)} \frac{dt}{da}= \frac{da}{\dot a a} = \frac{d\ln a}{\dot a} = \frac{d\ln a}{Ha} = \frac{da}{H a^2}\,.
\end{equation}
We can associate a comoving distance to this $d\chi=-d\eta=dz/H(z)$ where we used that $a=(1+z)^{-1}$. The particle momentum and velocity are related by $\vv=\v{p}/m=a d\vx/dt = \vx'$, where the prime is a derivative w.r.t. conformal time. As you have seen in the cosmology lectures by Pedro Ferreira \cite{LH_Ferreira2025}, in the absence of anisotropic stress, the Einstein equations imply that the two gravitational potentials are equal. The ‘relativistic’ Newton-Poisson equation relating the gravitational potential to the density reduces to the known Newton-Poisson equation on sub-Horizon scales. For looking at structure formation it will be beneficial to write the physical density as a product of the mean density and a normalised density
\begin{equation}
\rho(\vx,\eta)=\bar \rho(\eta) [1+\delta(\vx,\eta)]\,,
\end{equation}
where the density contrast $\delta$ encodes the local fluctuation away from the mean density.

\begin{figure}
\centering
\includegraphics[width=0.75\textwidth]{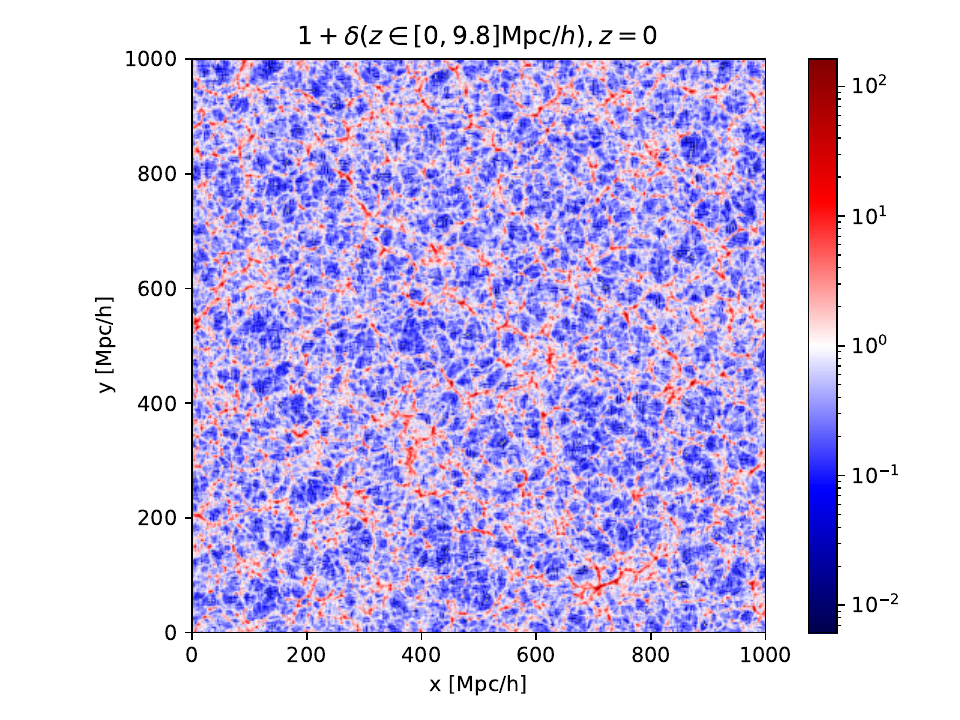}
\caption{Thin slice through the 3-dimensional dark matter distribution in the Quijote simulations at redshift $z=0$ (shown for realisation 0 of the fiducial cosmology).}
\label{fig:density_slice}
\end{figure}

\paragraph{Illustrations comparing theoretical predictions and simulations}
For key illustrations in these lecture notes, we rely on the publicly available Quijote Simulations \cite{QuijoteSims} - a large suite of $N$-body simulations with a box side length of 1 Gpc/$h$, for which the raw particle data and dark matter halo catalogs are available along with a selection of pre-computed summary statistics for matter and halos. A connected public Python package \href{https://github.com/franciscovillaescusa/Pylians3}{Pylians} can be used for performing basic simulation analysis tasks, such as creating density fields from particle or halo catalogs, computing displacement fields or including redshift space distortions, as well as extracting summary statistics like the two-point correlation function, power spectrum, bispectrum, one-point statistics and the halo mass function. As a starting point, Figure~\ref{fig:density_slice} illustrates a thin slice through the late-time matter density field, showing a characteristic web-like structure and illustrating that much of the volume is mildly underdense while matter tends to concentrate in regions of rather high densities.

\section{Modelling dark matter dynamics}
\label{sec:DMdynamics}

$N$-body simulations discussed in the `Numerical Cosmology' lectures and reviewed in \cite{AnguloHahnReview2022} have been established as the state-of-the-art benchmark for testing theoretical models and extracting effective fitting functions. The volume and the number of particles that can be simulated are steadily growing with the computational power of supercomputers. However, even the largest cosmological simulations to date with more than a trillion particles \cite{PKDGRAV3,HACCsimulations,DarkSky2014,OuterRim2019,AbacusSummit2021,Uchuu2021,EuclidFlagship2025} simulate particles with masses of the order of a billion solar masses and hence effective mass points representing huge conglomerates of dark matter rather than actual particles. Numerical artefacts intrinsic to N-body methods, such as sparse sampling and unphysical two-body effects, can pose a challenge for correctly resolving the dynamics of the coherent dark matter field down to small scales where one is sensitive to the particle properties of dark matter. This is why there is increasing activity in tackling the full phase-space dynamics numerically \cite{Colombi2008,Shandarin2012,Sousbie2015ColDICE,Hahn2016} or recreating the phase-space structure from $N$-body simulations \cite{Abel2012}.
A potential alternative is to use approximate field-based methods inspired by the quantum-classical correspondence which allows to reduce dynamics from phase-space to position space \cite{WidrowKaiser1993,Uhlemann2014schroedinger,Mocz2018,Uhlemann2019semiclassical}. Here we describe the phenomenology of the classical phase space dynamics closely following the description in \cite{Uhlemann_2018findingclosure}.

Understanding the formation of bound dark matter structures requires accessing the so-called multi-stream regime, in which dark matter cannot be treated as a perfect fluid described by just density and velocity. This regime emerges naturally when collisionless dark matter particles collapse onto an overdensity where they cross because of their collisionless nature and infall velocity. So far, the multi-stream regime is mostly the domain of numerical simulations, while only few analytical approaches for the treatment of shell-crossing \cite{Taruya2017,Rampf2017,Saga2018,Rampf2021Review} or its long-term limits \cite{FillmoreGoldreich1984,Pietroni2018} have been developed. Because bound structures are formed through a whole series of crossings that successively increases the number of streams, they require a method that can dynamically generate new streams. Here, we will approach this problem from a theoretical angle that looks at the phase-space distribution functions and possible expansions in terms of cumulants with respect to momentum.

\subsection{Phase space dynamics}

On scales that are small compared to the Hubble radius (and hence the observable universe), in the weak field regime and for non-relativistic velocities, one can use the Newtonian limit instead of the full Einstein equations to describe the time evolution of structures within the Universe \cite{ChisariZaldarriaga2011,GreenWald2012,Kopp2014}. If we are interested in the dynamics of a collection of dark matter particles which are very abundant in the Universe, collisional effects are negligible due to their suppression by the total number of particles \cite{Gilbert1968}. This means that we do not have to solve the Bogoliubov-Born-Green-Kirkwood-Yvon (BBGKY) hierarchy for the series of $n$-point phase-space distributions $f_n(\{\vx_i,\vp_i\}_{i=1,\ldots,n})$, but instead only a collisionless equation for the one-particle phase-space distribution $f=f_1(\vx,\vp)$. We will consider this in the continuum limit, where it is not a sum of Dirac delta-distributions located at individual particle positions, but instead a coherent field encoding the probability of finding a particle in a given phase-space volume.

\paragraph*{Vlasov-Poisson equation} 
The time evolution of collisionless dark matter subject to gravitational interaction is encoded in the Vlasov-Poisson equation for the one-particle phase-space density $f(t,\vx,\vp)$, in the absence of two-body interactions. 
This equation is consequence of the conservation of phase-space volume $df/dt=0$ and can be rephrased in terms of the Poisson-bracket $\{H,f\}$ of the system's Hamiltonian $H$ and the phase-space distribution $f$ 
\begin{subequations}
\label{VlasovPoissonEq}
\begin{align}
\label{VlasovEq}
\partial_t f&=   -\frac{\vp}{a^2 m}\cdot\vnabla_{\! \!  x} f + m \vnabla_{\! \!  x} V \cdot\vnabla_{\! \!  p} f =\{H,f\}\,,
\end{align}
where we adopted \textit{comoving} coordinates $\vx$ with the conjugate momentum $\vp=a^2 m d\vx/dt$. The gravitational potential $V$ is given by the Poisson equation in terms of the density, which in turn is the integral of the phase-space distribution
\begin{align}
\label{PoissonEq}
\Delta V &= 
\frac{4\pi G}{a}  \left(\int \vol{3}{p}\!\!f - \rho_0\ \right) = \frac{4\pi G}{a}  \left(\rho(\vx) - \rho_0\ \right)\,.
\end{align}
\end{subequations}
In comoving coordinates we have a constant background density $\rho(\vx,t)=\rho_0(1+\delta(\vx,t))$ related to the physical background density as $\bar \rho(t)=\rho_0 a(t)^{-3}$. The Vlasov equation is a partial differential equation involving cosmic time $t$ and 3+3-dimensional phase-space consisting of position and momentum $(\vx,\vp)$ as variables. The Vlasov-Poisson system is a coupled nonlinear, partial, integro-differential equation and thus in general difficult to solve.

\paragraph*{Initial conditions} 
Usually we rely on `cold' initial conditions  (with negligible initial velocities) for which $f(t=t_0,\vx,\vp)=\rho(\vx)\delta_D(\vp)$ describes an initially flat 3-dimensional phase-space sheet. In the course of the time evolution, this sheet deforms itself in 6-dimensional phase-space subject to strong conditions like avoiding self-intersections.

\begin{figure}
\includegraphics[width=\textwidth]{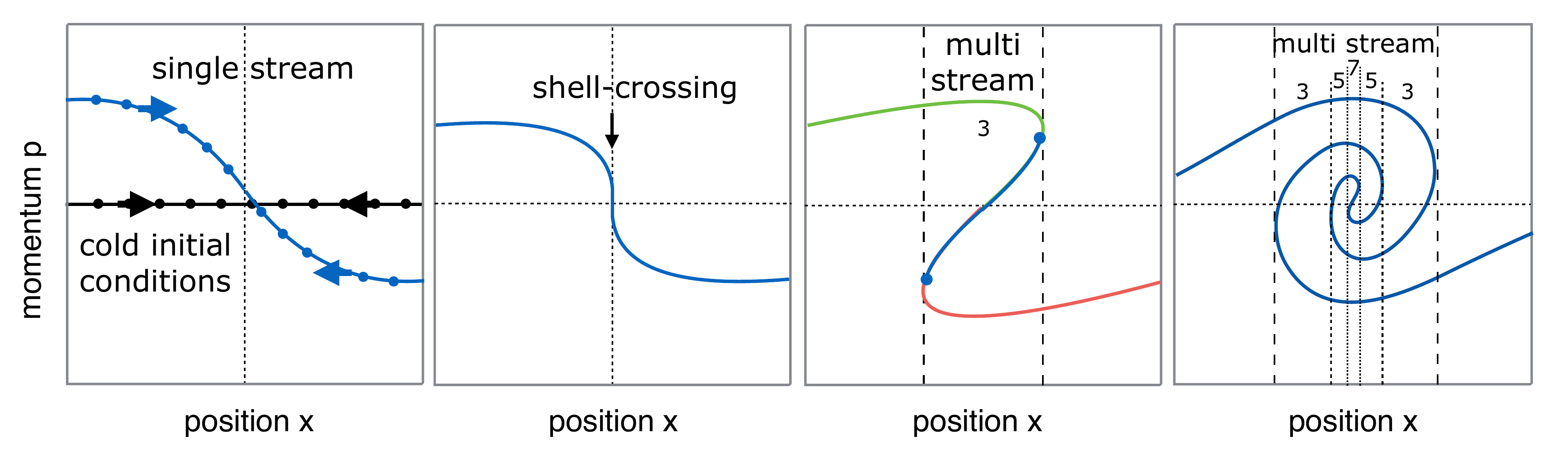} 
\caption{Schematic sketch of the time evolution of cold collisionless dark matter in $(1+1)$-dimensional phase-space. [Plot made similar to Figure~1 in \cite{Uhlemann_2018findingclosure}].
{\it Left:} Due to cold initial conditions the phase-space sheet is initially flat and slowly starts to bend due to the coherent infall caused by gravitational interaction. In this single-stream regime cold dark matter is well-described by a perfect pressureless fluid. 
{\it Middle left:} During shell-crossing the particle trajectories cross such that there are several momenta at one location. 
{\it Middle right:} Due to shell-crossing the single-stream has split into three fluid streams. 
{\it Right:} This process will happen repeatedly and result in a wound up phase-space sheet.}
\label{fig:pheno}
\end{figure}

\paragraph*{Phenomenology of time evolution}
Figure~\ref{fig:pheno} shows a sketch of the time evolution for the formation of a gravitationally bound structure starting from cold initial conditions (in black) in 1+1 dimensional phase space. A large ensemble of particles almost uniformly distributed in a slightly overdense region of space will create a gravitational potential that leads to a coherent infall to the central region. The particles accelerate and acquire velocities directed towards the center to which they move uniformly, at this stage the system is in the single stream regime. Eventually particles from both sides will reach the center and due to the collisionless nature they can overlapp (instead of colliding), a moment call shell-crossing. Due to the particle's nonzero velocities at the time they reach the center, they cross and overshoot which creates the characteristic S-shape of the phase-space sheet. After passing the center, the particles slow down due to the action of the gravitational potential that is deepest in the center. Their velocities decrease and they eventually reverse direction initiating a secondary infall onto the central region. The S-shape thus rotates around the center and the inner section of the S-shape now resembles the single-streaming stage of the earlier evolution. After a series of subsequent shell-crossings, the phase-space sheet will have wound up to an apparent whirl, but without any tears or self-intersections. While the outer regions might still be in the single-stream regime, an increasing number of streams appears in the regions closer to the center.  At some point the phase-space sheet is so tightly wound up, that it becomes effectively stationary and can be considered a bound structure. In general, collisionless self-gravitating systems are expected to evolve towards a steady state after a strong mixing phase such as violent relaxation \cite{Lynden-Bell1967}. 

\subsection{From phase space to fluids}

In practice, one is usually interested in the time evolution of properties of the spatial distribution, especially the density and mass-weighted velocity of the flow, rather than the fully fledged phase-space information encoded in the Vlasov equation. For this purpose, one extracts the relevant information from the phase-space distribution by computing momentum-weighted averages of the phase-space distribution, the so-called moments (with respect to momentum). In this subsection we suppress the time dependence of the phase space distribution and the corresponding generating functionals, moments and cumulants for brevity.
\paragraph*{Generating functional for phase-space moments and cumulants} For the purpose of computing moments and cumulants, it is useful to introduce a generating functional $G(\vx,\vJ)$, which can be viewed as a Fourier transform of the phase space distributon with respect to momentum
\begin{align}
&G(\vx,\vJ) = \int \vol{3}{p}\! \exp\left(\tfrac{i}{m}\vp\cdot\v{J}\right) f (\vx,\vp) \,, \label{genfun}
\end{align}
with the source $\vJ$. The moments $M^{(n)}(\vx)$ of the phase-space distribution function $f(\vx,\vp)$ are tensorial functions of the position obtained after integrating the phase space distribution weighted with a given number of momenta over momentum and thus can be obtained by taking derivatives of the generating functional with respect to the source
\begin{equation}
M^{(n)}_{i_1 \cdots i_n}(\vx):=\int \vol{3}{p} \frac{p_{i_1}}{m} \ldots \frac{p_{i_n}}{m} f (\vx,\vp)
=\frac{(-i)^n\del^n G(\vx,\vJ)}{\del J_{i_1} \ldots \del J_{i_n}} \Big|_{\v{J}=0}
\,.
\end{equation}
Higher order moments (from second order) will feature combinations of the lower-order ones, such that it is useful to define cumulants $C^{(n)}(\vx)$ as the connected parts of the moments, determined from derivatives of the natural logarithm of the generating functional
\begin{align}
C^{(n)}_{i_1 \cdots i_n}(\vx)= \frac{(-i)^n\del^n \ln G(\vx,\vJ)}{\del J_{i_1} \ldots \del J_{i_n}} \Big|_{\v{J}=0}\,.
\end{align}
The cumulants provide a good way to understand the prominent perfect pressureless fluid model which is a truncation of the Vlasov hierarchy at second order. One useful property of moments and cumulants alike is the total symmetry among all their indices. If the moment or cumulant generating functional exists, the probability distribution is uniquely determined by it \cite{lukacs1960book}. That does not necessarily mean that the moments or cumulants uniquely determine the probability distribution, because there are distributions where all moments exist and yet the limit that defines the generating functional does not (for 1-dimensional probability distributions a prominent example is the lognormal distribution). Notably, there are no distributions for which the cumulant generator is a nonlinear polynomial in $\vJ$. The linear case corresponds to a perfect pressureless fluid and describes the dynamics before shell-crossing. This suggests that for the dynamically generated multi-streaming behaviour, the cumulant generator might be more informative.

\paragraph*{Vlasov hierarchy} The evolution equation for the moment and cumulant generator are readily obtained from the Vlasov equation 
\begin{align}
\del_t G(\vx,\vJ) &= \frac{i}{a^2} \vnabla_J\cdot\vnabla_x G - iG \vJ \cdot\vnabla_x V \,,\\
\label{eq:evolutionlnG}
\del_t \ln G(\vx,\vJ) &= \frac{i}{a^2} \left(\vnabla_J\cdot\vnabla_x \ln G + \vnabla_J \ln G \cdot\vnabla_x \ln G\right) - i\vJ\cdot\vnabla_x V\,.
\end{align}
To determine evolution equations for moments and cumulants, respectively, one expands the corresponding generator in a power series in $\v{J}$. Because of the corresponding derivative in the $\vnabla_J$ term, the $n$-th order moment or cumulant will depend on the $(n+1)-th$ order one. For the cumulants, a mixing between different orders is evident that can only be avoided by a truncation at second order.

\paragraph*{Equations for density and velocity}
From the evolution equation for the cumulant generating function, we can the derive equations for the zeroth cumulant that is the logarithm of the density $C^{(0)}=\ln \rho$) and the first cumulant that is the velocity $C^{(1)}_i=v_i$
\begin{subequations}
\begin{align}
\label{eq:continuity}
\partial_t \ln \rho &= -\frac{1}{a^2}\left[\vnabla \cdot\vv + \vv\cdot \vnabla \ln \rho \right] \ \Leftrightarrow \ \partial_t \rho = -\frac{1}{a^2}\vnabla \cdot (\rho\vv)\,,\\
\label{eq:Euler}
\partial_t \vv &= -\frac{1}{a^2} \Bigg\{ (\vv\cdot\vnabla)\vv + \underbrace{\vnabla\cdot C^{(2)} + C^{(2)}\cdot \vnabla\ln \rho}_{\frac{\vnabla\cdot \left(\rho C^{(2)}\right)}{\rho}}\Bigg\} - \vnabla V\,.
\end{align}
Those equations are the continuity equation and the Euler equation, here written for a general velocity dispersion tensor $C^{(2)}$. Note that our definition of velocity is related to the conjugate momentum and hence not a peculiar velocity. It can be convenient to decompose the velocity into a gradient of a velocity potential and a vector potential. We can derive an evolution equation for the vorticity $\vw=\vnabla\times\vv$. For this it is useful to rewrite $(\vv\cdot\vnabla)\vv = \frac{1}{2}\vnabla{v^2} + \vw\times \vv$
\begin{align}
\label{eq:vorticity}
\partial_t\vw= -\frac{1}{a^2} \Bigg\{ \vnabla\times(\vv\times\vw)+ \vnabla\times \change{\left[\vnabla\cdot \v{C}^{(2)} + \v{C}^{(2)}\cdot\vnabla\ln n\right]}\Bigg\}\,.
\end{align}
\end{subequations}
The fluid equations for density and velocity are supplemented by the Poisson equation~\eqref{PoissonEq} rewritten as 
\begin{equation}
\label{PoissonEq_delta}
\Delta V = \frac{3}{2}\Omega_m(t)\dot a(t)^2  \delta(\vx,t)\,,
\end{equation}
where we used the defintion of the matter density $\Omega_m(t)=\frac{\bar\rho(t)}{\rho_c}$ with the critical density $\rho_c=3H^2/(8\pi G)$ and the background density $\bar \rho(t)=\rho_0 a^{-3}$ given in terms of the constant background comoving density $\rho_0$.

\paragraph{Second-order truncation: the perfect pressureless fluid model}
A truncation of cumulants at order $m$ is performed by setting all cumulants at order $m$ or above to zero, $C^{(n\geq m)}\equiv 0$. This has to be done in the evolution equations for the lower-order cumulants $C^{(n < m)}$ and neglects any possible sourcing of the higher-order cumulants in the course of time evolution. The truncated hierarchy is then a closed set of evolution equations for the lower-order cumulants $C^{(n < m)}$. For truncations that are {\it consistent} with time evolution,  the evolution equations for higher-order cumulants reduce to $\partial_t C^{(n\geq m)}=0$ and hence initially vanishing cumulants remain zero. The truncation at second order leads to the following moments and cumulants
\begin{subequations}
\begin{align}
M_\d^{(0)} &= \rho_\d\,, \qquad {M_\d}^{(1)}_i=\rho_\d v^{\d}_i\,, \quad { M_\d}^{(n \geq 2)}_{i_1\cdots i_n} = \rho_\d v^{\d}_{i_1} \cdots v^{\d}_{i_n}
\,,\\
C_\d^{(0)} &= \ln \rho_\d\,, \quad \ {C_{\d,i}}^{(1)}=v^{\d}_i  \,, \quad \quad \  {C_\d}^{(n \geq 2)}_{i_1\cdots i_n} = 0 \,.
\end{align}
\end{subequations}
From the finite number of cumulants, one can easily obtain the cumulant generator which is linear in $\v{J}$
\begin{align}
\ln G_\d(\vJ)=\ln \rho_\d + i\vJ\cdot\vv_\d \ \Rightarrow \
G_\d(\vJ) =\rho_\d \exp\left[i\v{J}\cdot\vv_{\d}\right] \,,
\end{align}
and all cumulants of order higher than one vanish identically. This corresponds to a phase-space distribution function of a perfect pressureless fluid (in part of the theoretical physics community also called dust)
\begin{align}
\label{eq:fdust}
f_{\d}(\vx,\vp)=\rho_{\d}(\vx) \delta_D\left(\vp - m\vv_{\d}(\vx)\right)\,.
\end{align}
This model does not include effects like velocity dispersion, encoded in the second cumulant $C^{(2)}$ or higher-order effects. Therefore for the perfect fluid ansatz $f_\d$, the Vlasov equation is equivalent to the pressureless fluid system consisting of the continuity equation~\eqref{eq:continuity} and Euler equation~\eqref{eq:Euler}. Since the perfect pressureless fluid has zero velocity dispersion, $C^{(2)}\equiv 0$, an initially irrotational velocity will stay irrotational until shell-crossing. By assuming the velocity is a gradient field $\vv=\vnabla\phi_\d$ one can reduce the Euler equation \eqref{eq:Euler} for the velocity to the Bernoulli equation for the velocity potential $\phi_\d$. The perfect pressureless irrotational fluid system reads
\begin{align}
\label{Fluid}
\del_t \rho_\d =   -\frac{1}{a^2}\vnabla\cdot( \rho_\d \vnabla\phi_\d) \,, \quad
\del_t \phi_{\d} = -\frac{1}{2a^2} (\vnabla\phi_\d)^2- V   
\end{align}
If $\rho_\d$ and $\vv_\d$ fulfill these equations and higher cumulants are set to zero identically, then all evolution equations of the higher moments are automatically satisfied. 

While from the equations for the cumulant generator, it appears that one can set $C^{(n\geq 2)} \equiv 0$ in a consistent manner,
shell-crossing dynamically produces multiple streams that source vorticity and velocity dispersion along with higher cumulants, see \cite{PueblasScoccimarro2009,Hahn2014}. The breakdown of the perfect pressureless fluid model description at shell-crossing is accompanied by the emergence of a singular density at the instant of shell-crossing. Note that the occurrence of this singularity can be rephrased when going to Lagrangian coordinates described in Section~\ref{subsec:LPT}.

\subsection{Perturbation Theory}
\label{subsec:PT}
Since it is generally difficult to solve the coupled nonlinear fluid equations, one often resorts to a perturbative treatment which is sufficient for the early stages of gravitational evolution, or capturing the behaviour on large, weakly nonlinear scales. In this section we will briefly review the basics of perturbation theory for the fluid model, both in the Eulerian (Standard Perturbation Theory, SPT) and the Lagrangian (LPT) framework and discuss their relation. A much more in-depth review can be found in \cite{BernardeauReview2002} from which we extracted some key results that we describe here.  In the Effective Field Theory of Large Scale Structure \cite{Baumann2012,Carrasco2012} one describes an effective fluid for which the impact of higher order cumulants are constructed from a whole series of all terms allowed by symmetry with free coefficients, which then have to calibrated against data and/or simulations. In the recently developed Vlasov Perturbation Theory \cite{garny2025vlasovPT} it has been demonstrated that the impact of velocity dispersion on commonly used summary statistics is relevant, while the third cumulant has a small quantitative impact and higher cumulants only have a minor effect on mildly non-linear scales.

\subsubsection{Eulerian Perturbation Theory}
\label{subsec:SPT}
In Eulerian Perturbation Theory quantities under consideration are the density contrast $\delta(\vx,\eta)$ which defines the physical density as $\rho(\vx,\eta)=\bar\rho(\eta)\left(1+\delta(\vx,\eta)\right)$ and the peculiar velocity field $\vv(\vx,\eta) = \v{u}(\vx,\eta)/a(\eta)$. In conformal time, the fluid equations can be recast in the following form
\begin{subequations}
\label{Fluid}
\begin{align}
\del_\eta \delta =   -\vnabla\cdot( (1+\delta) \vv) \,, \label{eq:continuity_delta}\\
\del_\eta \vv+\sH\vv +(\vv\cdot\vnabla)\vv =-\vnabla V   \,, \label{eq:Euler_conf}
\end{align}
\end{subequations}
where $\sH(\eta)=a^{-1}da/d\eta = da/dt=\dot a(t(\eta))$.
Those two coupled evolution equations for a cold dark matter fluid are supplemented by the Poisson equation~\eqref{PoissonEq_delta}. The velocity field can be decomposed into a velocity divergence $\theta=\vnabla\cdot \vv$ and a a vorticity $\vw=\vnabla\times\vv$ that vanishes identically due to the irrotational character. We then obtain for the velocity divergence in the Einstein-de Sitter case
\begin{equation}
\partial_\eta\theta+\sH\theta+\vnabla\cdot\left[(\vv\cdot\vnabla)\vv\right]=-\frac{3}{2}\Omega_m(\eta)\sH^2(\eta)\delta\,.
\end{equation}

\begin{figure}
\centering
\includegraphics[width=0.5\textwidth]{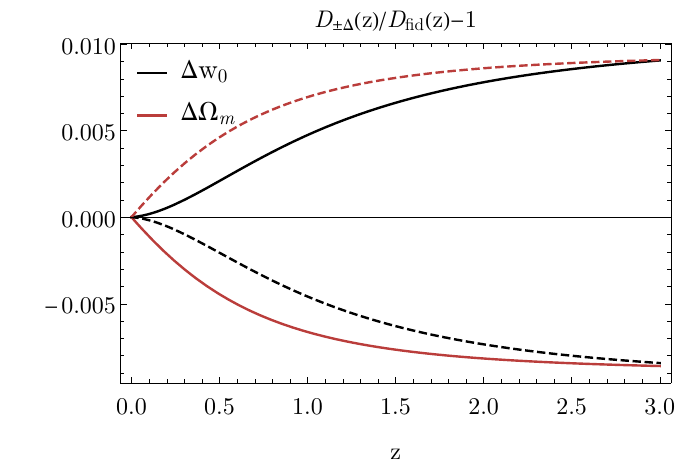}
\caption{Impact of two cosmological parameters on the normalised linear growth factor. For roughly 4\% changes in the matter density $\Delta\Omega_m=\pm 0.0125$ (red) and the dark energy equation of state $\Delta w_0=\pm 0.04$ (black) around a fiducial cosmology with $\Omega_m=0.3134$ and $w_0=-1$. We show the impact of positive and negative changes (solid and dashed lines), respectively. [Plot made in the style of Figure 19 in \cite{Uhlemann2020}]}
\label{fig:Dofzcomparison}
\end{figure}

\paragraph{Eulerian Linear Perturbation Theory}
Let us linearise the fluid-Poisson system, so neglect all terms that are quadratic in the density contrast or the velocity
\begin{align}
\label{Fluid_lin}
\del_\eta \delta =   -\vnabla\cdot \vv = -\theta \,, \quad
\partial_\eta\theta+\sH\theta =-\frac{3}{2}\Omega_m(\eta)\sH^2(\eta)\delta  \,. 
\end{align}
We can reduce this to a single second-order differential equation for the density contrast by computing a time derivative of the linearised continuity equation and inserting the linearised Bernoulli equation
\begin{equation}
\partial_\eta^2\delta+\sH\partial_\eta\delta =\frac{3}{2}\Omega_m(\eta)\sH^2(\eta)\delta \,. 
\end{equation}
Let us assume that the linear density contrast grows as $\delta^{(1)}(\vx,\eta) = D(\eta) \tilde\delta_1(\vx)$ given some initial conditions $\tilde\delta_1(\vx)$ and deduce an equation for the \emph{linear growth factor}
\begin{equation}
\label{eq:linear_growth}
D''(\eta) + \sH D'(\eta) = \frac{3}{2}\Omega_m(\eta)\sH^2(\eta) D(\eta)\,.
\end{equation}
The linear velocity dispersion then takes the form $\theta^{(1)}(\vx,\eta)=-D'(\eta) \tilde\delta_1(\vx)$. We can rewrite $D'(\eta)= D dln D/d\eta = D \sH d\ln D/d\ln a$ using equation~\eqref{eq:conformaltime} and define the \emph{growth rate} to express the first order velocity divergence in terms of the density
\begin{equation}
\label{eq:growth_rate}
f(\eta) = \frac{d\ln D}{d\ln a}\,,\quad \theta^{(1)}(\vx,\eta)=-\sH f \delta^{(1)}(\vx,\eta)\,.
\end{equation}
The growth is a sensitive probe of dark energy, in an EdS universe we have $D=a$, but with increasing $\Omega_\Lambda$ the growth decreases at lower redshifts. 
For a cosmology with dark energy beyond a cosmological constant, the growth function can be approximated as given in \cite{GlazebrookBlake2005}, 
\begin{align}
\label{eq:Dcosmo}
D(z)&=\frac{5\Omega_m}{2} \frac{H(a)}{H_0}\int_0^a \frac{{\rm d} a' H_0^3}{a'^3 H^3(a')}\,, \quad
\frac{H(a)}{H_0} = \sqrt{\frac{\Omega_m}{a^3}+ \Omega_\Lambda \exp\left(3 \int_0^z \frac{1+w(z')}{1+z'} {\rm d} z'  \right)}\,, 
\end{align}
 with the dark matter density, $\Omega_m$, the dark energy density, $\Omega_\Lambda$, the Hubble constant at zero redshift, $H_0$, the expansion factor $a\equiv1/(1+z)$ and the dark energy equation of state $w(z)$. For a flat wCDM universe with a constant equation of state, the linear growth of structure depends only on the matter density $\Omega_m$ and the equation of state parameter $w_0$ and following \cite{Silveira1994,Percival2005} can be written as
\begin{align}
\label{eq:Dsimp2}
D(z)=\frac{
    {}_{2}F_1\left[\frac{-1}{3w_0}, \frac{w_0 - 1}{2w_0}, 
     1 - \frac{5}{6w_0}, (1 + z)^{3w_0} \frac{ \Omega_m-1}{\Omega_m} \right]}{(1+z)
  {}_{2}F_1\left[\frac{-1}{3w_0}, \frac{w_0 - 1}{2w_0}, 
   1 - \frac{5}{6w_0}, \frac{ \Omega_m-1}{\Omega_m}\right]}\,,
 \end{align}
 where ${}_{2}F_1$ is the hypergeometric function. In Figure~\ref{fig:Dofzcomparison} we compare the growth for changes in the constant equation of state parameter $w_0$ and the matter density $\Omega_m$.

\paragraph{Eulerian Nonlinear Perturbation Theory: Basics}
Having solved the linear equations, we can now develop a strategy to solve the nonlinear equations - we can improve our approximation by inserting our linear solution into the nonlinear terms and collecting them at fixed order for powers of the fields. To make this efficient, let us rewrite the fluid equations by collecting all linear terms on the left hand side and all nonlinear terms on the right hand side.
\begin{align}
&\del_\eta \delta + \theta = - \delta \theta -\vnabla \delta\cdot \vv \,, \quad 
&\partial_\eta\theta+\sH\theta+\frac{3}{2}\Omega_m(\eta)\sH^2(\eta)\delta = - \vnabla(\vv\cdot\vnabla)\cdot\vv- (\vv\cdot\vnabla)\theta \,.
\end{align}
As a second step, let us switch  to Fourier space to convert derivatives into multiplications with the wave vector. In particular we have $\vnabla\delta(\vx)\rightarrow i\vk\delta(\vk)$ and $\theta(\vk)=i\vk\cdot \vv(\vk)$ such that for an irrotational velocity we have $\vv(\vk)=-i\vk/k^2\theta(\vk)$. The switch to Fourier space also means that products of fields will now be converted to convolutions, such that for the right hand side of the continuity equation we get
\begin{subequations}
\begin{align}
\delta \theta &\rightarrow \int \frac{\dd^3k_1}{(2\pi)^3} \int \frac{\dd^3k_2}{(2\pi)^3} (2\pi)^3\delta_D(\vk-\vk_1-\vk_2) \delta(\vk_1)\theta(\vk_2) \,,\\
\vnabla \delta\cdot \vv &\rightarrow 
\int \frac{\dd^3k_1}{(2\pi)^3} \int \frac{\dd^3k_2}{(2\pi)^3} (2\pi)^3\delta_D(\vk-\vk_1-\vk_2) \frac{\vk_1 \cdot \vk_2}{k_2^2} \delta(\vk_1) \theta(\vk_2)\,.
\end{align}
Similarly for the right hand side of the Bernoulli equation we have
\begin{align}
\vnabla(\vv\cdot\vnabla)\cdot\vv \rightarrow &
\int \frac{\dd^3k_1}{(2\pi)^3} \int \frac{\dd^3k_2}{(2\pi)^3} (2\pi)^3\delta_D(\vk-\vk_1-\vk_2) \frac{(\vk_1\cdot \vk_2)^2}{k_1^2k_2^2}  \theta(\vk_1) \theta(\vk_2)\,,\\
(\vv\cdot\vnabla)\theta \rightarrow & \int \frac{\dd^3k_1}{(2\pi)^3} \int \frac{\dd^3k_2}{(2\pi)^3} (2\pi)^3\delta_D(\vk-\vk_1-\vk_2) \frac{\vk_1\cdot \vk_2}{k_1^2} \theta(\vk_1) \theta(\vk_2)\,,
\end{align}
\end{subequations}
where we notice that in Fourier space the different contraction of indices for the derivatives translates into different scalar products of wave vectors.
This means we obtain in slightly shorter notation
\begin{align}
\label{eq:fluideq_nonlin_right}
&\del_\eta \delta + \theta =   -  \int \frac{\dd^3k_1\dd^3k_2}{(2\pi)^3} \delta_D(\vk-\vk_1-\vk_2) \underbrace{\left(1+ \frac{\vk_1 \cdot \vk_2}{k_2^2}\right)}_{\alpha(\vk_1,\vk_2)} \delta(\vk_1) \theta(\vk_2)\,, \\ 
&\partial_\eta\theta+\sH\theta+\frac{3}{2}\Omega_m\sH^2\delta = - \int \frac{\dd^3k_1 \dd^3k_2}{(2\pi)^3} \delta_D(\vk-\vk_1-\vk_2) \underbrace{\left(\frac{(\vk_1\cdot \vk_2)^2}{k_1^2k_2^2}  + \frac{\vk_1\cdot \vk_2}{k_1^2}\right)}_{\beta(\vk_1,\vk_2)} \theta(\vk_1) \theta(\vk_2)\,, \notag
\end{align}
where all fields on the left are with wavevector $\vk$. The functions $\alpha$ and $\beta$ stem from the nonlinearity and couple different modes, as evident from their dependence on $\vk_1$ and $\vk_2$ connected to the fields $\delta$ and $\theta$. We can see that to solve for the $\vk$ dependence of the fields, we do in principle need to integrate over all modes $\vk_1$ and $\vk_2$ such that $\vk=\vk_1+\vk_2$. For the following it will be useful to make the symmetry between switching $\vk_1\leftrightarrow\vk_2$ manifest in the expressions, by writing them as
\begin{equation}
\label{eq:SPT_alphabeta} 
\alpha(\vk_1,\vk_2)=1 + \frac{1}{2} \frac{\vk_1 \cdot \vk_2}{k_1k_2}\left(\frac{k_2}{k_1}+\frac{k_1}{k_2}\right)\,,\quad
\beta(\vk_1,\vk_2)=\frac{(\vk_1\cdot \vk_2)^2}{k_1^2k_2^2}  + \frac{1}{2}\frac{\vk_1\cdot \vk_2}{k_1k_2}\left(\frac{k_2}{k_1}+\frac{k_1}{k_2}\right)\,.
\end{equation}
A special property that will be useful later is to notice that when averaged over the angle between $\vk_1$ and $\vk_2$, we have $\bar\alpha=1$ and $\bar\beta=\frac{1}{3}$.

\paragraph{Second order Eulerian Perturbation Theory (2SPT)}
Now we are in the position to solve for the second order fields $\delta\approx \delta^{(1)}+\delta^{(2)}$ and $\theta\approx \theta^{(1)}+\theta^{(2)}$ using our knowledge about the first (linear) order, in particular it's time-dependence, we find
\begin{align}
\label{eq:fluideq_2SPT}
&\del_\eta \delta^{(2)} + \theta^{(2)} =   \sH f D^2 \int \frac{\dd^3k_1\dd^3k_2}{(2\pi)^3} \delta_D(\vk-\vk_1-\vk_2) \alpha(\vk_1,\vk_2) \tilde\delta_1(\vk_1) \tilde\delta_1(\vk_2)\,, \\ 
&\partial_\eta\theta^{(2)}+\sH\theta^{(2)}+\frac{3}{2}\Omega_m\sH^2\delta^{(2)} = - \sH^2 f^2 D^2\int \frac{\dd^3k_1 \dd^3k_2}{(2\pi)^3} \delta_D(\vk-\vk_1-\vk_2) \beta(\vk_1,\vk_2) \tilde\delta_1(\vk_1) \tilde\delta_1(\vk_2)\,,\notag
\end{align}
which shows we can still separate the dependence on time and space! If we assume matter domination ($f=1$, $\Omega_m=1$), we can show that $\delta^{(2)}(\vk,\eta)= D^2(\eta)\tilde \delta^{(2)}(\vk)$ and similarly $\theta^{(2)}(\vk,\eta)= (\sH D^2)(\eta)\tilde \theta^{(2)}(\vk)$ and then we can use $(D^2)'(\eta) = 2DD'(\eta) = 2D^2 \sH$ and $(\sH D^2)'=\sH' D^2+ 2\sH D D' \approx 3/2 \sH^2D^2$ and thus reduce the equations to purely spatial ones
\begin{align}
\label{eq:fluideq_2SPT_spatial}
&2\tilde\delta^{(2)}(\vk) + \tilde\theta^{(2)}(\vk)  =   \int \frac{\dd^3k_1\dd^3k_2}{(2\pi)^3} \delta_D(\vk-\vk_1-\vk_2) \alpha(\vk_1,\vk_2) \tilde\delta_1(\vk_1) \tilde\delta_1(\vk_2)\,, \\ 
&\frac{5}{2}\tilde\theta^{(2)}(\vk) +\frac{3}{2}\tilde\delta^{(2)}(\vk)  = - \int \frac{\dd^3k_1 \dd^3k_2}{(2\pi)^3} \delta_D(\vk-\vk_1-\vk_2) \beta(\vk_1,\vk_2) \tilde\delta_1(\vk_1) \tilde\delta_1(\vk_2)\,.\notag
\end{align}
By building linear combinations that isolate the density contrast and velocity divergence, respectively, we obtain the solutions
\begin{align}
\tilde\delta^{(2)}(\vk) &= \int \frac{\dd^3\vk_1 \dd^3\vk_2}{(2\pi)^3}\, \delta_D(\vk -\vk_1-\vk_2) 
\underbrace{\left(\frac{5}{7} \alpha(\vk_1,\vk_2) + \frac{2}{7}\beta(\vk_1,\vk_2)\right)}_{F_2(\vk_1, \vk_2)} 
\delta_0(\vk_1) \delta_0(\vk_2) \label{eq:delta2},\\
\tilde\theta^{(2)}(\vk) &= \int \frac{\dd^3\vk_1 \dd^3\vk_2}{(2\pi)^3}\, \delta_D(\vk -\vk_1-\vk_2)
\underbrace{\left(\frac{3}{7} \alpha(\vk_1,\vk_2)  + \frac{4}{7}\beta(\vk_1,\vk_2)\right)}_{ G_2(\vk_1,\vk_2)} 
 \delta_0(\vk_1)\delta_0(\vk_2) \label{eq:theta2},
\end{align}
where $F_2$ and $G_2$ are the second-order perturbation theory kernels built from the mode coupling functions $\alpha$ and $\beta$ defined in equation~\eqref{eq:SPT_alphabeta}. We can see that they are symmetric in the set of $\vk_1$ and $\vk_2$ (and thus homogeneous) and their degree is zero. We also notice that the kernels only depend on the relative angles encoded in $\vk_1\cdot \vk_2/(k_1k_2)$ and the ratio $k_1/k_2$, and if they are antiparallel then $F_2(\vk,-\vk)=0=G_2(\vk,-\vk)$, which will be useful later.

\paragraph{Eulerian Nonlinear Perturbation Theory: Beyond second order}
For an EdS universe and the fastest growing mode only, the Fourier space density contrast and the velocity divergence can in fact be written as a series with powers of the scale factor $a(\eta)$ 
\begin{equation}
\delta(\vk,\eta) = \sum_{n=1}^{\infty} a^n(\eta)\tilde\delta_n(\vk),\ \ \ \ \ 
\theta(\vk,\eta) = - {\cal H}(\eta) \sum_{n=1}^{\infty} a^n(\eta) \tilde\theta_n(\vk)
\label{ptansatz}\,.
\end{equation}
Remarkably it implies that the PT expansions are actually expansions with respect to the linear density field with time-independent coefficients. At small $a$ the series are dominated by their first term, and since $\tilde\theta_1(\vk) =-\tilde\delta_1(\vk) $ from the continuity equation, $\tilde\delta_1(\vk)$ completely characterizes the linear fluctuations. Similarly as done for second order, we can find solutions to higher orders by plugging in the lower order solution, which leads to a solution of the form
\begin{equation}
\tilde\delta_n(\vk) = \frac{\int \dd^3\vk_1 \ldots\dd^3\vk_n}{(2\pi)^{3(n-1)}}\, \delta_D(\vk -\vk_1\ldots -\vk_n) F_n(\vk_1, \ldots ,\vk_n) 
\tilde \delta_1(\vk_1) \ldots \tilde\delta_1(\vk_n) \label{ec:deltan},
\end{equation}
\begin{equation}
\tilde\theta_n(\vk) = \int \frac{\dd^3\vk_1 \ldots \dd^3\vk_n}{(2\pi)^{3(n-1)}}\, \delta_D(\vk -
\vk_1\ldots -\vk_n) G_n(\vk_1, \ldots ,\vk_n) \tilde\delta_1(\vk_1) \ldots \tilde \delta_1(\vk_n) \label{ec:thetan},
\end{equation}
where $F_n$ and $G_n$ are the perturbation kernels, which are homogeneous functions of the wave vectors \{$\vk_1, \ldots ,\vk_n $\} with degree zero. They are constructed from the fundamental mode coupling functions $\alpha(\vk_1, \vk_2)$ and $\beta(\vk_1, \vk_2)$ defined in equation~\eqref{eq:SPT_alphabeta} according to the recursion relations ($n \geq 2$, see~\cite{Goroff1986,JainBertschinger1994} for a derivation and equations~(43)-(44) in \cite{BernardeauReview2002}). We will show in Section~\ref{sec:SummaryStats} how those results can be used to compute the matter power spectrum, skewness and bispectrum. As an alternative, GridSPT can be used to generated perturbative density fields on a grid from which arbitrary summary statistics can be measured \cite{GridSPT}.
%
%
%

\subsubsection{Lagrangian Perturbation Theory}
\label{subsec:LPT}
\begin{figure}
\centering
\includegraphics[width=0.48\textwidth]{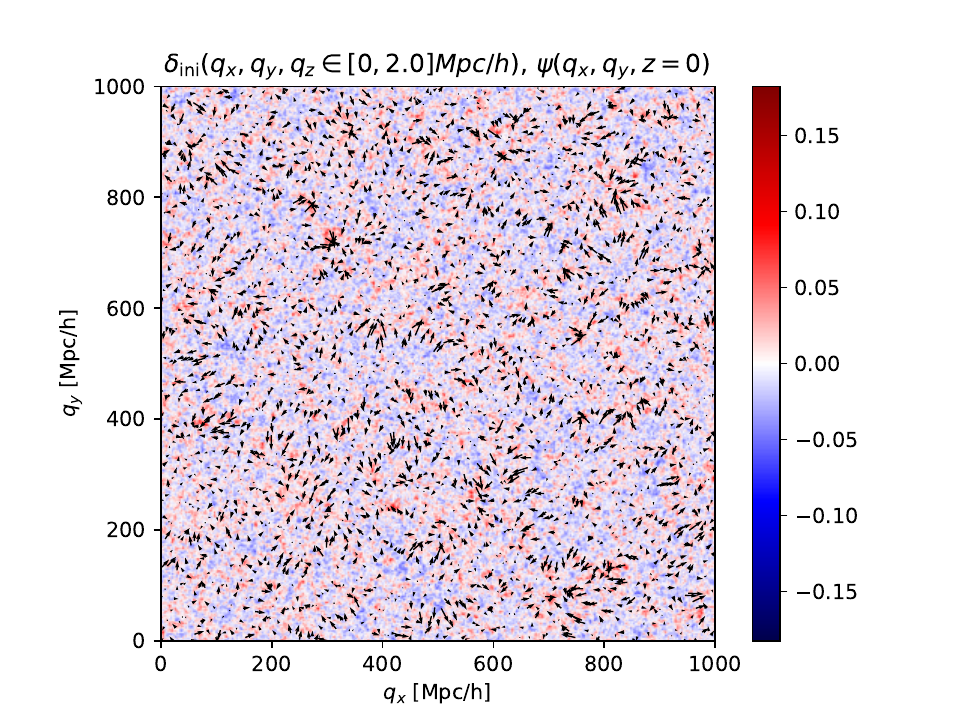}
\includegraphics[width=0.48\textwidth]{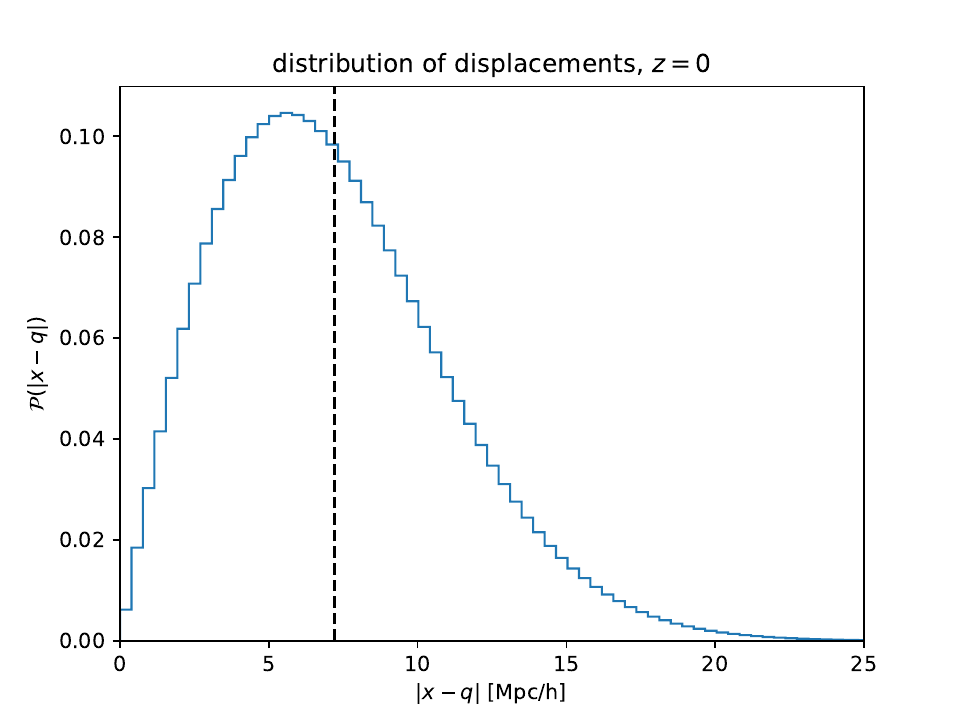}
\caption{(Left) Illustration of the initial density field at the starting redshift $z=127$ in a thin slice along with arrows indicating the matter displacement vector field $\v{\Psi}$ to redshift $z=0$ from the Quijote simulations for realisation 0 of the fiducial cosmology. (Right) Distribution of the absolute values of the displacement field $|\v{\Psi}|$.}
\label{fig:displacement_dist}
\end{figure}

In the Lagrangian framework the quantity under consideration is the displacement field $\v{\Psi}(\vq,\eta)$ which maps initial particle (or fluid element) positions $\vq$ into their final Eulerian position $\vx(\eta)$
\begin{equation}
\vx(\vq,\eta)=\vq+\v{\Psi}(\vq,\eta)\,.
\end{equation}
Figure~\ref{fig:displacement_dist} shows an example for the measured distribution of particle displacements in a simulation, highlighting that the mean displacement is below 10 Mpc$/h$.
The Jacobian $F_{ij}$ of the transformation from Eulerian to Lagrangian coordinates is given by
\begin{equation}
F_{ij}=\frac{\partial x_i}{\partial q_j}=\delta_{ij}+\Psi_{i,j}\,,\quad
J_F=\det\left[\delta_{ij}+\Psi_{i,j}\right] \,,
\end{equation}
where $\Psi_{i,j}$ is short hand notation for $\partial \Psi_i/\partial q_j$.
Mass conservation and the absence of decaying modes imply the following relations between the Jacobian determinant $J_F(\vq)$ and the density contrast
\begin{equation}
\label{eq:masscons}
[1+\delta(\vx)]\dd^3 x = \dd^3 q \quad \Leftrightarrow \quad 1+\delta = J_F^{-1}\,.
\end{equation}
Note that the relationship between the Eulerian density and the Lagrangian displacement field that determines the Jacobian determinant is nonlinear. This means that $n$-th order Lagrangian Perturbation Theory is not equivalent to $n$-th order Eulerian Perturbation Theory, unless this relationship is expanded perturbatively as well. In fact, Lagrangian perturbation theory turns out to be more effective in certain cases, in particular for predicting the correlation function around the baryon acoustic oscillation scale as illustrated in Figure~\ref{fig:ZA}.

The equation of motion for the Eulerian position $\vx$ can be used to obtain a closed-form equation for the displacement field $\v{\Psi}$
\begin{equation}
\label{eq:eom_Lagrangian}
\vx''+\sH\vx'=-\vnabla_x V 
\quad \Rightarrow \quad
\vnabla_x\cdot\left[\partial_\eta^2\v{\Psi}(\vq,\eta)+\sH \partial_\eta\v{\Psi}(\vq,\eta)\right]=\frac{3}{2}\sH^2\left[1-J_F^{-1}(\vq,\eta)\right]\,,
\end{equation}
which is supplemented by the irrotational constraint $\vnabla_x\times\vv=0$. The latter can be converted to Lagrangian coordinates by recalling that $\vv = \vx'=\partial_\eta\v{\Psi}(\vq)$. This equation is still not fully in Lagrangian space, as it involves Eulerian gradients, which can be converted to Lagrangian space using the chain rule
\begin{equation}
\label{eq:nablaEtoL}
\nabla_{x_i}=\frac{\partial}{\partial x_i} =\sum_j \frac{\partial q_j}{\partial x_i} \frac{\partial}{\partial q_j}= (F^{-1})_{ij}\nabla_{q_j}\,,
\end{equation}
where we require the inverse of the Jacobian.

\paragraph{Zeldovich approximation (ZA)}
Let us linearise equation~\eqref{eq:eom_Lagrangian} together with \eqref{eq:nablaEtoL}.
\begin{equation}
\label{eq:ZA}
\partial_\eta^2\vnabla_q\cdot\v{\Psi}^{(1)}(\vq,\eta)+\sH \partial_\eta\vnabla_q\cdot\v{\Psi}^{(1)}(\vq,\eta)=\frac{3}{2}\sH^2 \vnabla_q\cdot \v{\Psi}^{(1)}\,,
\end{equation}
where we can separate time and spatial dependence by writing $\vnabla_q\cdot\v{\Psi}^{(1)}(\vq,\eta)=D(\eta)\tilde\Psi_1(\vq)$ where the linear growth factor fulfills equation~\eqref{eq:linear_growth} and $\tilde\delta_1(\vq)$ is the density contrast field imposed by the initial conditions. At linear order, the displacement field is irrotational, such that it is fully defined by its divergence. The ZA \cite{Zeldovich1970} corresponds to \emph{ballistic motion} according to which particles (or fluid elements) move on straight lines with their initial velocity.  
A particularly nice feature of the ZA is that all statistical properties of $\v{\Psi}$ are inherited from the linear density field $\delta_L(\vq,z_i)$ which we assume to be a Gaussian random field. In Figure~\ref{fig:displacement_dist} we can see that the nonlinear displacement field has a non-Gaussian distribution, but with the ZA we can obtain a resonable Gaussian approximation.

Note that LPT correctly recovers SPT when the exact relation between the Eulerian density and Lagrangian displacement field, encoded in the continuity equation  is expanded in the same manner. The local Eulerian density field reads
\begin{equation}
        1+\delta(\vx(\vq,\eta),\eta) = J_F^{-1} \approx \left(1+ \vnabla_q\cdot \v{\Psi}^{(1)}\right)^{-1} \approx 1-\vnabla_q\cdot \v{\Psi}^{(1)}(\vq)
\end{equation}
such that from comparison from the linear Eulerian perturbation theory we can deduce that $\tilde\Psi_1(\vq)=-\delta_1(\vq)=-\nabla^2\varphi_{\rm ini}(\vq)$ meaning the displacements are given by the gradient of the initial gravitational potential. 
But in the ZA it is possible to keep this nonperturbative relation between the displacement field and the density contrast. This allows to partially resum perturbation theory in a physically motivated way by combining (i) an approximate law for the motion of particles from first order LPT with (ii) a proper determination of the density within a small volume as the sum of all particles divided by this volume. We can also write the Eulerian density as
\begin{equation}
        1+\delta(\vx(\vq,\eta),\eta) = J_F^{-1} = \frac{1}{[1- \lambda_{1} D(\eta)][1- \lambda_{2}D(\eta)][1- \lambda_{3}D(\eta)]} \,,
        \label{eq:dlag2}
\end{equation}
where $\lambda_{i}$ are the local eigenvalues of the Lagrangian tidal tensor $\tilde\Psi_{i,j}$. The relative size of these eigenvalues determines the type of collapse
\begin{itemize}
\item planar collapse occurs if one of the eigenvalues is positive and larger than the others,
\item filamentary collapse occurs if two equal positive eigenvalues are larger than the third,
\item spherical collapse occurs if all eigenvalues are positive and equal such that 
\begin{equation}
1+\delta(\vx(\vq,\eta),\eta) = \frac{1}{1-D(\eta)\delta(\vq)}=\left(1-\delta_L(\vq,\eta)\right)^{-1}\,.
\end{equation}
\end{itemize} 

\paragraph{Second order Lagrangian Perturbation Theory (2LPT)}
To determine the second order solution, we need to expand the inverse Jacobian and Jacobian determinant. Let us start with an expansion of the Jacobian matrix and determinant
\begin{align}
F_{ij} &\approx \delta_{ij}+\Psi^{(1)}_{i,j}+\Psi^{(2)}_{i,j}\\
 J_F\approx &\det\left(\delta_{ij}+\Psi^{(1)}_{i,j}+\Psi^{(2)}_{i,j}\right) 
\approx  1+ \Psi_{i,i}^{(1)} + \Psi_{i,i}^{(2)} +\frac{1}{2} \left(\Psi^{(1)}_{i,i} \Psi^{(1)}_{j,j} - \Psi^{(1)}_{i,j} \Psi^{(1)}_{j,i} \right)\,,
\end{align}
where we used the Einstein summation convention where repeated indices are summed over.
The inverse of the Jacobian multiplies terms of the displacement on the left in equation~\eqref{eq:eom_Lagrangian}, so we need it only at first order
\begin{align}
F_{ij}^{-1} &\approx \delta_{ij}-\Psi^{(1)}_{i,j}\,.
\end{align}
The inverse of the Jacobian appears on the right, so we need it at second order
\begin{align}
J_F^{-1} & \approx 1 - (J_F-1) +  (J_F-1)^2\\
&\approx  1- \Psi_{i,i}^{(1)} - \Psi^{(2)}_{i,i} -\frac{1}{2} \left(\Psi^{(1)}_{i,i} \Psi^{(1)}_{j,j} - \Psi^{(1)}_{i,j} \Psi^{(1)}_{j,i} \right) + \Psi^{(1)}_{i,i} \Psi^{(1)}_{j,j} 
\end{align}
Let us plug this into the equation and collect all second order terms (the first order terms we have already solved for) 
\begin{align}
&\partial_\eta^2\Psi_{i,i}^{(2)}+\sH \partial_\eta\Psi_{i,i}^{(2)} -\Psi^{(1)}_{i,j}\underbrace{\left[\partial_\eta^2\Psi_{i,j}^{(1)}+\sH \partial_\eta\Psi_{i,j}^{(1)}\right]}_{\frac{3}{2}\sH^2 \Psi_{i,j}^{(1)}}=\frac{3}{2}\sH^2\left[\Psi_{i,i}^{(2)} -\frac{1}{2} \left(\Psi^{(1)}_{i,i} \Psi^{(1)}_{j,j} + \Psi^{(1)}_{i,j} \Psi^{(1)}_{j,i} \right) \right]\,,\\
&\left(\partial_\eta^2+\sH \partial_\eta-\frac{3}{2}\sH^2\right)\Psi_{i,i}^{(2)} =\frac{3}{2}\sH^2\left[ -\frac{1}{2} \left(\Psi^{(1)}_{i,i} \Psi^{(1)}_{j,j} - \Psi^{(1)}_{i,j} \Psi^{(1)}_{j,i} \right) \right]\,.
\end{align}
We can see that we can separate the time and space dependence as $\Psi_{i,i}^{(2)}(\vq,\eta)=D_2(\eta) \tilde \psi_{i,i}^{(2)}(\vq)$. The time dependence we can determine from
\begin{equation}
D_2''(\eta)+\sH D_2'(\eta)-\frac{3}{2}\sH^2D_2 =\frac{3}{2}\sH^2D^2\,,
\end{equation}
where we find $D_2=cD^2$ and with EdS where $f=1$ and thus
\begin{equation}
c[2\underbrace{D'(\eta)^2}_{(\sH D)^2}+\underbrace{2DD''(\eta)+ 2\sH DD'(\eta)}_{3\sH^2D}-\frac{3}{2}\sH^2D^2] = \frac{3}{2}\sH^2D^2\quad \Rightarrow c= \frac{3}{7}\,.
\end{equation}
The spatial dependence is given by
\begin{equation}
\tilde \psi_{i,i}^{(2)}(\vq)=-\frac{1}{2} \left(\tilde\Psi^{(1)}_{i,i} \tilde\Psi^{(1)}_{j,j} - \tilde\Psi^{(1)}_{i,j} \tilde\Psi^{(1)}_{j,i} \right)\,.
\end{equation}
Recalling that the first order displacement field was given by the gradient of the initial gravitational potential, we can see that the second order displacement field is in fact sensitive to \emph{tidal terms}. We can actually make a similary to the second order SPT solution, where we already computed the solution for the density and velocity divergence. Given in EdS the linear order velocity divergence has an time-dependence of $\sH D=\dot a a = a^{3/2}$ and the higher order come with extra powers of the linear growth, we could redefine a velocity as $\theta=a^{3/2}\hat \theta$. The conformal time derivative would then be $\partial\eta\theta=a^{3/2}\partial\eta\hat\theta + \frac{3}{2}\sH a^{3/2}\hat\theta$ and we would obtain a Bernoulli equation with a difference $\delta+\hat\theta$. At leading order this vanishes, while at the second order we determined it will be given by 
\begin{equation}
F_2(\vk_1,\vk_2)-G_2(\vk_1,\vk_2)=\frac{2}{7}(\alpha(\vk_1,\vk_2)-\beta(\vk_1,\vk_2)) 
=\frac{2}{7}\left( \frac{k_1^2k_2^2-(\vk_1\cdot\vk_2)^2}{k_1^2k_2^2}\right)\,,
\end{equation}
which is the Fourier version of the tidal term, but in Eulerian space.
In 1D those are absent and actually the ZA is exact before shell-crossing, the moment when different initial positions $\vq_i$ are mapped to the same Eulerian position $\vx$. 

\paragraph{From Perturbation Theory to fast simulations} Lagrangian Perturbation Theory is not only indispensable for setting initial conditions for cosmological simulations, but can also be used to accelerate simulations by picking a suitable reference frame as done in COLA \cite{COLA2013}, or by expanding around solutions in the Zeldovich approximation as done in FastPM \cite{FastPM2016}, or more recently 2LPT as done in BullFrog \cite{Bullfrog2025}.

\paragraph{Higher orders} In an EdS universe, the exact displacement field can be expanded in a perturbative series with spatial parts and temporal coefficients using the scale factor $a(\eta)$
\begin{equation}
\v{\Psi}(\vq,\eta) = \sum_{n=1}^\infty a^n(\eta) \v{\Psi}^{(n)} (\vq)\,.
\end{equation}
Again, we can express the different orders in Fourier space with the help of perturbative kerneles $\v{L}^{(n)}$ in terms of powers of the linear density field $\delta_L$
\begin{equation}
\v{\Psi}(\vk) = i\int \frac{\dd^3p_1\cdots \dd^3p_n}{(2\pi)^{3(n-1)}} \delta_D\left(\vk-\sum_i \vp_i\right) \v{L}^{(n)}(\vp_1,\ldots,\vp_n) \delta_L(\vp_1)\cdots \delta_L(\vp_n)
\end{equation}
Although a separation ansatz with $a\rightarrow D$ on a $\Lambda$CDM background is not an exact solution to equation \eqref{eq:eom_Lagrangian} anymore, it is still a reasonably good approximation to use the time-independent kernels \cite{BernardeauReview2002}.

\paragraph{Relation between Eulerian and Lagrangian picture} 
In the following we will briefly describe the mapping between the Eulerian framework based on the Eulerian density contrast $\delta(\vx)$ and velocity divergence $\theta(\vx)$ and the Lagrangian description relying on the displacement field $\v{\Psi}(\vq)$. A more detailed description of the connection between the series in Lagrangian and Eulerian perturbation theory can be found in \cite{RampfBuchert2012}. Employing mass conservation~\eqref{eq:masscons} we can write the Eulerian density in terms of the displacement field
\begin{equation}
1+\delta(\vx) = \int \dd^3 q \delta_D\left(\vx-\vq-\v{\Psi}(\vq)\right)\,.
\end{equation}
Rewriting the density contrast in Fourier space and gives 
\begin{equation}
\delta(\vk) = \int \dd^3x \exp(-i\vk\cdot\vx) \delta(\vx) = \int \dd^3q \exp(-i\vk\cdot\vq) \left[\exp\left(-i\vk\cdot\v{\Psi}(\vq)\right)-1\right]\,.
\end{equation}
Similarly, the velocity divergence $\theta=\vnabla_x\cdot\vv$ can be transformed from Lagrangian space using that $\vv=\vx'=\partial_\eta\v{\Psi}(\vq,\eta)$
\begin{equation}
\frac{\theta(\vk)}{\sH} = \int \dd^3q  \exp\left(-i\vk\cdot(\vq+\v{\Psi}(\vq)\right) J_F \frac{\theta(\vx(\vq))}{\sH} \,.
\end{equation}
Note that within the mapping from the Eulerian to the Lagrangian frame the absence of Eulerian vorticity $\vnabla_x\times \vv(\vx) =0$ implies a constraint for the transverse parts of $\v{\Psi}$.

\subsection{Spherical collapse}
\label{sec:SC}As the name suggests, the spherical collapse model describes the evolution of a spherically symmetric overdense region. In an Einstein-de Sitter universe, which is a flat, matter-dominated cosmological model this can be easily solved. To arrive at a solution, we consider the mass $M$ within the spherical region of uniform density $\rho_i$ and initial radius $R_0$ as $M(R_0) = \frac{4}{3} \pi R_0^3 \rho_i$.

The equation of motion for the radius is determined by a Newtonian force law 
\begin{equation}
\label{eq:SC_R}
\ddot{R} = -\frac{GM(R)}{R^2}= -\frac{4\pi G}{3} \rho(t)R\,,
\end{equation}
where $M(R)$ is the constant mass inisde the shell.  To solve the equation, we can multiply \eqref{eq:SC_R} by $2\dot R$ and integrate once to obtain
\begin{equation}
\dot R^2 = \frac{2GM}{R}-C\,,
\end{equation}
with an integration constant $C$. This is the equation of a cycloid, for which we can find a parametric solution - a solution for which the two variables $R$ and $t$ are given in terms of a parameter $\varphi \in [0,2\pi]$
\begin{equation}
R=GM(1-\cos\varphi)/C\,,\quad t=GM(\varphi-\sin\varphi)/C^{3/2}\,.
\end{equation}

For pressureless matter the background density evolves as 
\begin{equation}
\rho_0(t) =\frac{3M(R_0)}{4\pi (R_0 a(t))^{3}}=\frac{3M(R_0)}{4\pi (R_0 a_0)^{3}}(t_0/t)^{2}\,,
\end{equation}
such that we can define a density contrast in the spherical shell as
\begin{equation}
1+\delta=\frac{\rho}{\rho_0}=\left(\frac{a(t)R_0}{R}\right)^3\,,
\end{equation}
and outside we have $\delta=0$. Since we assumed a uniform density, the density contrast is a top-hat function, which allows to cancel all spatial derivatives. We can now obtain an evolution equation in conformal time
\begin{equation}
\label{eq:delta_ODE_SC}
\delta''+\left(1+\frac{\sH'}{\sH}\right)\delta'-\frac{3}{2}\Omega_m\delta = \frac{4}{3}\frac{\delta'}{1+\delta} + \frac{3}{2}\Omega_m\delta^2\,,
\end{equation}
where on the left hand side we can see the ingredients of the linear equation of motion and nonlinear terms on the right hand side.
We can now translate this result to the nonlinear and linear density contrasts, using the time-evolution for de-Sitter $a(t)=a_0(t/t_0)^{2/3}$
\begin{equation}
\label{eq:SphericalCollapseEdS}
1+\delta = \frac{9(\varphi-\sin\varphi)^2}{2(1-\cos\varphi)^3}\,,\quad
\delta_L = \frac{3}{5} \left[\frac{3}{4}(\varphi-\sin\varphi)\right]^{2/3}\,.
\end{equation}
Here $\delta_L$ is the solution to the linear equation, the left hand side in equation \eqref{eq:delta_ODE_SC} and we assumed an overdensity $\delta_L>0$. 
Since we have assumed mass conservation, our analysis is only valid until shell-crossing.

\paragraph{Critical value for collapse} The evolution of the radius of the overdensity has different stages, first it expands, as small perturbations expand with the cosmological expansion. Then it reaches a turnaround point and then collapses under its own gravity. In this simple model it will reach $R=0$ which is unphysical and due to the assumptions being violated by having a sufficiently large density, shell-crossing will invalidate the treatment as perfect pressureless fluid and non-radial fluctuations will develop. Even the collissionless dark matter will undergo a process called ``violent relaxation'' which will establish virial equilibrium. 

An interesting result from this model is a critical threshold for collapse of the linear fluctuation $\delta_L$. When $\varphi=0$ the perturbations are zero and the density contrast $\delta$ reaches a turnaround at $\varphi=\pi$ for which we have the nonlinear and linear density contrast 
\begin{equation}
\delta_T=\delta(\pi)=(3\pi/4)^2-1\approx 4.55\,,\quad \delta_L(\pi)\approx 1.063\,.
\end{equation}
For the final $\varphi=2\pi$ the nonlinear density contrast becomes singular, which occurs at 
\begin{equation}
\label{eq:delta_c}
\delta(2\pi)\rightarrow\infty \Rightarrow \delta_c= \delta_L(2\pi)=\frac{3}{5} (3\pi/2)^{2/3}\approx 1.686
\end{equation} at a time that is twice the turnaround time. Interestingly, this value is time-independent such that in an EdS universe, a spherical perturbation collapses to a singularity whenever the linear density exceeds this critical threshold $\delta_c$.

\paragraph{Virialisation} Of course, a realistic density perturbation will neither be spherical nor homogeneous, and we know that shell-crossing will occur at some point. In the long run the collapse does not proceed to a point (and hence a singular density), but it reaches  virial equilibrium such that the potential and kinetic energies are related as $U=-2K$ with total energy $E=U+K=U(r_{\rm max})=1$. As $U\propto R^{-1}$ we have $R_{\rm vir} = R_{\rm max}/2$. As $R(\varphi)\propto 1-\cos\varphi$ we had the maximum at $\varphi=\pi$ and thus half the value is found at $\varphi_{\rm vir}=3\pi/2$ after which $R(\varphi)$ should be considered constant. Let us now look at the overdensity at virialisation, obtained from compressing the overdensity at turnaround in half the radius while the background dilutes with $a^{-3}\propto t^{-2}$
\begin{equation}
\frac{(1+\delta(\pi))\cdot 2^3}{(t(\pi)/t(2\pi))^2}=18\pi^2 \approx 178.
\end{equation}
This provides a motivation for a rounded threshold density $\Delta=200$ for defining the mass and radius of a halo, often denoted by $M_{200}$ and $R_{200}$.

\paragraph{Closed-form approximations} For simplicity, one can rely on an approximate explicit parameterisation for spherical collapse in an EdS universe, given by
\begin{equation}
\label{eq:SCnu}
\rho_{{\rm NL},\nu}(\delta_L)=\left(1-\frac{\delta_L}{\nu}\right)^{-\nu}\,,
\end{equation}
where $\delta_L$ is the linear density at redshift zero. The parameter $\nu$ controls the amplitude of the skewness before smoothing $3(1+1/\nu)$ and can be matched to the prediction in equation~\eqref{eq:S3pred}, yielding $\nu=21/13$. Originally, this parametric form has been suggested in \cite{Bernardeau1994}, with approximately $\nu=1.5$, which becomes exact for $\Lambda$=0 in the limit of $\Omega_m\rightarrow 0$ and drives the shape of the PDF in low density regions. In excursion-set inspired models, usually the critical linear density for collapse is used, setting $\nu=\delta_c=1.686$ as advocated in \cite{LamSheth2008}. We can also see that the Zeldovich approximation (which is exact in 1D) corresponds to $\nu_{\rm ZA}=1$, which suggests a dependence on the dimensionality of the collapse. Indeed in cylindrical collapse, $\nu_{\rm 2D}\approx 1.4$ is a good approximation \cite{Bernardeau1995} and useful to predict angular statistics related to photometric clustering and weak lensing. 

\begin{figure}
\centering
\includegraphics[width=0.5\textwidth]{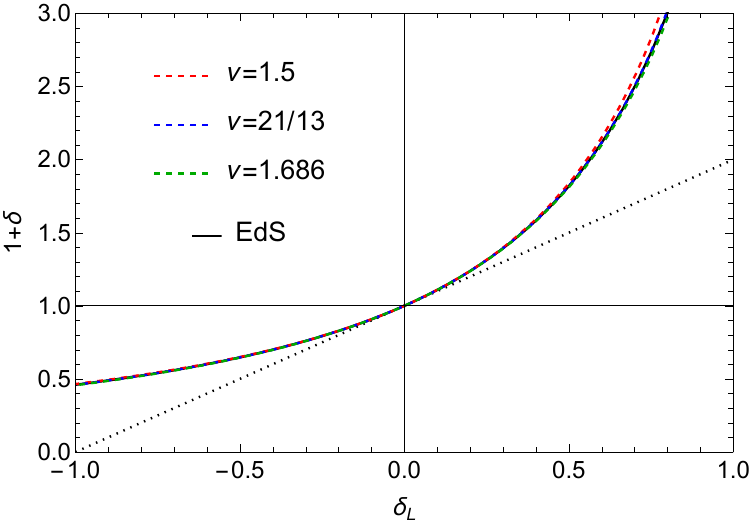}
\caption{Relationship between the linear density contrast $\delta_L$ and the nonlinear normalised density $1+\delta$, showing the parameteric spherical collapse solution for Einstein-deSitter (black solid line) and linear theory (black dotted line) with its underestimation of overdensities and overestimation of underdensities. Also shown are simple power law approximations with different power-law indices $\nu$ (coloured dashed lines), with the simplest one being $\nu=3/2$ (red dashed), the one matching the skewness $\nu=21/13$ (blue dashed) and the one matching the diverging density $\nu=\delta_c=1.686$ (green dashed). Residuals of those approximations with respect to the full $\Lambda$CDM spherical collapse solution can be found in Figure~2 in \cite{Uhlemann2020}.}
\label{fig:SCcomparison}
\end{figure}

In Figure~\ref{fig:SCcomparison}, we compare the parametric EdS form~\eqref{eq:SphericalCollapseEdS} (black line) to the parametrisations~\eqref{eq:SCnu} with different $\nu$ (blue, green and red lines). Note that the parametric solution for EdS approximates the numerical $\Lambda$CDM solution extremely well, with sub-percent residuals in the range of relevant densities $\rho\in [0.1,10]$ \cite{Uhlemann2020}. 

\subsection{Halo mass function and halo model}

One of the reasons why spherical collapse is useful is that it provides a way to estimate the number of collapsed objects that are formed in a Gaussian random field by simply counting at any given time how many regions have an overdensity above the collapse threshold given by $\delta_c$. At a given redshift $z$, we can smooth the Gaussian linear density field with a spherical top-hat filter of radius $R$ (each containing on average the mass $M=\frac{4\pi}{3}\rho_0 R^3$ according to the background density). Smoothing is a linear operation, so a Gaussian field will remain Gaussian and the variance is changed to $\sigma^2_M(z)$. If we now assume all cells with $\delta>\delta_c$ undergo collapse and virialisation, we can estimate the fraction of collapsed regions above mass M as
\begin{equation}
\mathcal P(M,z)_{\delta>\delta_c} = \frac{1}{\sqrt{2\pi\sigma_M^2(z)}}\int_{\delta_c}^\infty  \exp\left(-\frac{\delta_M^2}{2\sigma_M^2(z)}\right)\, \dd\delta_M
= \frac{1}{2}\rm{erfc}\left(\frac{\delta_c}{\sqrt 2\sigma_M(z)}\right)\,.
\end{equation}
As the error function approaches 1 for large arguments, in the limit where we send the threshold mass to zero, still only half of the regions will collapse, because only the overdensities do so. If we want all the masses to end up in some object, we need to insert an additional factor of two into this oversimplified model. We can now look at the fraction of objects in a given mass bin, so between $M$ and $M+dM$
\begin{align}
d\mathcal P(M) &= \left|\frac{\partial \mathcal P(M,z)_{\delta>\delta_c} }{\partial M}\right| dM 
=  \frac{1}{\sqrt{\pi}}\exp\left(-\frac{\delta_c^2}{2\sigma_M^2(z)}\right) \frac{\delta_c}{\sqrt{2}} \left(\frac{-1}{\sigma_M(z)^2}\right)\frac{\partial \sigma_M(z)}{\partial M}\\
&=  \frac{1}{M} \underbrace{\frac{1}{\sqrt{2\pi}}\exp\left(-\frac{\delta_c^2}{2\sigma_M^2(z)}\right) \frac{\delta_c}{\sigma_M(z)}}_{ f(\nu)} \left(-\frac{\partial \ln\sigma_M(z)}{\partial\ln M}\right)
\end{align}
Although the regions above the critical threshold for collapse are certainly not in the linear regime, we want to use linear theory to estimate which fraction of regions have collapsed. We thus implicitly assume that the variance $\sigma_M(z)\ll 1$ is in the linear regime, and can be computed from the linear power spectrum, which grows with the linear growth factor $\sigma_M(z)=D(z)\sigma_M(z=0)$.

\begin{figure}
\centering
\includegraphics[width=0.75\textwidth]{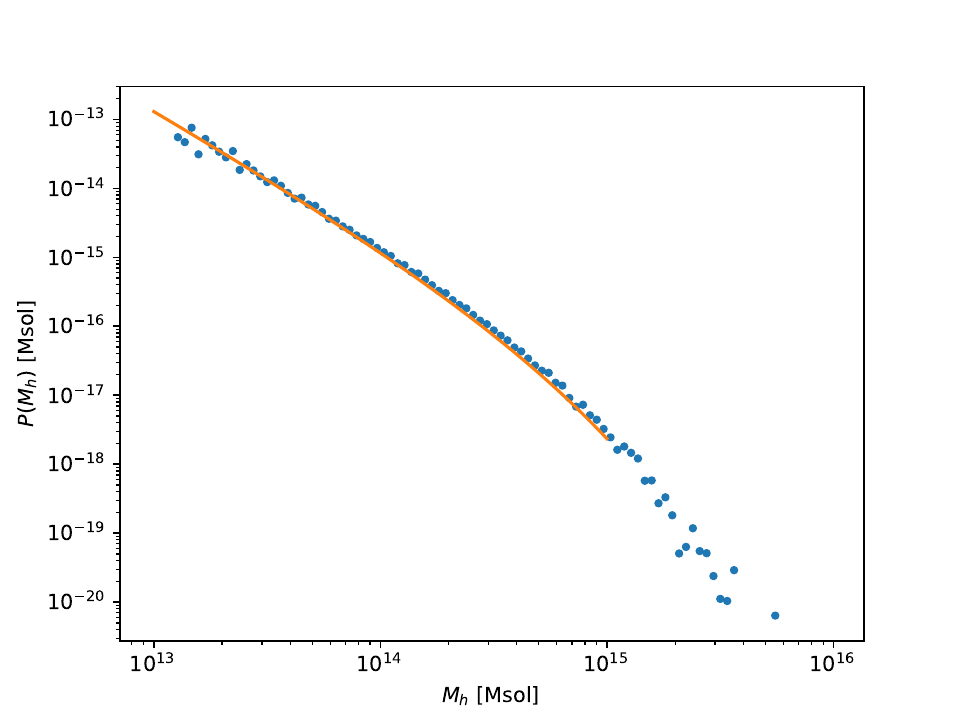}
\caption{Double-logarithmic plot of the distribution of halo masses in the simulation (data points), normalised such that the integral is one, in comparison to the theoretical Sheth-Tormen mass function (orange line).}
\label{fig:HMF}
\end{figure}

The volume occupied by single collapsed object is $V_M=M/\rho$, so if we have $N$ objects it is $NV_M$ and or in terms of the occupied volume fraction $d \mathcal P V=NV_M$. We can easily solve this for the number density $dn$ of collapsed objects in mass bin of width $dM$ as
\begin{equation}
dn=\frac{N}{V}=\frac{d \mathcal P}{V_M}=\frac{\rho}{M} \left|\frac{\partial \mathcal P(M,z)_{\delta>\delta_c} }{\partial M}\right| dM\,.
\end{equation}
If we now integrate over all masses, we find that only half of the regions (namely the overdense ones) end up on a bound structure. This is a flaw of the model that is remedied in refined models of the gravitational collapse. As a quick ad-hoc fix, we can introduce a fudge factor of 2 for normalisation to define $f_{\rm PS}$ or look at an extension -- the Sheth-Tormen mass function
\begin{equation}
f_{\rm PS}(\nu)=\sqrt{\frac{2}{\pi}}\nu\exp\left(-\frac{\nu^2}{2}\right)\,,\quad
f_{\rm ST}(\nu)=A\sqrt{\frac{2a}{\pi}}\nu\left[1+\left(\frac{1}{a\nu^2}\right)^p\right]\exp\left(-a\frac{\nu^2}{2}\right)\,,
\label{eq:massfunc}
\end{equation}
where $\nu=\delta_c/\sigma$ and the parameters are calibrated against simulations $A=0.3222$, $a=0.707$, $p=0.3$. Figure~\ref{fig:HMF} illustrates its comparison to a measured mass function from the Quijote simulation suite. From a given halo mass function, one can also compute approximations for the bias of dark matter halos, examples of which are given in equation~\eqref{eq:halobias_PS_ST}.

In the \emph{halo model} \cite{CooraySheth2002,Asgari2023}, all matter is assumed to reside in gravitationally bound dark matter halos of different masses. Each halo is a virialised
object containing a certain mass of dark matter, and the entire matter distribution can be built up by considering the population of halos and their properties, including their characteristic density profiles. This allows predictions for multiple clustering statistics including power spectra reviewed in Pedro Ferreira's lecture notes. Many fitting formulas for the power spectrum (e.g. halofit \cite{Halofit2003} and its extensions \cite{TakahashiHalofit2012,MeadHalofit2015}) and bispectrum (e.g. bihalofit \cite{TakahashiBihalofit2020}) are based on the halo model with additional input from $N$-body simulations, providing an indispensable tool in particular for predicting weak lensing observables that are highly sensitive to nonlinearities and difficult to predict from perturbation theory.

\section{Clustering statistics}
\label{sec:SummaryStats}
While galaxy survey data consists of maps of galaxy positions and shapes, performing cosmological inference using their full 3-dimensional distribution is computationally challenging, as outlined in the lecture notes for field-level inference by Florent Leclercq. The standard analysis consists of compressing the catalogs down to the two-point correlation function or the power spectrum, which would be optimal for a Gaussian field, which is a good approximation at early times and in particular at the epoch of the CMB. Nonlinearities in the gravitational clustering induce non-Gaussianity which leaks information into beyond 2-point statistics, as we will demonstrate using two exemplary examples consisting of the one-point distribution and its skewness, and the three-point correlation function or bispectrum.

\subsection{Statistics of the Gaussian initial density field}
If the initial density field is Gaussian (as we will assume in most of our lectures), then the initial one-point distribution function of the density contrast $\delta_0(R)$ smoothed scale $R$ will be zero-mean Gaussian
\begin{equation}
\mathcal P(\delta_0(R),R) =\frac{1}{\sqrt{2\pi\sigma^2(R)}} \exp\left(-\frac{\delta_0(R)^2}{2\sigma^2(R)}\right)\,,
\end{equation}
and thus fully determined by the variance $\sigma^2(R)$. Note that since smoothing is a linear operation, a Gaussian field will remain Gaussian and just its variance will change. The corresponding moments are given by
\begin{align}
\langle(\delta_0(R))^2(\vx)\rangle&=\int \dd\delta_0 \delta_0^2 \mathcal P(\delta_0,R)=\sigma^2(R) \\
\langle(\delta_0(R))^{2n+1}(\vx)\rangle&=0 \quad \forall n\in\mathbb{N}\\
\langle(\delta_0(R))^{2n}(\vx)\rangle&=(2n-1)!! \sigma^{2n}(R) \quad \forall n\in\mathbb{N}\,,
\end{align}
where we see that all odd moments vanished and for the even moments we used the double factorial defined as a product of the subsequent terms in steps of two $n!!=n(n-2)(n-4)\cdot ... $. For example, the 4th moment is given by $\langle(\delta_0(R))^4(\vx)\rangle =3\sigma^4(R)$.

The Fourier transform is linear such that the Fourier transformed density contrast is also a Gaussian random field, although it will be complex. Since the density contrast in physical space is real, the Fourier-space density contrast is Hermitian $\delta^*(-\vk) = \delta(\vk)$. Similarly to the one-point distribution, all odd Fourier-space correlators vanish and all even ones can be written as sums of products of power spectra according to Wick's theorem. So explicitly we have for the third order (initial bispectrum) and the fourth order (initial trispectrum)
\begin{align}
\label{eq:GaussianICs}
\langle \delta_0(\vk_1)\delta_0(\vk_2)\rangle = & (2\pi)^3 \delta_D(\vk_1+\vk_2) P_0(k_1) &\\
\langle \delta_0(\vk_1)\delta_0(\vk_2)\delta_0(\vk_3)\rangle = & 0 \\
\langle \delta_0(\vk_1)\delta_0(\vk_2)\delta_0(\vk_3)\delta_0(\vk_4)\rangle = & (2\pi)^6 
\left[\delta_D(\vk_1+\vk_2)\delta_D(\vk_3+\vk_4)P_0(k_1)P_0(k_3) \right.\notag\\
&\qquad + \delta_D(\vk_1+\vk_3)\delta_D(\vk_2+\vk_4)P_0(k_1)P_0(k_2)\\
&\qquad + \left.\delta_D(\vk_1+\vk_4)\delta_D(\vk_2+\vk_3)P_0(k_1)P_0(k_2)\right]\,.\notag
\end{align}
\subsection{Two-point correlations and the power spectrum}

\begin{figure}
\includegraphics[width=0.48\columnwidth]{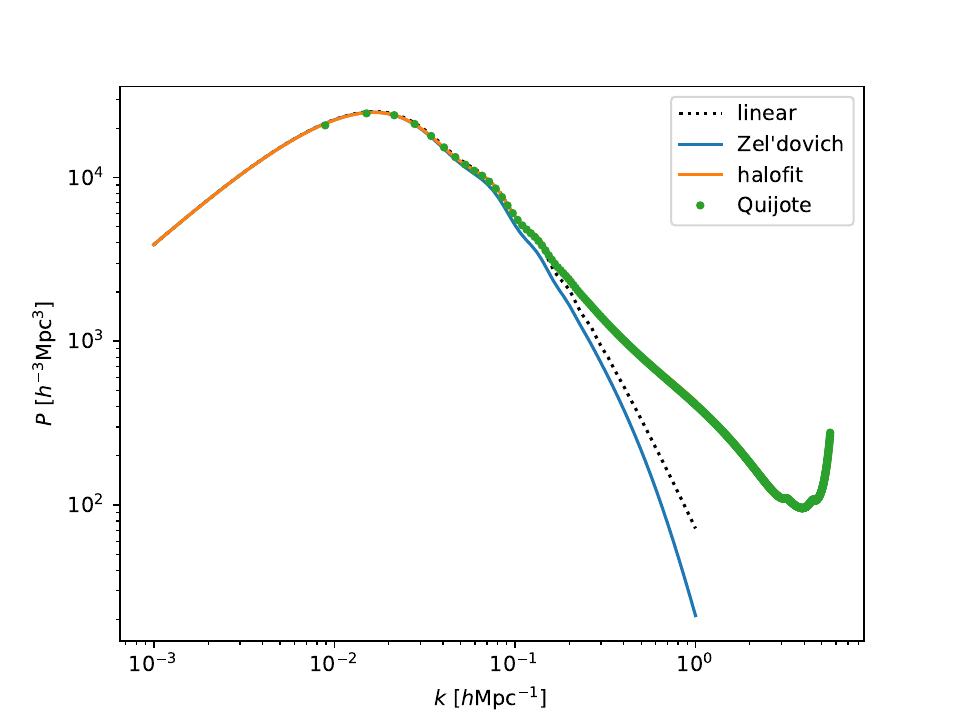}
\includegraphics[width=0.48\columnwidth]{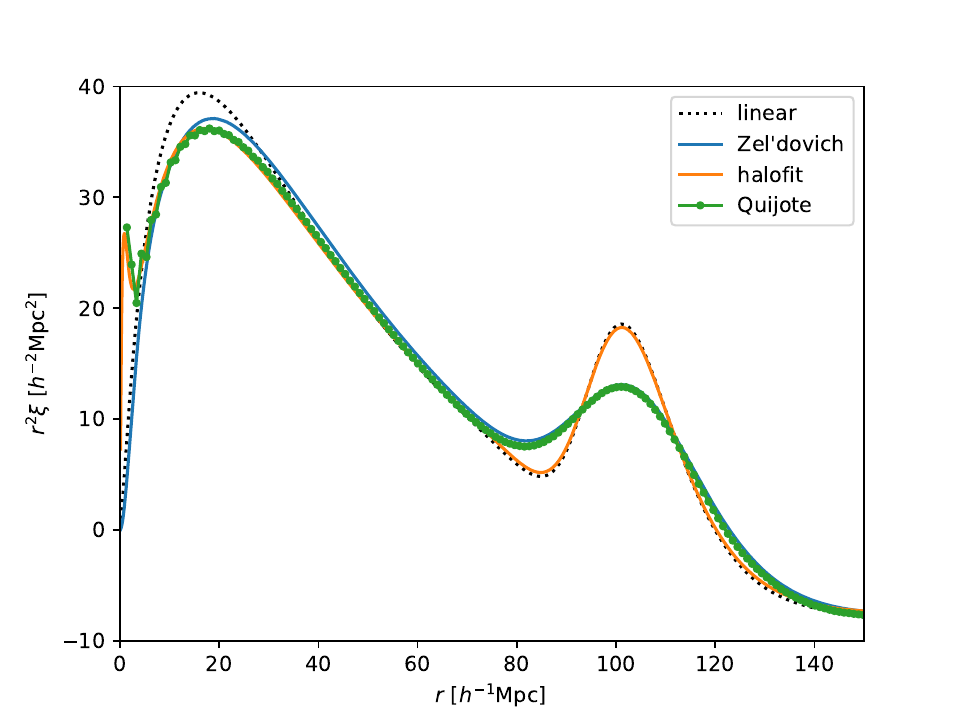}
   \caption{Comparison of the matter power spectrum (left) and matter correlation function (right) for linear theory (black dotted), the Zeldovich approximation (blue) and the halofit power spectrum (orange) and measurements from the Quijote simulations averaged over 1000 realisations (green points). We can see that while the Zeldovich approximation lacks power on small scales, it predicts a broadening around the baryon acoustic oscillation scale that is crucial for a match with simulations.} 
   \label{fig:ZA}
\end{figure}

\paragraph{Correlation functions} Crucial clustering information is encoded in correlation functions which can be determined from galaxy catalogs from redshift surveys or mock catalogs generated from cosmological simulations (for example generated from halo catalogs by using a halo occupation distribution). One simple quantity to characterise the clustering is the two-point correlation function of the tracer $\xi_t$, measuring the excess probability (compared to a random Poisson distribution) of finding a tracer at a distance $r$ from another one. So far, we are working in real space consisting of 3D positions $\vr$
\begin{equation}
\xi_t(r=|\vr_2-\vr_1|)=\langle \delta_t(\vr_1)\delta_t(\vr_2) \rangle\,.
\end{equation}
Due to spatial statistical homogeneity the correlation function is only a function of the separation $\vr=\vr_2-\vr_1$ and due to statistical isotropy the real space correlation function depends only on the distance $r=|\vr|$. Note that the latter property does not hold true for the redshift space correlation functions $\xi_t(\vs=\vs_2-\vs_1)$ due to anisotropies arising from the mapping between real and redshift space that we will discuss in \ref{sec:RSD}. In the right panel of Figure~\ref{fig:ZA} we illustrate the dark matter correlation function, which generally displays that correlations are decaying at large distances, but showing a characteristic peak around $r\approx 100$ Mpc$/h$ that is an imprint of the baryon acoustic oscillations discussed in the CMB lectures.

\paragraph{Power spectra}
Working in Fourier space is natural in particular in perturbation schemes which straightforwardly provide power spectra, the Fourier analog of correlation functions. In principle, the two descriptions are equivalent, although for example survey windows, scale cuts and systematics can affect statistics differently. In Fourier space, modes of different wave-vectors decouple at linear order, which is advantageous for the analytical treatment in SPT. The dark matter power spectra of density $\delta$ and the velocity divergence $\theta$ are defined as
\begin{equation}
\label{eq:powerspectra}
\langle \delta(\vk) \delta^*(\vk')\rangle = (2\pi)^3 \delta_D(\vk-\vk') P_{\delta\delta}(k)\,,\quad
\langle \theta(\vk) \theta^*(\vk')\rangle = (2\pi)^3 \delta_D(\vk-\vk') P_{\theta\theta}(k) \,,
\end{equation} 
and in the absence of vorticity the velocity power spectrum (defined from a scalar product) is given by $k^2P_{vv}(k)=P_{\theta\theta}(k)$. In physical space, the density contrast is real, such that in Fourier space we have $\delta^*(\vk')=\delta(-\vk')$. We can see that the Dirac delta distribution will enforce $\vk=\vk'$. For simplicity we will often directly evaluate correlators of the form $\langle \delta(\vk) \delta^*(\vk)\rangle=\langle \delta(\vk) \delta(-\vk)\rangle$ while suppressing the $\delta_D$ part.

We can obtain the correlation function from the power spectrum (and vice versa)
\begin{equation}
\label{eq:corr_from_PS}
\xi(r) = \int \frac{\dd^3 k}{(2\pi)^3} \exp(-i\vk\cdot \vr) P(k) = \int \dd \ln k \Delta^2(\vk) \frac{\sin(kr)}{kr}\,,
\end{equation}
where the second equality follows from rotational invariance and we defined the dimensionless power spectrum is defined as $\Delta ^{2}(k)\equiv \frac{4\pi k^3P(k)}{(2\pi )^{3}}$. Figure~\ref{fig:ZA} shows the matter power spectrum and correlation function next to each other, highlighthing that the characteristic BAO peak in the correlation function translates to a series of wiggles in the power spectrum. While in principle power spectrum and correlation function contain the same information, theoretical predictions of different flavours have different strengths and weaknesses, as we will see when comparing Eulerian Perturbation Theory at linear and 1-loop order, the Zeldovich approximation as first order Lagrangian Perturbation theory and the nonlinear halofit function.

\begin{figure}
\centering
\includegraphics[width=0.48\columnwidth]{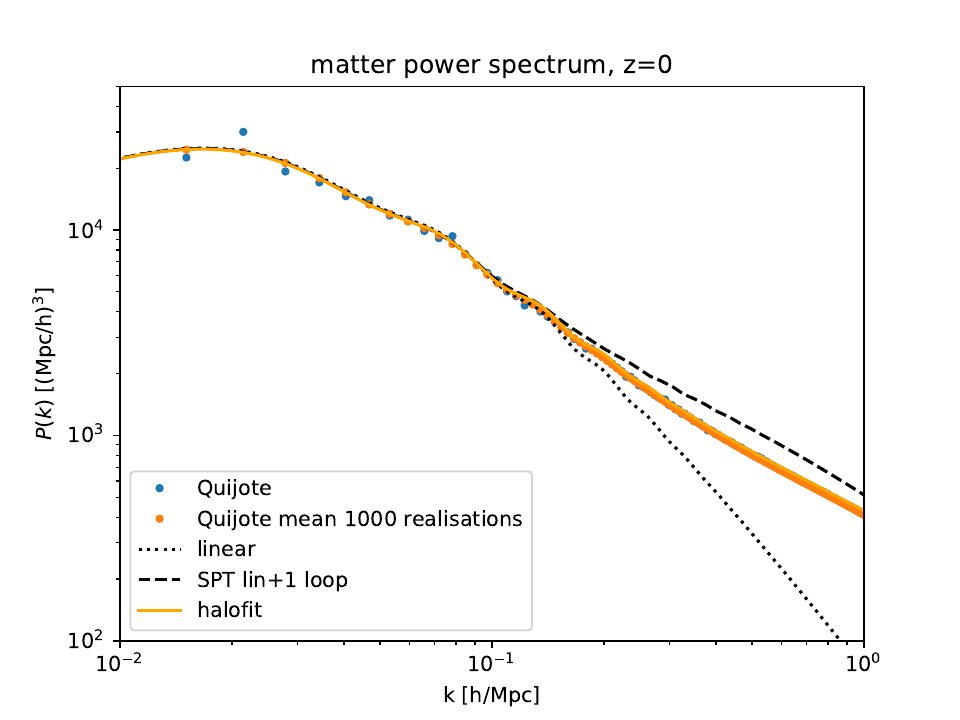}
\includegraphics[width=0.48\columnwidth]{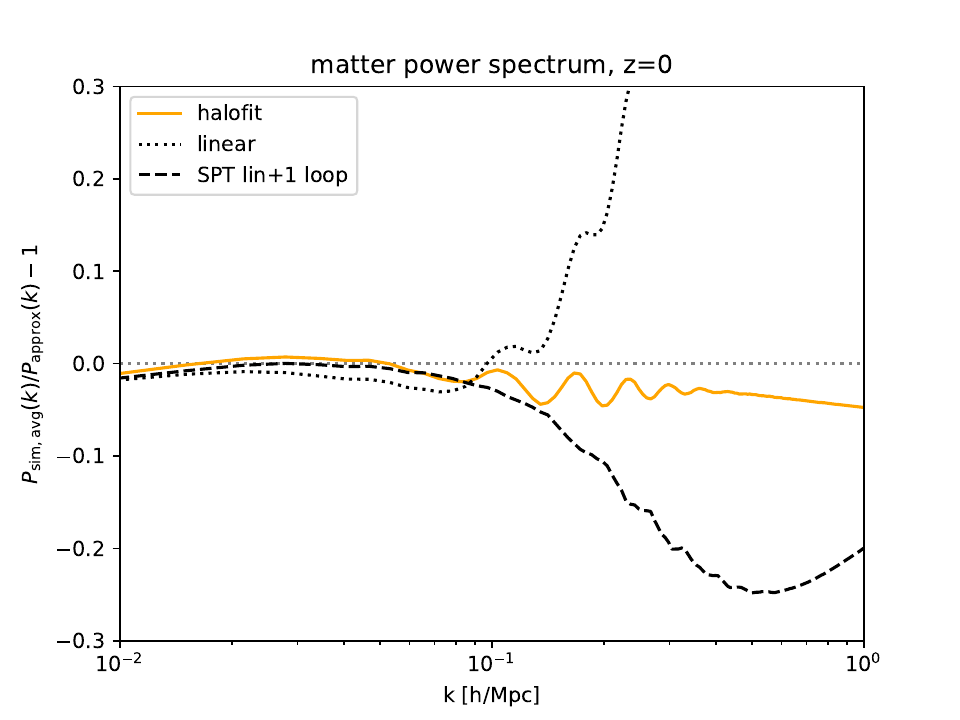}
\caption{(Left) Comparison between the measured nonlinear power spectrum from one realisation of the Quijote simulations (blue points), an average over 1000 realisations (orange points) and the predictions from linear theory (black dotted line), Eulerian Perturbation Theory including the linear and 1-loop contributions (black dashed line) and the empirical halofit fitting formula (orange line). (Right) residuals between the simulated nonlinear power spectrum averaged over 1000 realisations and the different approximations.}
\label{fig:Pk_sim_SPT_Halofit}
\end{figure}

\paragraph{Power spectra from Eulerian Perturbation Theory}
\begin{figure}
\centering
\includegraphics[width=0.5\columnwidth]{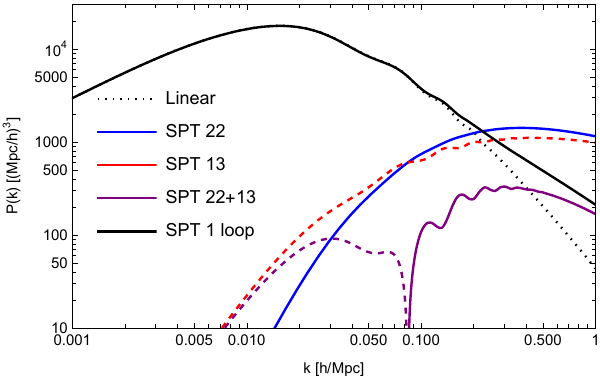}
\caption{Illustration of the predicted linear power spectrum (black dotted) vs. the 1-loop contributions in Eulerian perturbation theory (SPT) consisting of the combination of the positive $P_{22}$ (blue solid) and the negative $P_{13}$ (red dashed) leading to the negative/positive combination (purple dasehd/solid).}
\label{fig:Pk_SPT}
\end{figure}
Using our perturbative solution, we can compute the corresponding power spectra. Let us look at the computation for the matter density power spectrum
\begin{align}
\langle\delta_m(\vk)\delta_m(\vk')\rangle \approx &\langle \left(\delta^{(1)}+\delta^{(2)}+\delta^{(3)}\right)(\vk) \left(\delta^{(1)}+\delta^{(2)}+\delta^{(3)}\right)(\vk') \rangle \\
\approx & \underbrace{\langle \delta^{(1)}(\vk)\delta^{(1)}(\vk')\rangle}_{\rightarrow P_L(k,\eta)} + \underbrace{\langle \delta^{(2)}(\vk)\delta^{(2)}(\vk')\rangle + 2\langle \delta^{(1)}(\vk)\delta^{(3)}(\vk')\rangle}_{\rightarrow P_{\rm NLO}(k,\eta)}\,,
\end{align}
where we collected all terms up to $\langle\delta^{(k)}\delta^{(4-k)}\rangle$ which will be of a similar order, but neglected terms where the total order is larger than $4$. We have seen that the higher perturbative orders are made from convolutions of $n$ initial densities and depend on higher powers of the growth rate $\delta^{(n)}\propto D^n\delta_0^n$. Let us plug in the expression for the second and third order density contrast using the kernels $F_2$ and $F_3$
\begin{align*}
\langle \delta^{(2)}(\vk)\delta^{(2)}(\vk')\rangle &= \int \frac{\dd^3\vk_1 \dd^3\vk_3}{(2\pi)^6}
F_2(\vk_1, \vk-\vk_1) F_2(\vk_3, \vk'-\vk_3)\underbrace{\langle \delta_0(\vk_1) \delta_0(\vk-\vk_1) \delta_0(\vk_3) \delta_0(\vk'-\vk_3)\rangle}_{2\langle \delta_0(\vk_1) \delta_0(\vk_3) \rangle\langle\delta_0(\vk-\vk_1) \delta_0(\vk'-\vk_3)\rangle}
\end{align*}
where we used the property of iniital Gaussian fields~\eqref{eq:GaussianICs} for which there are 3 different pairings of the 4 wavevectors. We need to pair them in a way such that the two paired wave vectors can sum to zero, hence the term including $\langle \delta_0(\vk_1) \delta_0(\vk-\vk_1) \rangle$ vanishes. Also the expression is symmetric in exchanging $\vk_1\leftrightarrow\vk_3$ such that we obtain the same term twice. We obtain the corresponding power spectrum
\begin{align}
P_{22}(k,\eta) 
%
&= 2\int \frac{\dd^3\vk_1}{(2\pi)^3} [F_2(\vk_1, \vk-\vk_1)]^2 P_L(k_1,\eta) P_L(|\vk-\vk_1|,\eta)\,.
\label{eq:P22}
\end{align}
We can do the same computation for the second combination
\begin{align}
\langle \delta^{(1)}(\vk)\delta^{(3)}(\vk')\rangle &= \int \frac{\dd^3\vk_1 \dd^3\vk_2}{(2\pi)^6}
F_3(\vk_1,\vk_2,\vk'-\vk_1-\vk_2)
\underbrace{\langle \delta_0(\vk) \delta_0(\vk_1) \delta_0(\vk_2) \delta_0(\vk'-\vk_1-\vk_2)\rangle}_{3\langle \delta_0(\vk) \delta_0(\vk_1)\rangle\langle \delta_0(\vk_2) \delta_0(\vk'-\vk_1-\vk_2)\rangle }\,,
\end{align}
where we used that $F_3$ is cyclic in the arguments
\begin{align}
P_{13}(k,\eta) = 3P_L(k,\eta)  \int \frac{\dd^3\vk_2}{(2\pi)^3} F_3(-\vk,\vk_2,-\vk_2) P_L(k_2,\eta)\,.
\label{eq:P13}
\end{align}
It is important to note that $P_{22}$ and $P_{13}$ are of the same perturbative order, so have to be considered in combination, and in fact have opposite signs leading to a large cancellation. We illustrate the different SPT contributions in Figure~\ref{fig:Pk_SPT}, the linear power spectrum (dotted black), the $P_{22}$ term (blue solid) and the negative $P_{13}$ term (red dashed) and their sum (purple for positive contributions, purple dashed for negative contributions), along with the full so-called 1-loop result. In Figure~\ref{fig:Pk_sim_SPT_Halofit} we compare the SPT predictions at linear and 1-loop order against simulation measurements, illustrating that going to 1-loop order improves the agreement on large scales. Beyond the perturbative regime at around $k\gtrsim 0.2h/$Mpc, both predictions diverge from the measurements, motivating the formulation of other approaches. 

\paragraph{Range of applicability} To get an intuition on which scales we can apply perturbation theory, we can look at the variance of the dark matter density field when filtered with a spherically symmetric filter function $W$, this variance tells us about the typical size of density fluctuations. In physical space, filtering a density field corresponds to a convolution while in Fourier space this is simply a multiplication with the Fourier transform of the filter function
\begin{equation}
\delta_W(\vx) =\int \dd^3x' W(|\vx-\vx'|) \delta(\vx') \quad\Leftrightarrow\quad \hat\delta_W(\vk) = \hat W(\vk) \hat\delta(\vk)\,.
\end{equation}
In the following, we will denote the Fourier-transformed fields like the density as $\delta(\vk)$ instead of $\hat\delta(\vk)$. Note that $\delta$ is a real field, while its Fourier transform is complex, with the contraint that $\delta^*(\vk) = \delta(-\vk)$.
\begin{align}
\label{eq:var_from_PS} 
\sigma^2_W &= \langle\delta_W^2(\vx)\rangle = \int \frac{\dd^3 k}{(2\pi)^3} \int \frac{\dd^3 k'}{(2\pi)^3} \langle \delta_W(\vk) \delta_W^*(\vk')\rangle \exp\left(i(\vk-\vk')\cdot\vx\right)\\
&= \int \frac{\dd^3 k}{(2\pi)^3} |W(\vk)|^2 P(k) = \frac{1}{2\pi^2} \int \dd k k^2 P(k) |W(k)|^2
\end{align}
The variance is indeed small for large smoothing scales, with a scale around $R\sim 5-10$Mpc/$h$ giving a variance of order one at redshift zero. 


We can define a rough nonlinear Fourier scale, where the matter density field will become fully nonlinear using the power spectrum
\begin{equation}
\label{eq:kNL}
k_{\rm NL}^{-2} = \int \frac{\dd^3k}{(2\pi)^3} k^{-2} P(k) = \int \frac{\dd k}{2\pi^2} P(k)
\end{equation}

\paragraph{Halofit matter power spectrum}
The halofit fitting formula \cite{Halofit2003,TakahashiHalofit2012} is an empirical model used to calculate the nonlinear matter power spectrum. It was developed to replicate the results of N-body simulations, which model the gravitational evolution of structures on small scales. Halofit is particularly useful for accounting for the nonlinear effects that occur when density contrasts become large and linear theories are no longer sufficient. The halofit model starts with the linear matter power spectrum, which is described by cosmological parameters and the growth of structures. The challenge is to extrapolate this spectrum into the nonlinear regime, where structures collapse to form galaxies and clusters. Halofit employs a set of empirically determined parameters to model the transition from linear to nonlinear scales. These parameters are chosen to reproduce the results of detailed N-body simulations. The foundation of halofit is the halo model of large-scale structure. This model assumes that all matter is distributed in halos of various masses and that the density profiles of these halos are universal. The model is flexible enough to be adapted for different cosmological models, including those with varying dark energy or neutrino masses. The halofit fitting formula is widely used in cosmology because it provides a practical way to model the effects of nonlinear gravitational forces on the large-scale distribution of matter without the need for exhaustive simulations for every cosmological scenario. Another major advantage over perturbative approaches is that halofit can be applied to predict projected statistics that probe small scales, in particular relevant for weak gravitational lensing discussed in Section~\ref{sec:WL}.

\paragraph{Power spectra and correlation function from Lagrangian Perturbation Theory}
We can use the expression for the Fourier density contrast from the displacement field to obtain the power spectrum
\begin{align}
1+\langle\delta(\vk)\delta(-\vk)\rangle &=\int \dd^3q_1 \dd^3q_2 \exp(-i\vk(\cdot\vq_1-\vq_2)) \langle\exp\left(-i\vk\cdot(\v{\Psi}(\vq_1)-\v{\Psi}(\vq_2))\right)\rangle\,,\\
P(\vk)&=\int \dd^3q \exp(-i\vk\cdot\vq) \left[\langle\exp\left(-i\vk\cdot(\v{\Psi}(\vq_1)-\v{\Psi}(\vq_2))\right)\rangle\Big|_{\vq=\vq_2-\vq_1}-1\right]\,,
\end{align}
where we can notice that the expectation value can only depend on the difference $\vq_2-\vq_1$ and thuse we could remove one integral. The expectation value of the exponent can be expanded with the cumulant expansion theorem, thus converting it to an exponential of a sum of cumulants $\kappa_n$, which are given by expectation values of powers of the exponent
\begin{equation}
\Big\langle\exp\Big(i \underbrace{\vk\cdot(\v{\Psi}(\vq_2)-\v{\Psi}(\vq_1))}_{\phi}\Big)\Big\rangle 
= \exp\Bigg[ \sum_n \frac{1}{n!} \kappa_n(\phi) \Bigg] \,.
\end{equation}
For the first order of Lagrangian Perturbation Theory (the Zeldovich approximation) we can benefit from the fact that the displacement field is a zero-mean Gaussian and hence we only need the second cumulant to obtain the power spectrum as
\begin{equation}
P_{\rm ZA}(k) = \int \dd^3q \exp(i\vk\cdot\vq) \exp\left[-\frac{1}{2}k_ik_j \Big\langle (\Psi_i(\vq_2)-\Psi_i(\vq_1))  (\Psi_j(\vq_2)-\Psi_j(\vq_1))\Big\rangle\right]\,.
\end{equation}
The ZA well describes the early stages of the evolution and is very good on large scales, particularly for the BAO peak in the correlation function as shown in the right panel of Figure~\ref{fig:ZA}. On the other hand, the ballistic motion of particles causes an erasure of structure on small scales that hampers its accuracy in the power spectrum, as illustrated on the left.

\subsection{Skewness and one-point distribution}
The nonlinear gravitatioanl dynamics discussed in the last Section induce non-Gaussianity, which can be quantified using beyond 2-point statistics. We will discuss the  leading order contributions to the skewness, which quantifies the asymmetry in the one-point distribution.
\paragraph{Spherical average of Eulerian PT kernels}
Let us compute the skewness of the (unsmoothed) density field in an EdS universe, which at leading order will be given by the second order perturbation kernel
\begin{align}
\label{eq:skewness_start}
\langle\delta^3(\vx,\eta)\rangle &\approx \langle(\delta^{(1)}(\vx,\eta)+\delta^{(2)}(\vx,\eta))^3\rangle 
\approx \langle(\delta^{(1)}(\vx))^3\rangle + 3 \langle(\delta^{(1)}(\vx))^2 \delta^{(2)}(\vx)\rangle \\
\label{eq:skewness_Fourier}
&= 3\int \frac{\dd^3k_1 \dd^3k_2 \dd^3k_3}{(2\pi)^9} \exp\left[i\vx\cdot\left(\vk_1+\vk_2+\vk_3\right)\right]
\langle \delta^{(1)}(\vk_1) \delta^{(1)}(\vk_2) \delta^{(2)}(\vk_3)\rangle \\
&= 3a^4 \int \frac{\dd^3k_1 \dd^3k_2 \dd^3k_3 \dd^3k_4}{(2\pi)^9} \exp\left[i\vx\cdot\left(\vk_1+\vk_2+\vk_3+\vk_4\right)\right]
F_2(\vk_3,\vk_4) \\
&\qquad \times\langle \tilde\delta_1(\vk_1) \tilde\delta_1(\vk_2) \tilde\delta_1(\vk_3)\tilde\delta_1(\vk_4)\rangle\,.\notag
\end{align}
In this expansion we used that for Gaussian initial conditions the initial skewness vanishes, and that higher orders of the density contrast come with higher powers of the scale factor.

If the initial field is Gaussian, it can only correlate in pairs such that we have 3 pairings for $\vk_1$. If $\vk_3$ is paired with $\vk_4$, meaning $\vk_3+\vk_4=0$, then we get zero because $F_2(\vk,-\vk)=0$. The other two contributions are the same such that we get 
\begin{align}
\langle\delta^3(\vx,\eta)\rangle &\approx 6a^4 \int \frac{\dd^3k_1 \dd^3k_2}{(2\pi)^6} P(k_1)P(k_2)F_2(-\vk_1,-\vk_2)\\
&= 6a^4 \int \frac{\dd k_1 \dd k_2}{(2\pi^2)^2} k_1^2 P_1(k_1) k_2^2P_1(k_2) \underbrace{\int \frac{d\Omega_1 d\Omega_2}{(4\pi)^2} F_2(-\vk_1,-\vk_2)}_{\frac{17}{21}} \\
&= \frac{34}{7} \langle\delta_1^2(\vx,\eta)\rangle^2\,.
\end{align}
Interestingly, we obtain a characteristic \emph{reduced skewness} \cite{Peebles1980} that is close to constant
\begin{equation}
\label{eq:S3pred}
S_3=\frac{\langle\delta^3\rangle}{\langle\delta^2\rangle^2} \approx \frac{34}{7}\,.
\end{equation}
This result is not yet practical as we considered the unsmoothed density field, while in practice we can only access a smoothed density field from simulations or observations. But before turning to the smoothing, let us investigate the connection to spherical collapse

\paragraph{Spherical collapse}
We can also use our approximate parametrised relation between linear and nonlinear densities that we obtained from spherical collapse to obtain the unsmoothed skewness
\begin{align}
\langle\delta^3\rangle \approx \underbrace{\langle \delta_L^3\rangle}_{0} + 3 \underbrace{\langle \delta_L^2 \frac{1}{2}\delta''(\delta_L=0) \delta_L^2\rangle}_{(1+1/\nu)\langle\delta_L^4\rangle} = 3 \left(1+\frac{1}{\nu}\right)\langle\delta_L^2\rangle^2\,.
\end{align}
If we seek to match this to the tree-level perturbation theory result, we obtain $\nu=21/13$.

\paragraph{The skewness of the smoothed density field}
If we consider the smoothed density $\delta_R(\vx)$ obtained from spherically symmetric filter on a given scale $R$ with a Fourier kernel $W_R(k)$, then in equation~\eqref{eq:skewness_Fourier} we have to replace $\delta^{(n)}(\vk)\rightarrow W_R(k)\delta^{(n)}(\vk)$, which then leads to
\begin{align}
\langle\delta_R^3(\vx)\rangle &\approx 6a^4 \int \frac{\dd^3k_1 \dd^3k_2}{(2\pi)^6} P(k_1)P(k_2)W_R(k_1)W_R(k_2) W_R(|\vk_1+\vk_2|)F_2(-\vk_1,-\vk_2)\,.
\end{align}
A spherical top hat filter has special properties that allow to re-express the angular average of $W_R(|\vk_1+\vk_2|)F_2(\vk_1,\vk_2)$ to be re-expressed in terms of a sum of products $W_R(k_1)W_R(k_2)$ and $W_R(k_1)k_2W_R'(k_2)$, where the derivative with respect to the argument of $W$ can be written as a derivative with respect to the radius $R$. With this, one can show that the skewness acquires a mild scale-dependence \cite{Bernardeau94skewkurt} determined by the variance $\sigma^2(R)$ and hence the power spectrum as in equation~\eqref{eq:var_from_PS} 
\begin{equation}
\label{eq:S3pred_smoothing}
S_3(R)=\frac{\langle\delta_R^3\rangle}{\langle\delta_R^2\rangle^2} \approx \frac{34}{7} + \frac{d\log\sigma^2(R)}{d\log R}\,.
\end{equation}

\begin{figure}
\centering
\includegraphics[width=0.48\textwidth]{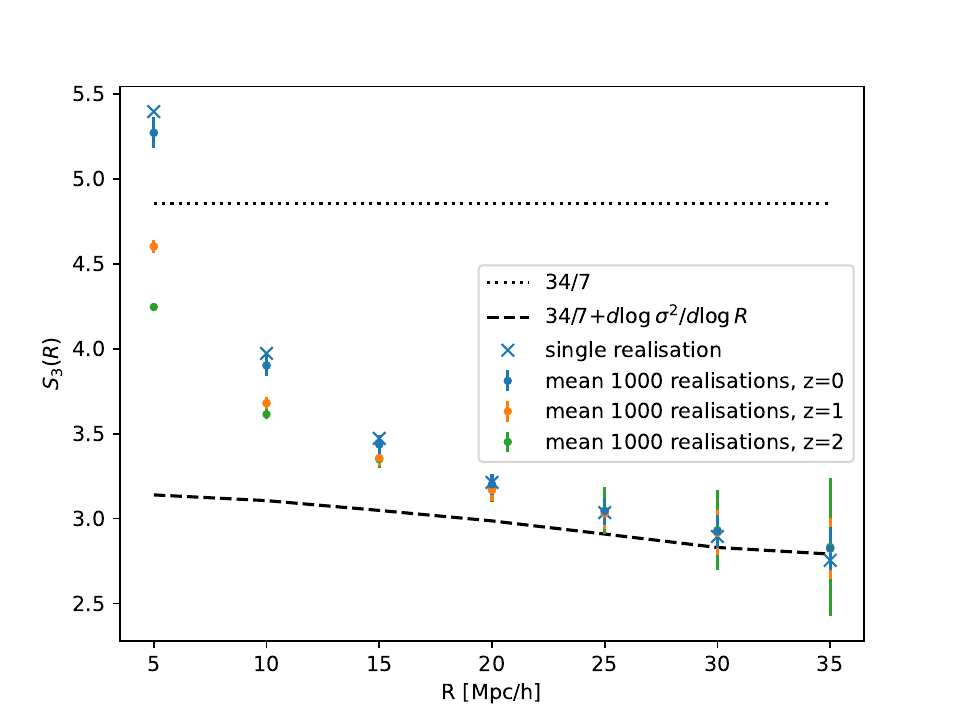}
\includegraphics[width=0.51\textwidth]{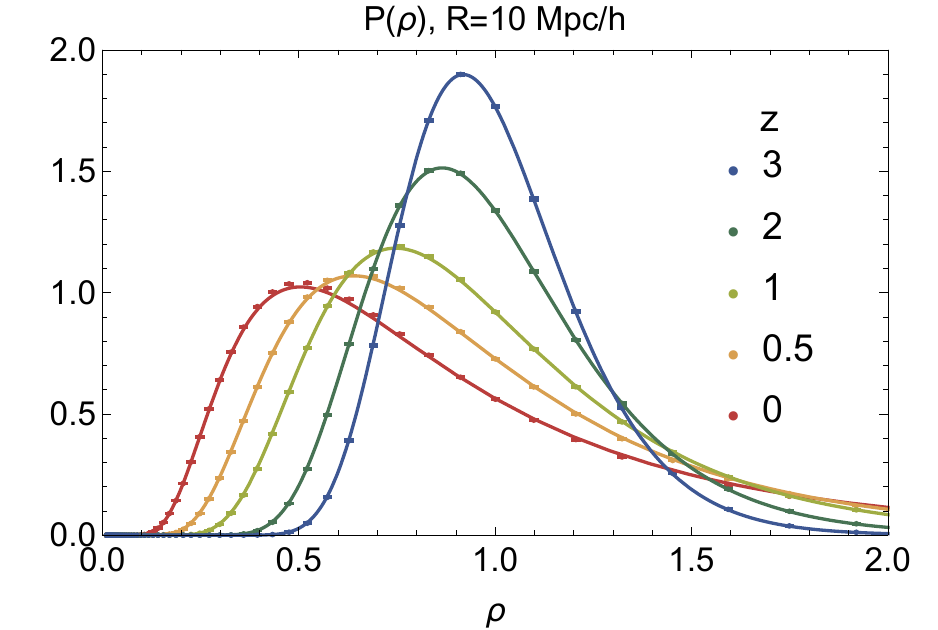}
\caption{(Left) Reduced skewness of the smoothed matter density field $S_3(R)$ as a function of scale as predicted from tree-order perturbation theory (black dashed) and measured in 1000 realisations of the Quijote simulations at redshift $z=0$ (blue), $z=1$ (orange), $z=2$ (green). (Right) Evolution of the probability density function of the normalised dark matter density $1+\delta$ at a fixed physical scale over cosmic time showing an increasing amount of non-Gaussianity (from blue to red). Measurements averaged over all realisations of the fiducial Quijote simulations (data points) are compared with  predictions from large-deviation statistics that use spherical collapse together with the nonlinear variance (solid lines). [Plot made similar to Figure~6 in \cite{Uhlemann2020}].}
\label{fig:S3_PDF}
\end{figure}

In the left panel of Figure~\ref{fig:S3_PDF} we show a comparison between this prediction of tree-order perturbation theory (black dashed line) including the correction compared to the constant value (black dotted) and simulation measurements of density fields smoothed with a spherical top hat kernel (data points) with the colour indicating the redshift. We can see that at fixed physical scale, higher redshifts are less nonlinear and hence in better agreement with the prediction.

The skewness is a result of the non-Gaussianity of the final density distribution (induced by nonlinearity even for Gaussian initial density fields), it measures the asymmetry between underdense and overdense regions. While matter gets concentrated in increasingly large overdensities, most of the volume of the Universe becomes underdense. To make this more quantitative, we can look at the shape of the one-point probability of finding a given matter density in spheres of radius $R$ across different redshifts $z$. Over time, underdense regions expand while overdense regions contract, so the maximum of the probability distribution wanders to negative density contrasts. The right panel of Figure~\ref{fig:S3_PDF} illustrates this, with the data points obtained from an average over the Quijote simulations at the fiducial cosmology. Using spherical collapse, mass conservation and a recipe to compute the nonlinear variance (e.g. from a power spectrum model), we can extend the predictions we just made for the skewness to predict the full one-point probability distribution (in the mildly nonlinear regime where the nonlinear variance is small enough) using large-deviation theory, as outlined in \cite{Uhlemann2016,Uhlemann2020}. Those predictions are shown as solid lines, obtained from the simpliedfied saddle-point approximation on the log-density, with the measured variance as an input. Predictions for the matter PDF and related observables can be obtained from the public codes \href{https://github.com/OliverFHD/CosMomentum}{CosMomentum} \cite{Friedrich2020}, and predictions can be extended to modified gravity as done in \href{https://github.com/mcataneo/pyLDT-cosmo}{pyLDT-cosmo} \cite{Cataneo2022}.

\subsection{Three-point correlations and the bispectrum}

\begin{figure}
\centering
\includegraphics[width=0.75\textwidth]{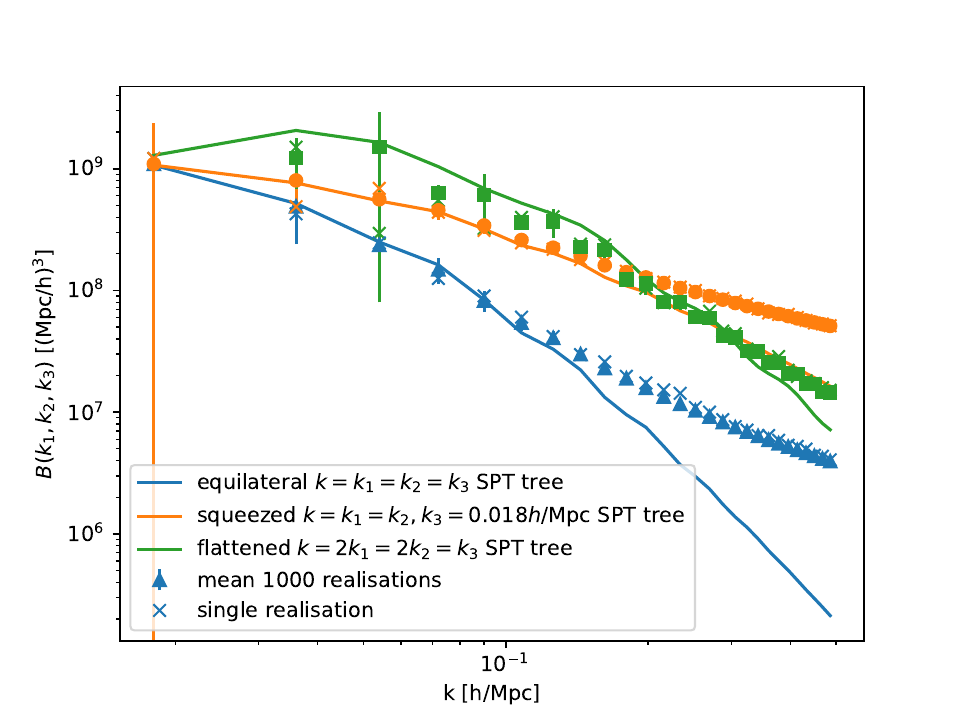}
\caption{Matter bispectra in equilateral (blue), squeezed (orange) and flattened (green) configurations from the Quijote realisations at the fiducial cosmology and redshift $z=0$ as measured from a single realisation (crosses) and averaged over 1000 realisations (markers with error bars) and the prediction by tree-order perturbation theory~\eqref{eq:bispectrum_tree} (solid lines).}
\label{fig:bispectrum}
\end{figure}

Let us compute the leading order correction to the matter bispectrum, which is zero for a Gaussian (linear) density field
\begin{equation}
\langle\delta(\vk_1)\delta(\vk_2)\delta(\vk_3)\rangle = (2\pi)^3 \delta_D(\vk_1+\vk_2+\vk_3) B(\vk_1,\vk_2)
\end{equation}
Using our perturbative solution, we can also compute a bispectrum induced by the nonlinearity of gravitational evolution
\begin{align}
\langle\delta_m(\vk_1)\delta_m(\vk_2)\delta_m(\vk_3)\rangle \approx &
\Big\langle \left(\delta^{(1)}+\delta^{(2)}\right)(\vk_1) \left(\delta^{(1)}+\delta^{(2)}\right)(\vk_2) \left(\delta^{(1)}+\delta^{(2)}\right)(\vk_3)  \Big\rangle \\
\approx & \langle \delta^{(1)}(\vk_1)\delta^{(1)}(\vk_2)\delta^{(2)}(\vk_3)\rangle + \text{cyclic}\\
= & \int\frac{\dd^3p}{(2\pi)^3} F_2(\vp,\vk_3-\vp) \langle \delta^{(1)}(\vk_1)\delta^{(1)}(\vk_2)\delta^{(1)}(\vp)\delta^{(1)}(\vk_3-\vp)\rangle + \text{cyclic}
\,,
\end{align}
where we collected all terms up to $\langle\delta^{(k)}\delta^{(l)}\delta^{(4-k-l)}\rangle$, which means only one of the deltas is second order. As opposed to the power spectrum, the leading order contribution is only requiring $F_2$. Let us now plug in Wick's theorem \eqref{eq:GaussianICs}
\begin{align}
\frac{\langle \delta^{(1)}(\vk_1)\delta^{(1)}(\vk_2)\delta^{(2)}(\vk_3)\rangle}{(2\pi)^3}
&=  \int \dd^3p F_2(\vp,\vk_3-\vp) 
\delta_D(\vk_1+\vk_2)\delta_D(\vk_3)P_0(k_1)P_0(p) \\
\notag &+\int \dd^3p F_2(\vp,\vk_3-\vp)\delta_D(\vk_1+\vp)\delta_D(\vk_2+\vk_3-\vp)P_0(k_1)P_0(k_2)\\
\notag &+ \int \dd^3p F_2(\vp,\vk_3-\vp) \delta_D(\vk_1+\vk_3-\vp)\delta_D(\vk_2+\vp) P_0(k_1)P_0(k_2)\\
& = 
\int\dd^3p \underbrace{F_2(\vp,-\vp)}_{0} \delta_D(\vk_1+\vk_2)\delta_D(\vk_3)P_0(k_1)P_0(p) \\
\notag &+\delta_D(\vk_1+\vk_2+\vk_3)\left[ F_2(-\vk_1,-\vk_2)+F_2(-\vk_2,-\vk_1) \right] P_0(k_1)P_0(k_2)
\\
B_{112}(k_1,k_2,k_3;\eta)&=2\left[F_2(\vk_1,\vk_2)P_0(k_1,\eta)P_0(k_2,\eta)+\text{cyclic}\right]\,.
\label{eq:bispectrum_tree} 
\end{align}
This result is called the tree-level bispectrum, because only one of the fields is at second order, which corresponds to a loop in the diagrammatic representation, or an integral over a free momentum as we have encountered for the matter power spectrum orders $P_{13}(k)$~\eqref{eq:P13} and $P_{22}(k)$~\eqref{eq:P22}. 
Due to statistical homogeneity, the bispectrum probes triangle configurations for which $\vk_1+\vk_2+\vk_3=0$, which means the wave vectors are in one plane. Statistical isotropy implies that the value should be independent of the direction and rotation of the plane, which means that we only need to specify the triangle configurations itself. They can be labelled by the three absolute values of the wavevectors $k_1,k_2,k_3$, or equivalently the scalar products of the two vectors spanning the triangle $k_1, k_2$ and the angle between $\mu=\vk_1\cdot \vk_2/(k_1k_2)$, or in spherical coordinates using an absolute value $k_{\rm sph}=
\sqrt{(k_1^2+k_2^2+k_3^3)/3}$ and two angles $\theta_{\rm sph}=\arctan(k_3/\sqrt{k_1^2+k_2^2})$ and $\phi_{\rm sph}=\arctan(k_1/k_2)$. In Figure~\ref{fig:bispectrum} we show three different configurations of the bispectrum (chosen following the presentation of \cite{Lazanu2018}): equilateral where $k=k_1=k_2=k_3$, squeezed $k=k_1=k_2$ with fixed small $k_3=0.018$ h/Mpc and flattened configuration where $k_1=k_2=k,k_3=k$.

\section{From dark matter to large-scale structure observables}
\label{sec:DMtoObs}
Investigating the clustering properties of dark matter related-tracers like galaxy positions or shapes allows to build the bridge between observational and theoretical cosmology and infer the dark matter distribution from galaxy surveys. Relating actual observations to the underlying dark matter distribution demands addressing two basic challenges: On the one hand, one has to relate the locations of tracers to the underlying dark matter - a problem called tracer bias and stochasticity, and on the other hand one has to take into account how redshifts are inferred from observations, in particular including the impact of redshift-space distortions induced by peculiar velocities and in particular for photometric surveys the impact of redshift errors due to inferring redshifts from just a few colour bands.

Large-scale structure probes broadly fall into two categories: Clustering probes tracers like galaxies, clusters, quasars, or gas in their vicinity - through emission (e.g. by intensity mapping) or absorption (Lyman-alpha forest). For the Lyman-alpha forest and how it can constrain dark matter properties see the lecture notes by Eric Armengaud. Traditionally, there has been a focus on galaxy clustering making use of optical and near infrared surveys probing late times, although radio surveys provide interesting avenues for probing a wide range of redshifts. Weak lensing probes the overall matter, largely consisting of dark matter, and thus is a cosmological probe insensitive to galaxy bias. Because the weak lensing kernel is sensitive to the full distribution of matter between the lensing sources and the observer, it is more sensitive to nonlinearities. Combining weak lensing and photometric clustering, in particular in the context of 3x2point statistics can benefit from breaking degenarcies between cosmology and the galaxy-matter connection and thus help to jointly constrain cosmological and astrophysical parameters. 

%
%
%
%

\subsection{Tracer bias and stochasticity}

While theoretical models predict the dark matter density field and its clustering, large-scale redshift surveys measure tracer densities (from galaxies or their emission) whose clustering properties need not precisely mirror the clustering of the dark matter. This phenomenon is called tracer bias, a terminology that has been introduced by \cite{Kaiser1984}. It is due to the physics of galaxy formation causing the galaxy distribution to be a possibly non-local and stochastic function on the underlying total matter field dominated by dark matter. Although galaxies are the objects relevant for observations, theoretical models often first study the bias of dark matter halos, which are accessible from dark matter only simulations as an intermediate step to link the dark matter to the galaxy distribution. Simplified galaxy populations can be modelled once the halo model \cite{CooraySheth2002} is equipped with occupation statistics that specify how galaxies populate the dark matter halos, as done in the halo occupation distribution discussed in the end.

\begin{figure}
\centering
\includegraphics[width=0.75\textwidth]{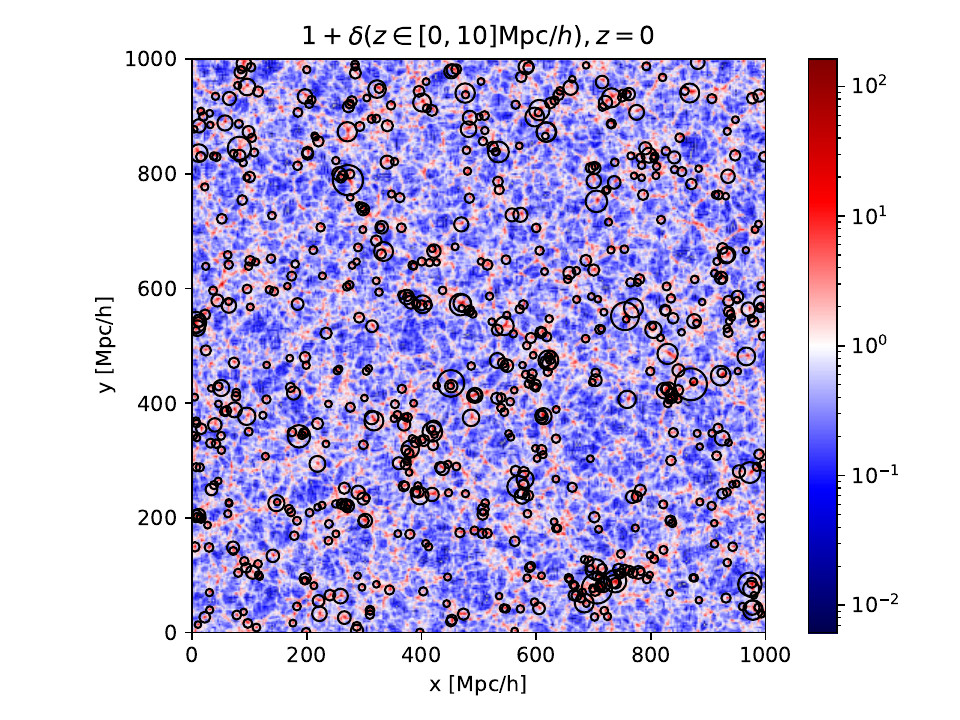}
\caption{Thin slice through the 3-dimensional dark matter density distribution in the Quijote simulations at redshift $z=0$ (shown for realisation 0 of the fiducial cosmology) as shown in Figure~\ref{fig:density_slice}, but overlaid with dark matter halos with more than 100 particles, corresponding to a minimum halo mass of $6.6\times 10^{13} M_\odot$ (black circles, with the size being proportional to the particle number in the halo).}
\label{fig:density_slice_halos}
\end{figure}

\subsubsection{Eulerian bias: from local to nonlocal}

\begin{figure}[h!]
    \centering
    \includegraphics[width=0.75\columnwidth]{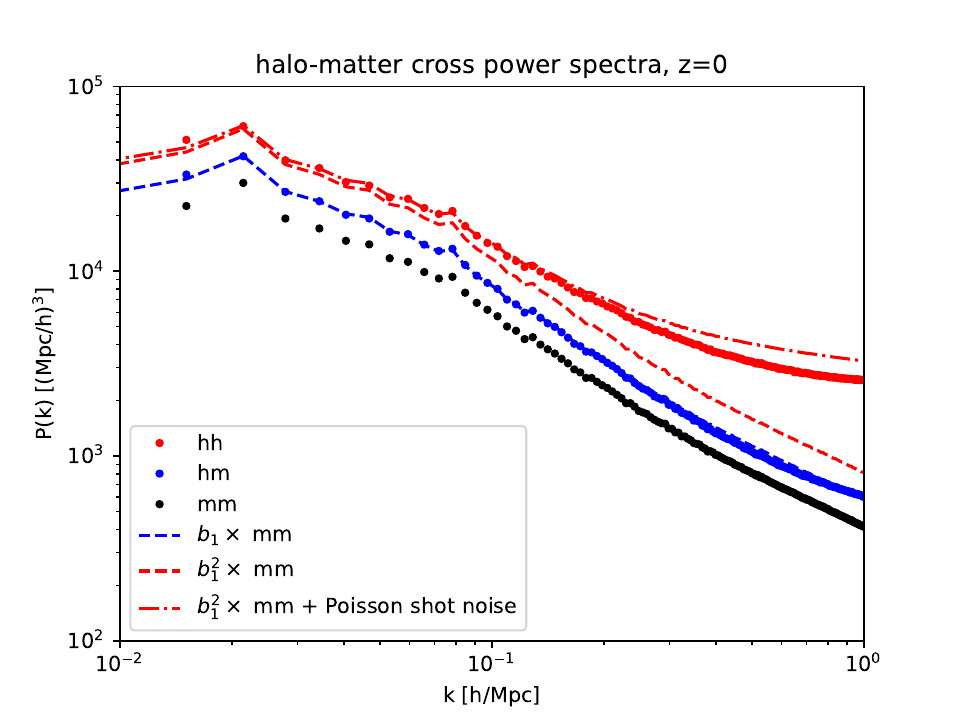}
    \caption{Cross power spectra of the halo and matter density fields, comparing the measured matter power spectrum (black points), with the halo-matter cross-power spectrum (blue points) and the halo auto power spectrum (red points). We can see that on large scales, the halo-matter cross power spectra is proportional to the matter power spectrum with a constant of priortionality given by the linear bias $b_1^E$. The halo auto power spectrum on large scales is also proportional to this but with a square $(b_1^E)^2$. Additionally we see that the halo auto power spectrum has significantly more power going to smaller scales, which is a sign of stochasticity stemming from the small number density of halos - a phenomenon called shot noise. }
    \label{fig:linearbias}
\end{figure}

\paragraph{Linear bias} The simplest model is the deterministic \emph{linear local bias} for which the tracer and matter densities are simply proportional
\begin{equation}
\delta_t(\vx,\eta)\approx b_1\delta_m(\vx,\eta)\,.
\end{equation}
This model is valid on large scales, where additional complications are negligible, and directly translates into Fourier space. The most common tracers (such as massive dark matter halos and the galaxies residing in them) have a bias above unity, meaning their clustering is enhanced compared to the dark matter. This is expected for objects residing in larger overdensities, while the bias of objects residing in underdensities can be well below unity or even negative. Linear bias predicts that the tracer-matter cross power spectrum and the tracer auto power spectrum are proportional to the matter power spectrum 
\begin{align}
P_{tm}(k) &=\langle\delta_t(\vk)\delta_m(-\vk)\rangle \approx b_1 P_{mm}(k) \\
P_{tt}(k) &= \langle\delta_t(\vk)\delta_t(-\vk)\rangle \approx b_1^2 P_{mm}(k)\,,
\end{align}
for which we demonstrate the large-scale validity for dark matter halos in Figure~\ref{fig:linearbias}. 
There we can see that the measured halo-matter cross-power spectrum matches the simplistic prediction remarkably well even on more nonlinear scales, while the halo auto power spectrum displays significantly more power than expected. Those deviations result from the relatively low tracer number density $n_t$ that causes a stochasticity, which is uncorrelated with the matter density thus not affecting the cross power spectrum. If we assume the stochasticity to be Poissonian, then we can simply add the constant shot noise power spectrum $n^{-1}$, which leads to a good match on large scales. More generally, we could assume a white noise, which is still constant but potentially of a different amplitude $\alpha n^{-1}$. Assuming a linear bias model, we can even reconstruct a scale-dependent linear bias and shot noise 
\begin{equation}
\label{eq:linear_bias_SN}
b_1(k) =\frac{P_{tm}(k)}{P_m(k)} \,,\quad
\frac{\alpha(k)}{n}=P_{tt}(k) - b_1(k)^2 P_m(k) = \frac{P_{tm}(k)^2}{P_m(k)}
\end{equation}

\paragraph{Nonlinear bias} The key idea of our simplistic bias model was to express the local number density of the tracer in terms of the underlying matter. While there are many complications, as tracers are generally not conserved (as opposed to matter), can be subject to non-gravitational forces including pressure, and their formation given just the local matter density is not deterministic but stochastic. For a perturbative modelling of tracer bias, we write the tracer density in terms of physically allowed building blocks, which are the matter density contrast $\delta_m$, the velocity $u^i$ and the gravitational potential $\psi$, along with its derivatives. Terms proportional to $\psi$, $\partial_i \psi$ and $u^i$ are not locally observable, as they can be removed by a coordinate transformation -- thus they cannot appear in local galaxy formation. Following the equivalence principle, galaxies and matter fall at the same rate under a large-scale gravitational acceleration. Because of this the actual building blocks are the matter density contrast $\delta_m$, first derivative of the matter velocity $\partial_j u^i$ and second derivatives of the gravitational potential $\partial_i\partial_j \psi$. When going to Fourier space, it is easy to see that higher-derivative terms such as $\nabla\delta(\vx)\rightarrow i\vk \delta(\vk)$, $\nabla^2\delta(\vx)\rightarrow k^2\delta(\vk)$ come with extra powers of the wavevector $k$ and are thus more suppressed on large scales.

As an instructive example, let us determine how a Eulerian quadratic bias model manifests in the tracer cross and auto power spectra. We start with the deterministic tracer density (omitting shot noise for now)
\begin{subequations}
\begin{align}
\label{eq:quad_bias_field}
\tilde\delta_t(\vec{x})=&b_1\delta_m(\vec{x})+\frac{b_2}{2}\left(\delta_m(\vec{x})^2-\langle \delta_m(\vec{x})^2\rangle\right) +\frac{b_{s^2}}{2} \left(s_{ij}(\vec{x})^2-\langle s_{ij}(\vec{x})^2\rangle\right)
\,,
\end{align}
where $b_{s^2}$ describes the non-local bias associated with the tidal shear 
\begin{equation}
s_{ij}=\left(\partial_i\partial_j-\partial^2/3\right)\psi=(\partial_i\partial_j/\partial^2-\delta_{ij}/3)\delta_m(\vec{x})\,,
\end{equation}
where second derivatives of the gravitational potential were rewritten in terms of the density. In Fourier space the squares translate to convolutions
\begin{align}
\label{eq:quad_bias_field_k}
\tilde\delta_t(\vec{k})=&b_1\delta_m(\vec{k})+\frac{b_2}{2} \underbrace{\int\! d^3q\, \delta_m(\vec{k}-\vec{q})\delta_m(\vec{q})}_{(\delta_m* \delta_m)(\vec{k})} +\frac{b_{s^2}}{2}\underbrace{\int\! d^3q\, \delta_m(\vec{k}-\vec{q})\delta_m(\vec{q})S_2(\vec{q},\vec{k}-\vec{q})}_{(s_{ij}*s_{ij})(\vec{k})}\,,
\end{align} 
where $S_2(\vec{q_1},\vec{q_2})=\vec{q_1}\cdot\vec{q_2}/(q_1^2q_2^2)-1/3$ and we omitted the $\delta$-function terms that arise from the subtraction of the constant expectation values $\langle \cdot \rangle$. With this we obtain the cross power spectrum
\begin{align}
    P_{tm}(k)=\langle\delta_t(\vec{k})\delta_m(-\vec{k})\rangle
\label{eq:Pk_tracer_cross_Eulerian}
    =b_1 P_m(k)+\frac{1}{2} \int\! d^3q \left(b_2+b_{s^2} S_2(\vec{q},\vec{k}-\vec{q})\right)B_m(\vec{q},\vec{k}-\vec{q},-\vec{k}) \,,
\end{align}
which depends on the matter power spectrum $P_m$ and an integral over the matter bispectrum $B_m$, where the first term is proportional to the so-called skew-spectrum. The scale-dependent linear bias~\eqref{eq:linear_bias_SN} would then be
\begin{align}
\label{eq:bias_Pk_Eulerian}
b_1(k) = \frac{P_{tm}(k)}{P_m(k)}&= b_1+ \frac{b_2}{2} \frac{\int\! d^3q B_m(\vec{q},\vec{k}-\vec{q},-\vec{k})}{P_m(k)}
+\frac{b_{s^2}}{2} \frac{\int\! d^3q S_2(\vec{q},\vec{k}-\vec{q}) B_m(\vec{q},\vec{k}-\vec{q},-\vec{k})}{P_m(k)}\,,
\end{align}
where the quadratic bias terms term would be evaluated perturbatively and lead to either constant terms that add to the scale-independent $b_1$ or combine to a leading-order scale dependence of $k^2$.
\end{subequations}
For the case of just local bias (thus neglecting the tidal bias term $b_{s^2}$), the tracer auto power spectrum is obtained as
\begin{subequations}
\begin{align}
\label{eq:Pk_tracer}
    P_{t}(k)&=\langle\delta_t(\vec{k})\delta_t(-\vec{k})\rangle\notag\\
    &=b_1^2 P_m(k)+b_1b_2\langle(\delta_m*\delta_m)(\vec{k})\delta_m(-\vec{k})\rangle +\frac{b_2^2}{4} \langle(\delta_m*\delta_m)(\vec{k})(\delta_m*\delta_m)(\vec{-k})\rangle + P_\epsilon(k) \\
    &=b_1^2 P_m(k)+b_1 b_2 \int\! d^3q B_m(\vec{q},\vec{k}-\vec{q},-\vec{k})+\frac{b_2^2}{4}\!\!\int\!\!\!\!\int\!\!d^3q_1 d^3q_2 T_m(\vec{q_1},\vec{q_2},\vec{k}-\vec{q_1},-\vec{k}-\vec{q_2}) + P_\epsilon(k)\notag\,,
\end{align}
which depends on the matter power spectrum, the skew-spectrum and an integrated version of the trispectrum $T_m$ and the shot noise power spectrum $P_\epsilon(k)=\langle\epsilon(\vec{k})\epsilon(-\vec{k})\rangle$. The tidal bias terms will involve similar integrals with additional weightings with one $S_2$ kernel for the bispectrum and two for the trispectrum term.
The shot noise defined in~\eqref{eq:linear_bias_SN} would then be given by
\begin{align}
\label{eq:shotnoise_Pk_Eulerian}
\frac{\alpha(k)}{\bar n}
&=\frac{b_2^2}{4} \Bigg[\!\!\int\!\!\!\!\int\!\!d^3q_1 d^3q_2 T_m(\vec{q_1},\vec{q_2},\vec{k}-\vec{q_1},-\vec{k}-\vec{q_2}) 
- \frac{\left(\int\! d^3q B_m(\vec{q},\vec{k}-\vec{q},-\vec{k})\right)^2}{P_m(k)} \Bigg] + P_\epsilon(k)\,,
\end{align}
where the terms including the trispectrum and square of the bispectrum are of higher perturbative order. As expected, nonlinear bias does contribute to the shot noise which we had defined relative to the linear bias. 
\end{subequations}

\begin{figure}
    \centering
    \includegraphics[width=0.5\columnwidth]{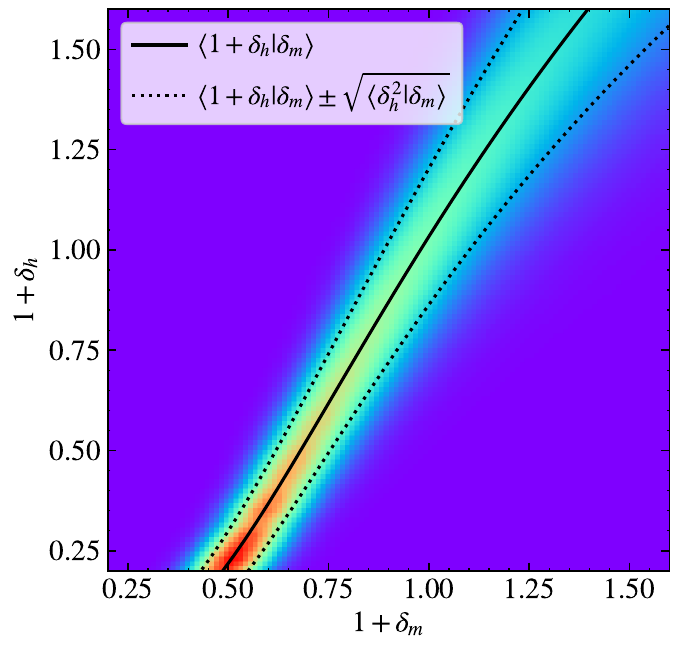}
    \caption{Conditional PDF of halo given matter density $\mathcal P(\delta_h|\delta_m)$ using 8000 realisations of the Quijote fiducial cosmology at redshift $z=0.0$ and smoothing scale $R=25$Mpc$/h$. The black line shows the conditional mean roughly following the `ridge' of the conditional PDF, while there is also some scatter around the conditional mean, i.e. shot noise (black dotted lines). [Adapted from Figure~2 in \cite{Gould2025} (CCBY-4.0)]}
    \label{fig:condPDF}
\end{figure}
\subsubsection{A local view of tracer bias and stochasticity} Instead of looking at the power spectrum, we can also investigate the local relation between tracer and matter densities at a given smoothing scale $R$. A tracer density contrast can be defined through $\delta_t=N_t/\bar N_t -1$  with the mean number of tracers per cell $\bar N_t$. We can then look at the joint distribution of the local tracer and matter densities, for example through performing a `scatter plot' where we plot a point for each pair of densities. Instead of looking at the density of those points, we can directly look at the joint probability of the tracer and matter density $\mathcal P(\delta_t,\delta_m)$, or the conditional probability of finding a given tracer density given a matter density $\mathcal P(\delta_t|\delta_m)=\mathcal P(\delta_t,\delta_m)/\mathcal P(\delta_m)$. We show an example of the latter for the conditional distribution of halo densities given a certain matter density in spheres in Figure~\ref{fig:condPDF}.

We see that there is a clear correlation between matter and tracer densities in cells (with correlation coefficients of around $0.9$), but also some scatter. We describe this conditional PDF with two ingredients, the conditional mean tracer count $\overline{N_t}(\delta_m)=\langle N_t|\delta_m\rangle$ and the conditional variance of tracer counts at fixed matter density contrast $\langle N_t^2|\delta_m\rangle_c$. The conditional mean -- shown by the solid black line -- follows the `ridge' of the joint PDF, while the conditional variance captures the scatter around the expectation value -- indicated by the dotted lines. For a Poisson distribution, the conditional variance agrees with the conditional mean, and as such we express the model in terms of the ratio $\alpha(\delta_m)=\langle N_{\rm t}^2|\delta_m\rangle_c/\langle N_{\rm t}|\delta_m\rangle$.

\subsubsection{Lagrangian bias} If we assume local Lagrangian bias between the halo density field and dark matter, then the following relation holds between Eulerian and Lagragian space
\begin{equation}
\left[1+\delta_t(\vr,t)\right]\dd^3r = F[\delta_R(\vq),t]\dd^3q
\end{equation}
This means that proto-halos identified in the linear initial conditions, depending only on the smoothed initial linear density field $\delta_R(\vq)$ are conserved until they form a proper halo at time $t$. The proto-halo initial density field is assumed to be a local functional, $F$, of the initial linear density field smoothed over some scale $\delta_R$ related to the Lagrangian size $R(M)$ of the proto-halo. 

All proto-halos are assumed to move along perfect fluid trajectories $\vx(\vq,t)$ with initial conditions  $\delta_R(\vq)$  and we assume zero velocity bias between halos and dark matter $\vv_t=\vv$. Within the Lagrangian framework, we can furthermore replace $\vv=a\v{\dot\Psi}$ which follows from the fact that the integral line of the single-stream velocity $\vv$ is the displacement field $\v{\Psi}(\vq,t)=\vr-\vq$ such that we obtain
\begin{equation}
1+\delta_t(\vr,t) = \int \dd^3q F[\delta_R(\vq),t]\delta_D\left(\vr-\vq-\v{\Psi}(\vq,t)\right)\,.
\end{equation}
The Lagrangian bias parameters are obtained as expectation values of derivatives of the Lagrangian halo density field $b_n^L=\langle F^{(n)}\rangle$ leading to an expansion in terms of the initial density contrast $F[\delta_R(\vq)]=1+\sum_n b_n (\delta_R(\vq))^n$. Note that while the Lagrangian bias operates on the basis of initial density fields, the Eulerian bias is based on the dynamically evolved density field. Since the dynamical evolution can be non-local, those two local biasing schemes are in general not equivalent to each other, except for special cases such as spherical collapse or linear evolution, see \cite{Matsubara2008}.

\paragraph{Lagrangian halo bias} Dark matter halos are gravitationally bound objects of dark matter that result from the nonlinear collapse of initial density perturbations and host galaxies. In the initial dark matter density-field, so-called proto-halos, regions that will later form halos, can be identified as density peaks above a criticial collapse threshold. It has been shown in \cite{BBKS1986} that for an initial Gaussian distribution of mass density fluctuations, high peaks are more clustered than the underlying mass distribution such that proto-halos are biased tracers of dark matter. As outlined by \cite{PressSchechter1974}, the fraction of mass in collapsed objects more massive than a certain mass $M$ at a given redshift $z$ is related to the fraction of volume in which the smoothed initial density fluctuation $\delta_R(\vq)$ is above some threshold $\delta_c$. That region will collapse to a halo of mass $M(R)=4\pi/3 \rho_0R^3$ forming at redshift $z_c>z$, where $\rho_0$ is the comoving background matter density. Typically, the critical density threshold is about $\delta_c=1.686$ as we derived in the spherical collapse model discussed in Section~\ref{sec:SC}. Dark matter particles that reside in one of the final halos can be assigned to proto-halos in the initial conditions whose fluctuation field $F[\delta_R(\vq)]$ is a functional of the smoothed initial density fluctuation $\delta_R(\vq)$. The halo distribution at redshift $z$ can then be obtained by mapping the proto-halos from their initial to their final position $\vq\rightarrow\vx$ using the displacement field $\v{\Psi}(\vq,z)$ which encapsulates gravitational dynamics. This displacement field could be obtained from a perturbative approach, or alternatively from an N-body simulation, which when paired with a perturbative bias expansion has been dubbed `Hybrid Effective Field Theory' \cite{HEFTModi2020}.

The excursion set approach \cite{Bond1991} allows to compute the halo mass function, describing the abundance of dark matter halos as a function of their mass and formation redshift, by extending the formalism of \cite{PressSchechter1974} with the requirement to consider only the largest possible smoothing scale $R(M)$ for which the density threshold $\delta_c$ is exceeded. Since furthermore any background density field enhances the peak and hence the probability of forming a halo, it is useful to perform a peak background split \cite{MoWhite1996} to implement the intuitive idea that a large-scale density field reduces the local threshold for collapse. Based on this idea, one can determine the conditional mass function \cite{LaceyCole1993} which determines halo bias and encodes merger histories characteristic to hierarchical clustering. Well known yet simplistic models are known by the name of the authors Sheth-Tormen \cite{ShethTormen1999} and Sheth-Mo-Tormen  \cite{ShethMoTormen2001} based on mass functions calculated from an extension to the Press-Schechter excursion set approach \cite{PressSchechter1974} using ellipsoidal collapse equations and fits to numerical simulations. This allows for statistical predictions of the bias for halos and HOD galaxies as summarised below.

\subsubsection{Halo Bias and the Halo Mass Function}

Halo bias describes how the spatial distribution of dark matter halos is enhanced compared to the overall matter distribution in the universe. It can be derived from the halo mass function, which describes the number of halos as a function of mass.

The mass function $n(M)$ provides the number of halos per unit volume per unit mass interval. The Lagrangian bias $b(M)$ of a halo with mass $M$ can be obtained from the halo mass function, generally as a function of the rarity $\nu=\delta_c/\sigma(M)$ where the variance $\sigma^2(M)$ represents the linear density fluctuations on the mass scale $M$. The Lagrangian bias (relating to the initial density field) can then be related to Eulerian biases, for the first order bias parameter the general relation is $b_{1}^{\rm E}=1+b_1^{\rm L}$ which leads to the following expressions for Press-Schechter and Sheth-Tormen mass functions
\begin{align}
b_{PS}^{\rm E}(M) &= 1 +\frac{\nu^2-1}{\delta_c}\,,\quad
b_{ST}^{\rm E}(M) = 1 +\frac{q\nu^2-1}{\delta_c} + \frac{2p/\delta_c}{1+(q\nu^2)^p}\,,\quad
\label{eq:halobias_PS_ST}
\end{align} 
Those formulas reflect how the clustering of halos depends on their mass. More massive halos (lower $\sigma(M)$ and hence higher $\nu$) tend to be more strongly biased compared to the underlying matter distribution. The inclusion of the extra parameters  $p$ and $q$ allows the Sheth-Tormen bias to capture the effects of ellipsoidal collapse, which leads to a more accurate prediction of halo clustering properties across different mass scales.
From the bias as function of mass, we can obtain the bias for any halo selection from an average over all selected masses weighted by their probability $\mathcal P(M_h)$ via
\begin{equation}
\label{eq:bias_massint}
    \bar b_{{\rm h},n}= \int \mathrm{d}M_h b_n^{\rm E}(M_h) \mathcal P(M_h) \,.
\end{equation}

\subsubsection{Halo occupation distributions: from halos to galaxies}
Halo occupation distributions (HODs) are used as simplistic recipes to translate halo catalogs to mock galaxy catalogs. The key idea is that halos are occupied by central and satellite galaxies, in the standard HOD model the mean number of galaxies is given by their sum
\begin{subequations}
\label{eq:HOD}    
\begin{equation}
\label{eq:HOD_total}
    \langle N_g (M_h)\rangle = \langle N_{\rm cen}(M_h)\rangle + \langle N_{\rm sat}(M_h)\rangle \,.
\end{equation}
The mean occupation of central galaxies is parameterised by
\begin{equation}
\label{eq:HOD_cental}
    \langle N_{\rm cen}(M_h)\rangle  = \frac{1}{2} \left[1 - {\rm erf}\left(\frac{\log (M_h/M^{\rm min}_{h})}{\sigma_{\log M }}\right)\right]\,,
\end{equation}
where $M^{\rm min}_{h}$ is the minimum mass for which half of the halos host a central galaxy above the luminosity threshold and $\sigma_{\log M}$ is related to the scatter of the central galaxy luminosity in halos of mass $M_h$. Every individual halo has either no or one central galaxy with the probability indicated above. The mean occupation of satellites follows a power law as
\begin{equation}
\label{eq:HOD_satellite}
    \langle N_{\rm sat}(M_h)\rangle  =  \langle N_{\rm cen}(M_h)\rangle \left(\frac{M_h - M_0}{M_1}\right)^{\alpha}\,.
\end{equation}
\end{subequations}
with $M_0$ is the halo mass cut-off for satellite occupation, $M_1$ is given such that $M_h = M_0 + M_1$ is the typical mass scale for halo to host one satellite and $\alpha$ is the slope at high halo mass.  

If the bias of dark matter halos is known as a function of mass (see \cite{DesjacquesBiasReview2018} for a review of how to determine it from the halo mass function), the bias for HOD galaxies can be obtained from a re-weighting given by \cite{Zheng2007}
\begin{equation}
\label{eq:bias_SMT_HOD}
    \bar{b}_{{\rm g}, n} = \frac{\int  \, dM_{h} \langle N_g (M_h)\rangle  b_n^{\rm E}(M_h)\mathcal P(M_h)}{\int \, dM_{h}\langle N_g (M_h)\rangle \mathcal P(M_h)}\,,
\end{equation}
where the expected number of galaxies per halo mass $\langle N_g (M_h)\rangle$ is determined by the HOD~\eqref{eq:HOD}. While theoretical models based on mass functions such as \cite{ShethMoTormen2001} are insufficiently precise for real data, they can be useful for performing consistency checks and informing the choice of priors in cosmological analysis.

\subsection{Redshift space distortions}
\label{sec:RSD}
The observed position of a tracer $\v{s}_{\rm obs}$ -- its angle on the sky $\hat{\v{n}}_{\rm obs}$ and its observed redshift $z_{\rm obs}$ -- corresponds to a point on the observer's past light cone. As a first step towards calculating tracer correlations on the past light cone, we will make two common simplifying assumptions. First, since we are interested in equal-time correlation functions, we will approximate the light cone in the neighbourhood of $t$ by the $t$=const slice. Secondly, we use the distant observer approximation, where the line of sight is assumed to be a fixed direction $\hvz$ which is without loss of generality chosen as the direction of the $z-$axis, to relate the observed redshift-space position $\v{s}$ of a dark matter tracer to its real-space position $\v{r}$ . Those approximations are despite their simplicity sufficient even for modern wide-area surveys within the level of current error bars \cite{Yoo2013}. For a general definition of redshift space and a discussion of wide-angle effects in linear perturbation theory we refer to \cite{Matsubara2000}. In the distant observer approximation the observed comoving distance in redshift space $\v{s}$ is affected by the peculiar velocity $\v{v}\cdot \hvz =v_z$ of the tracer along the line of sight via
\begin{equation}
\label{eq:RSspace}
\v{s} = \v{r}+ \sH^{-1} (\v{v}\cdot \hvz)\ \hvz \,,\quad \v{s}_\perp = \v{r}_\perp \,, \quad s_{||}=\v{s}\cdot \hvz =r_{||} + \sH^{-1} v_z
\end{equation}
where $\mathcal{H} = a H = \dot{a}$. The observed position of the tracer perpendicular to the line of sight $\v{s}_\perp$ remains unaffected if we neglect gravitational lensing. In contrast, its coordinate $s_{||}$ parallel to the line of sight $\hvz$  depends on the peculiar velocity $v_z$.
Assuming that all objects remain observable, we get the following relation between the densities in real and redshift space
\begin{align}
(1+\delta_t(\v{s},t))\, \varvol{3}{s} &= (1+\delta_t(\v{r},t))\, \varvol{3}{r} \,.\label{realtoredshift}
\end{align}
Although the correction to the real space position in redshift space is very small $\sH^{-1} v_z \ll r_{||}$, the clustering is affected considerably since the change of volume measure between real and redshift space, given by the Jacobian between $\varvol{3}{s}$ and $\varvol{3}{r}$, involves the gradient of $v_z$ in linear perturbation theory \cite{Kaiser1987}.
In the distant observer approximation and for a perfect fluid, the tracer density fluctuation in redshift space \eqref{realtoredshift} can be equivalently written as
\begin{align} 
1+\delta_t(\v{s},t) = &\int \vol{3}{r} (1+\delta_t(\v{r},t))\, \delta_{\rm D}\left(\v{s} -\v{r} -  \frac{\v{v}(\v{r},t)\cdot\hvz}{aH} \hvz\right)  \,. \label{deltasr}
\end{align}

\subsubsection{Kaiser effect}

\begin{figure}
\centering
\includegraphics[width=0.9\textwidth]{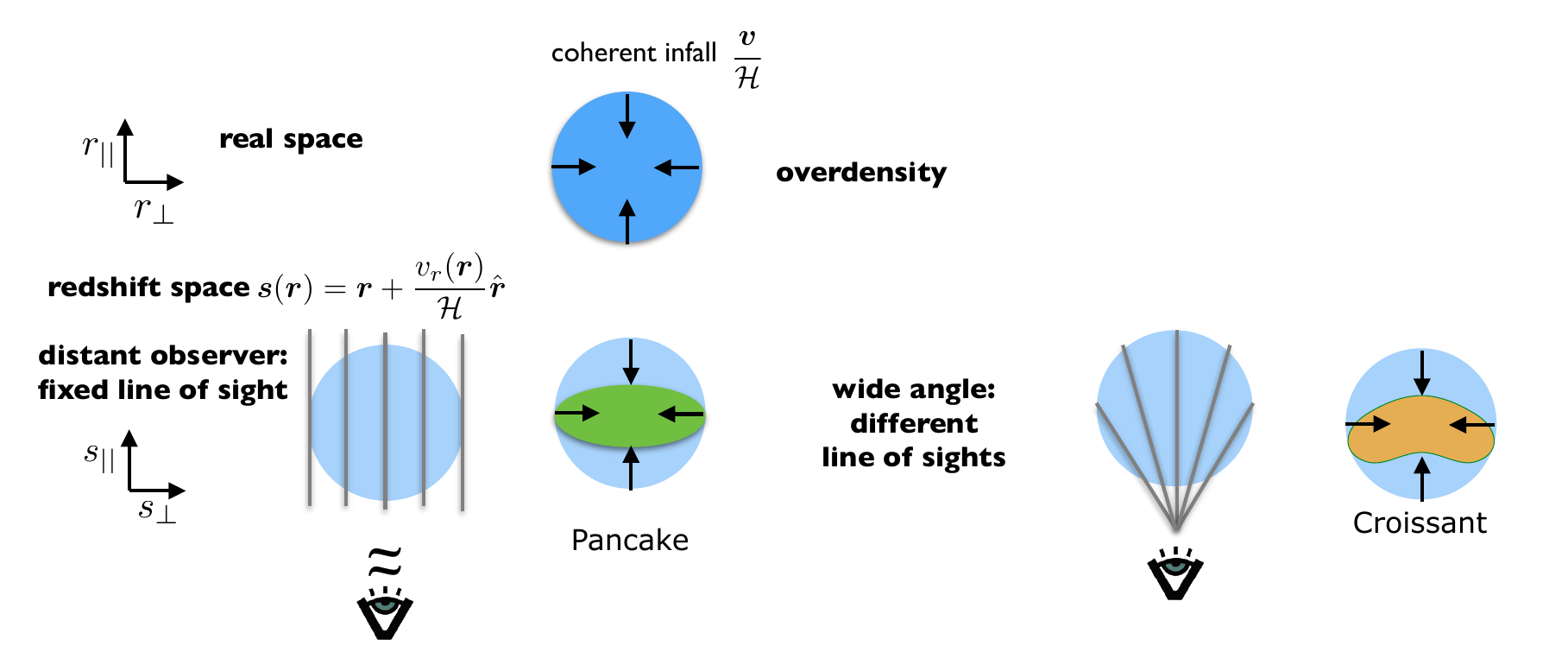}
\caption{Sketch of the linear impact of redshift space distortions in the distant observer approximation (left) or including wide angle effects (right).} 
\label{fig:RSD}
\end{figure}
\begin{subequations}
To get some quantitive understanding of the impact of RSD on large scales, we will retrace the idea behind the seminal work by \cite{Kaiser1987}. To derive the leading order, linear effect we start from mass conservation and determine the Jacobian determinant of real to redshift space mapping in the distant observer approximation
    \begin{equation}
    (1 + \delta_s) \, d^3s = (1 + \delta_r) \, d^3r\,, \quad J = \frac{d^3s}{d^3r} \approx 1 + \sH^{-1} \frac{\partial v}{\partial z}\,,
    \end{equation}
   where $z$ indicates the $z$-component of the position $\vr$ (instead of the redshift). Now we linearise the density relation to obtain
    \begin{equation}
    1 + \delta_s= J^{-1}(1 + \delta_r) \Rightarrow \delta_s \approx \delta_r - \sH^{-1}\frac{\partial v}{\partial z}
    \end{equation}
We assume a potential flow for the velocity and thus express it in terms of the velocity divergence, which we then in turn obtain from the linearised continuity equation
\begin{equation}
    \frac{\partial v}{\partial z}\approx -\nabla^{-2} \frac{\partial^2 \theta}{\partial z^2}\,,\quad f \sH \delta = \theta = -\nabla \cdot \mathbf{v}
    \end{equation}
    with growth function $f$ as introduced before. If we now write the density in Fourier space and build the power spectrum, we obtain for dark matter:
    \begin{equation}
    P_m^s(k, \mu) = (1 + f \mu^2)^2 P_L^r(k)\,,\quad \mu=\cos\theta=\cos \angle(\vk,\hat\vz)
    \end{equation}
    which depends on the angle between the wave-vector $\vk$ and the line of sight $\hat\vz$. As we can see from this expression, the clustering in redshift space appears anisotropic, as it has a dependence on both the wave-number $k$ and its angle $\mu$ to the line of sight.

Adding tracer bias affects densities, assuming a local linear (Eulerian) remapping in the absence of velocity bias:
 \begin{equation}
  P_t^s(k, \mu) = (b_E^1 + f \mu^2)^2 P_L^r(k) = \left(1 + \frac{f}{b_E^1} \mu^2\right)^2 (b_E^1)^2 P_L^r(k)
  \label{eq:Pk_anisotropic_linearbias}
 \end{equation}
 \end{subequations}
 
   \begin{figure}
   \centering
\includegraphics[width=0.75\textwidth]{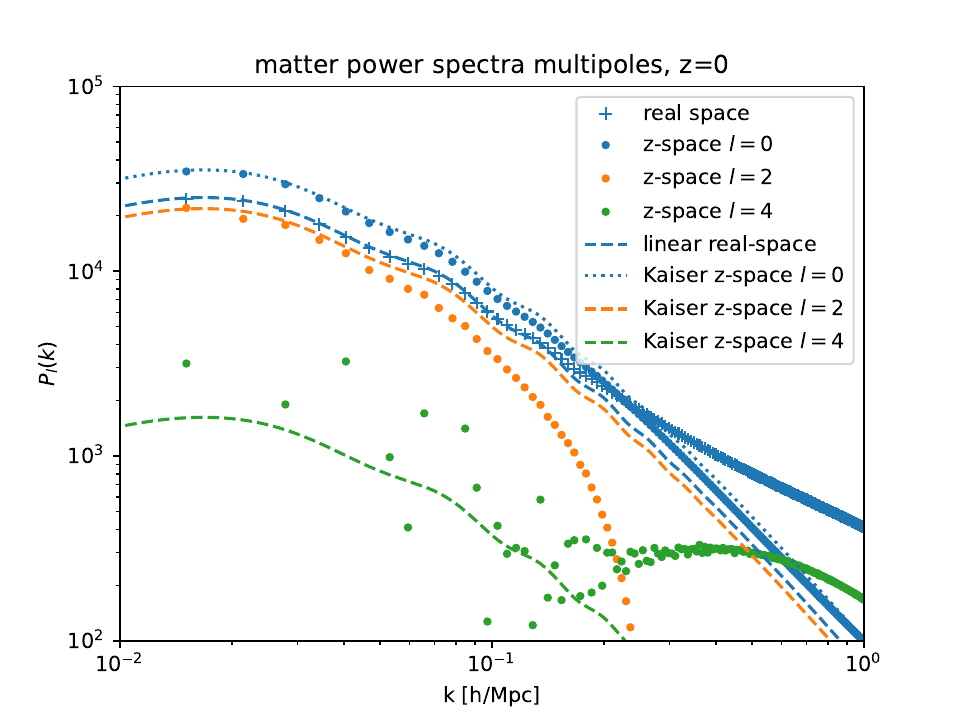}
\caption{Redshift space matter power spectrum multipoles in the Quijote simulation (averaged over 1000 realisations at the fiducial cosmology and $z=0$) showing the monopole (blue points), quadrupole (orange points), and hexadecapole (green points), in comparison to the real space power spectrum (blue plusses). We also show the predictions by linear theory \eqref{eq:Pk_multipoles_linear} in dashed lines of the respective color.} 
\label{fig:RSD_lin}
\end{figure}

\paragraph{Multipole decomposition} To capture the anisotropies in the redshift space distribution of  density fluctuations, we can expand the anisotropic power spectrum in terms of Legendre polynomials $\mathcal L_l(\mu)$ as follows:
\begin{equation}
P(k,\mu)=\sum_l P_l(k)\mathcal L_l(\mu)\,,
\quad P_l(k) = \frac{2l+1}{2} \int_{-1}^{1}d\mu\, P(k,\mu)\,\mathcal L_l(\mu)\,,
\end{equation}
where $P_l(k)$ are the multipole moments of the power spectrum. Only even orders contribute, in particular the monopole ($l=0$), which represents the isotropic part, the quadrupole ($l=2$) which captures the leading anisotropies, and the hexadecapole ($l=4$). These multipole moments are computed by integrating over the angle $\mu$, which using the linear result \eqref{eq:Pk_anisotropic_linearbias} gives
\begin{align}
\notag
P_{t,0}(k) &= 
\left( (b_1^E)^2 + \frac{2}{3} b_1^E f + \frac{1}{5} f^2 \right)  P_L(k)\\
  \label{eq:Pk_multipoles_linear} P_{t,2}(k)& = \left(  \frac{4}{3} b_1^E f + \frac{4}{7} f^2 \right) P_L(k)  \\
\notag  P_{t,4}(k)& = \frac{8}{35} f^2P_L(k)    \,.
   \end{align}
In Figure~\ref{fig:RSD_lin} we illustrate the accuracy of the linear predictions for the matter power spectrum multipoles on large scales.
 
Similarly, the redshift-space correlation function is anisotropic  $\xi_t(s_{||},s_\perp,t)$ and it is useful to expand it into Legendre polynomials using $s^2 = s_{||}^2+s_\perp^2$ and $\mu = s_{||}/s$  
\begin{subequations}
\label{ximultipoles}
\begin{align} 
\xi_t(s,\mu)= \sum_{l=0}^\infty \mathcal L_l(\mu) \xi_{X,n}(s)\,,\quad
\xi_{X,n}(s) =\frac{2 l+1}{2}\, \int_{-1}^1 \xi_t(s,\mu,t) \mathcal L_{l}(\mu) d \mu \,.
\end{align}
\end{subequations}
$\xi_l$ vanishes for all odd $l$. In linear perturbation theory, the only non-zero multipoles are the monopole $\xi_0$, quadrupole $\xi_2$ and hexadecapole $\xi_4$ and similarly as done for the power spectrum \eqref{eq:Pk_multipoles_linear}, one can obtain the multipoles 
\begin{align}
\notag \xi^L_0(s) &= 
\left( (b_1^E)^2 + \frac{2}{3} b_1^E f + \frac{1}{5} f^2 \right)  \underbrace{\frac{1}{2\pi} \int d k\, k^2 P_L(k) j_0(ks)}_{\xi^L(r=s)}  \\
 \xi^L_2(s)& = 
 - \left(  \frac{4}{3} b_1^E f + \frac{4}{7} f^2 \right)\frac{1}{2\pi}\int d k\, k^2 P_L(k) j_2(ks)  \\
\notag  \xi^L_4(s)& = 
  \frac{8}{35} f^2 \frac{1}{2\pi}\int d k\, k^2 P_L(k) j_4(ks)     \,,
   \end{align}
where $f$ is the linear growth rate and $j_l(x)$ are the spherical Bessel functions related to an integral of the Legendre polynomials
\begin{equation}
j_l(x) = \frac{i^l}{2} \int_{-1}^{1} \dd\mu \exp(ix\mu) \mathcal L_l(\mu)\,.
\end{equation}
Even in the nonlinear regime the magnitude of $\xi_l$ rapidly decreases with $l$. 

\subsubsection{Nonlinearities}

\begin{figure}
\includegraphics[width=0.9\textwidth]{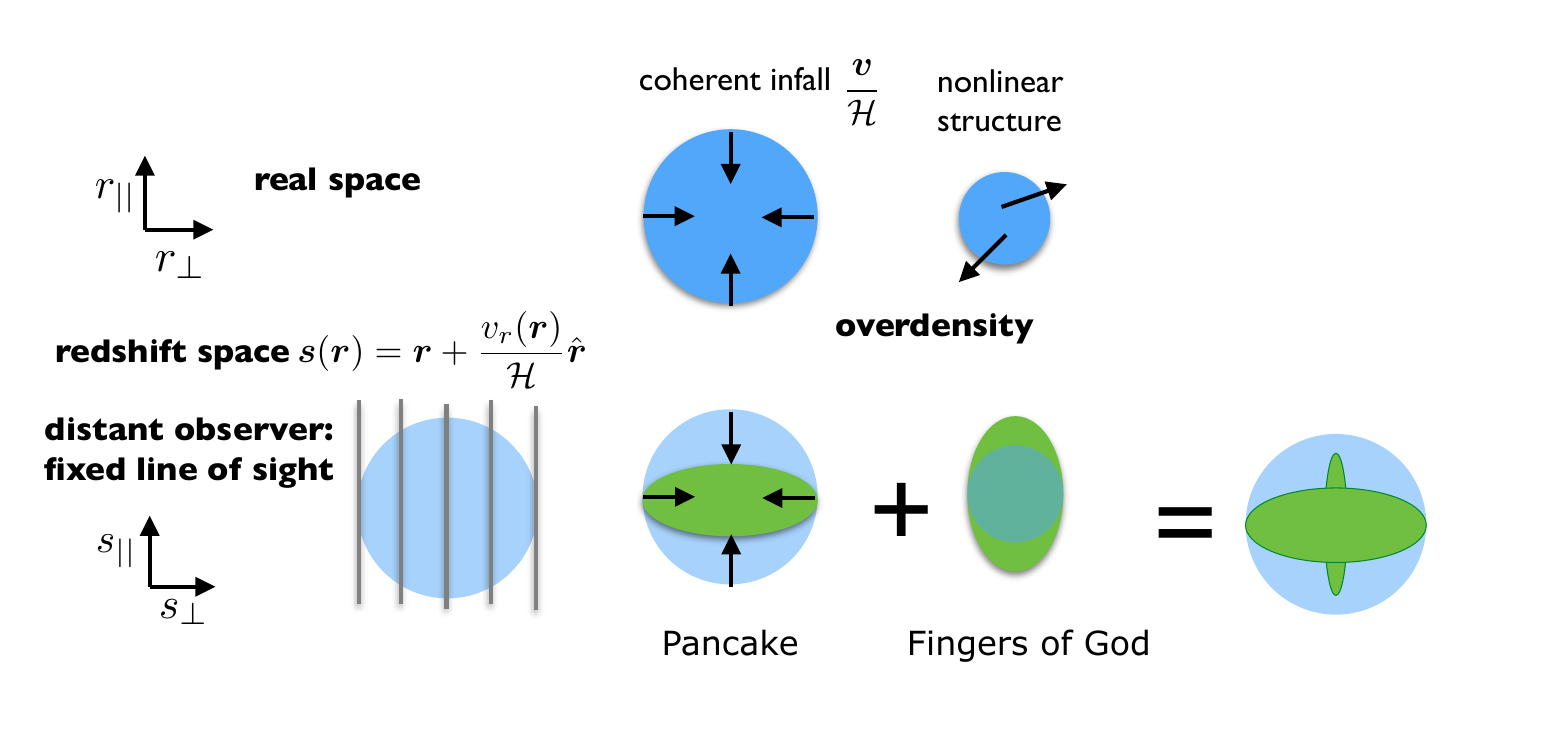}
\caption{Sketch of the impact of redshift space distortions in the distant observer approximation with the linear Kaiser squashing on large scales (left) along with the `Fingers of God' elongation on small scales (middle) and their combination (right).} 
\label{fig:RSD_nonlin}
\end{figure}

There are two main effects that affect the correlation function on large and small scales, respectively. On large scales, the peculiar velocity associated with the coherent infall into overdense regions apparently squashes structures and enhances the correlation function along the line of sight. This effect is captured by linear theory and known as the Kaiser effect \cite{Kaiser1987}. On small scales, the apparent elongation of nonlinear structures along the line-of-sight, the so-called `Fingers of God' effect as they seem to point towards the observer and lead to a suppression of the correlation function. The basic principle underlying those two effects is sketched in Figure~\ref{fig:RSD_nonlin}.

%
%
%
%
%

For a better model of the redshift-space correlation function, we can turn to one of various models for nonlinear redshift-space distortions, e.g. the Taruya-Nishimichi-Saito (TNS) model \cite{TNS2010} or the streaming model \cite{Scoccimarro2004}. 

\paragraph{TNS model}
The TNS model posits a general redshift-space power spectrum of the form
\begin{equation}
\begin{split}
    P_{\rm TNS}(k,\mu)=&\frac{
    P_{\delta\delta}(k)+2f\mu^2P_{\delta\theta}(k) +f^2\mu^4P_{\theta\theta}(k)+A(k,\mu;f)+B(k,\mu;f)}{1+k^2\mu^2f^2\sigma_{v,\rm{eff}}^2}\,,
\end{split}
\end{equation}
where $P_{\delta\delta}$, $P_{\theta\theta}$ and $P_{\delta\theta}$ represent the density and velocity divergence power spectra and their cross-power spectrum, respectively. This form is heavily inspired from Eulerian Perturbation Theory, as evident from the appearance of the 
The denominator accounts for the Finger-of-God (FoG) effect using the velocity dispersion $\sigma_{v,{\rm eff}}^2$, which is generally left as a free parameter but for which a first guess can be made using the linear approximation

\begin{equation}
    \sigma_{v,L}^2 = \frac{1}{3}\int\frac{\textrm{d}^3q}{(2\pi)^3}\frac{P_{\theta\theta}}{q^2}.
\end{equation}\label{eq:sigma_v}

Code with which to calculate the $A$ and $B$ correction terms is publicly available.\footnote{\url{http://www2.yukawa.kyoto-u.ac.jp/~atsushi.taruya/}} The TNS model has been routinely used for the interpretation of observational data, in particular SDSS, BOSS and eBOSS\cite{GilMarin2020,Brieden2022BOSS}. It also informed updated prescriptions to augment perturbation theory with non-perturbative RSD models push cosmological inference to smaller scales \cite{Eggemeier2025}.

\paragraph{Streaming models} Based on the observation of the `Fingers of God' effect one of the first streaming models was developed in \cite{Peebles1980} by assuming an exponential pairwise velocity distribution with a scale-independent dispersion. To reunite the two disparate results for large and small scales, the so-called Gaussian streaming model was introduced \cite{Fisher1995}. To obtain the streaming model, the matter correlation function in redshift space was derived by considering the joint probability distribution of density and velocity. Assuming that the density is a Gaussian random field and that the velocity is related to density as in linear perturbation theory one obtains a simple expression for the redshift space correlation function. It is given by a convolution of the real space correlation function and an approximately Gaussian pairwise velocity distribution with a scale-dependent mean and variance of the pairwise velocity. 

We are interested in the redshift space two-point correlation function
\begin{equation}
\label{2ptcorrfctgeneral}
1+\xi_t(\v{s},t) =  \Big\langle (1+\delta_t(\v{s}_1)) (1+\delta_t(\v{s}_2)) \Big\rangle \,,
\end{equation}
where $\v{s}=\v{s}_2-\v{s}_1$. By inserting \eqref{deltasr} in \eqref{2ptcorrfctgeneral} and re-expressing the delta functions in Fourier space and integrating over $\v{R}=\v{r}_1+\v{r}_2$ and one momentum variable the correlation function can be brought into the following form
 \begin{subequations} 
 \label{xizspacegeneral}
\begin{align}
\label{xizspaceexprgeneral}
1+\xi_t(\v{s},t) &=\int \vol{3}{r}\!\! \int \frac{\vol{3}{k}}{(2\pi)^3} e^{i \v{k}\cdot (\v{r} - \v{s}) } Z\Big(\v{r}, \v{J} = (\v{k}\cdot\hvz)\ \hvz , t\Big)\,,\\
 Z(\v{r}, \v{J}, t) &= \langle (1+\delta_t(\vx_1))(1+\delta_t(\vx_2)) \rangle (\vx_1-\vx_2=\v{r},t)  \exp\left[i \frac{(\vv_2 - \vv_1) \cdot \v{J}}{a H}\right]\label{defZgeneral}\,,
\end{align}
\end{subequations}
where $Z$ is the pairwise generating function.
Next we Taylor expand $W(\v{J}):= \ln Z$ around $\v{J}=0$
up to second order
\begin{subequations}
\label{Wexpansion}
\begin{align}
W(\v{J}) \simeq &\ln (1+ \xi_t(r,t))+ i \v{v}_{12} \cdot \v{J} - \frac{1}{2}\v{J}^{T} \v{\sigma}^2_{12} \v{J} \label{W2}
\end{align}
\end{subequations}
with the cumulants $\v{\kappa}_n$ as expansion coefficients 
\begin{subequations}
\label{GSMparam}
\begin{align}
1+ \xi_t(r,t) &:= \exp \kappa_0 = Z\, |_{J=0} \,,\label{xirel}\\
\v{v}_{12}(\v{r},t) & := \v{\kappa}_1  = \frac{\frac{\partial Z}{(\partial i \v{J})}\big|_{J=0}}{(1+ \xi_t(r,t))} \,,\label{v12rel}\\
\v{\sigma}^2_{12}(\v{r},t) & :=  \v{\kappa}_2  = \frac{\frac{\partial^2 Z}{(i\partial \v{J})^2}\big|_{J=0}}{(1+ \xi_t(r,t))} -\frac{\frac{\partial Z}{(i\partial \v{J})} \frac{\partial Z}{(i\partial \v{J})}\big|_{J=0}}{(1+ \xi_t(r,t))^2}  =  \tilde{\v{\sigma}}^2_{12}(\v{r},t) -\v{v}_{12}(\v{r},t) \v{v}_{12}(\v{r},t)\label{sigma12rel}\,.
\end{align}
\end{subequations}
Since we have to evaluate all expressions in a scalar product with $\v{J} = (\v{k}\cdot\hvz)\ \hvz$ we project the cumulants $\v{\kappa}_n$ onto the line of sight $\kappa_n= \kappa_{n}^{i_1\cdots i_n} \hat{z}_{i_1}\cdots \hat{z}_{i_n}$. Expanding $W$ in \eqref{W2} up to second order in $\v{J}$ implies that all redshift space distortion induced clustering is encoded in the scale dependent mean and variance given by the pairwise velocity $v_{12}$ and its dispersion $\sigma_{12}^2$. This corresponds to the Gaussian streaming model for which the redshift space correlation function can be written as
\begin{align} 
1+\xi_t(s_{||},s_\perp,t)&\approx \int^{\infty}_{-\infty} \frac{ \vol{}{r_{||}\, (1+\xi_t(r,t))}}{\sqrt{2 \pi} \sigma_{12}(r,r_{||},t)}   \exp\left[-\frac{\left(s_{||}- r_{||} - v_{12}(r,t) r_{||}/r\right)^2}{2 \sigma_{12}^2(r,r_{||},t)}\right] \label{GSM} \,.
\end{align}
The streaming model can be extended to include higher order cumulants using an Edgeworth expansion \cite{UhlemannESM2015}. For predictions of the ingredients of the streaming model, the real-space correlation function along with the scale-dependent pairwise velocity and its dispersion, often Lagrangian Perturbation Theory is applied. In Lagrangian space, we can include a tracer bias functional $F$, a displacement due to the gravitational collapse and a redshift space displacement in a single expression by expressing the velocity in terms of the displacement $\v{v} = a \dot{\v{\varPsi}}$ 
\begin{align} \label{deltasq}
1+\delta_t(\v{s},t) &= \int \vol{3}{q} F[\delta_R(\v{q}),t] \delta_{\rm D}\left(\v{s} -\v{q} - \v{\varPsi}(\v{q},t) - \frac{\dot{\v{\varPsi}}(\v{q},t)  \cdot \hvz}{H}\hvz\right)\,.
\end{align}

\subsection{Photometric clustering and weak lensing}
\label{sec:WL}

\begin{figure}
\centering
\includegraphics[width=0.48\textwidth]{figures_Quijote/Quijote_matter_density_slice_z0.pdf}
\includegraphics[width=0.48\textwidth]{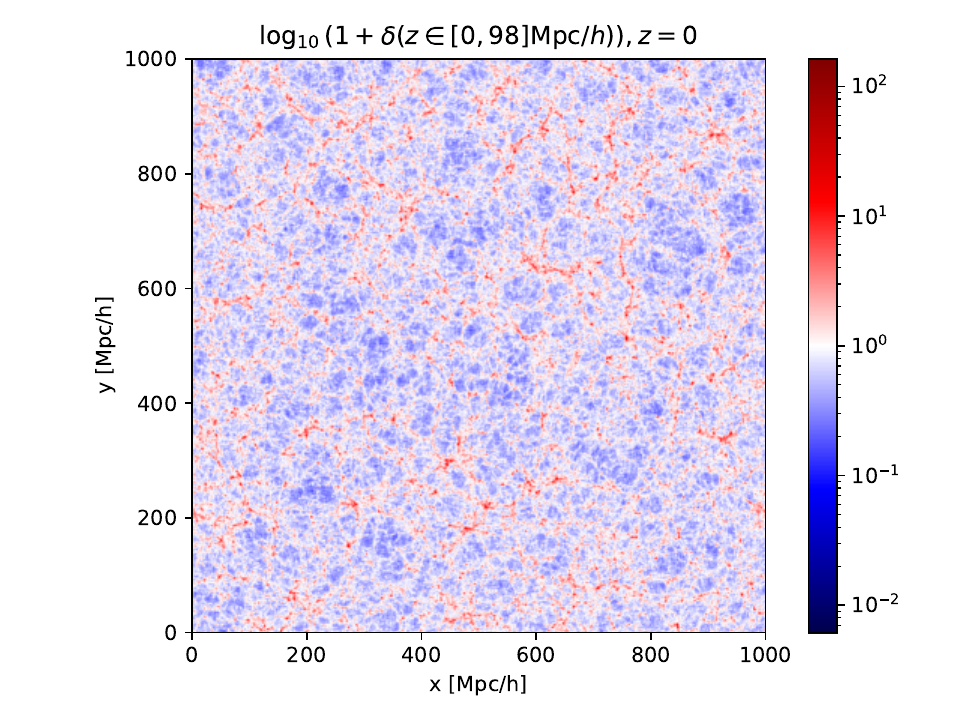}
\caption{Thin slice through the 3-dimensional dark matter distribution in the Quijote simulations at redshift $z=0$ (shown for realisation 0 of the fiducial cosmology) as shown in Figure~\ref{fig:density_slice} (left) in comparison with a 10 times thicker slice as probed by photometric clustering (right).}
\label{fig:density_slice_thick}
\end{figure}

\begin{figure}
\includegraphics[width=0.3\textwidth]{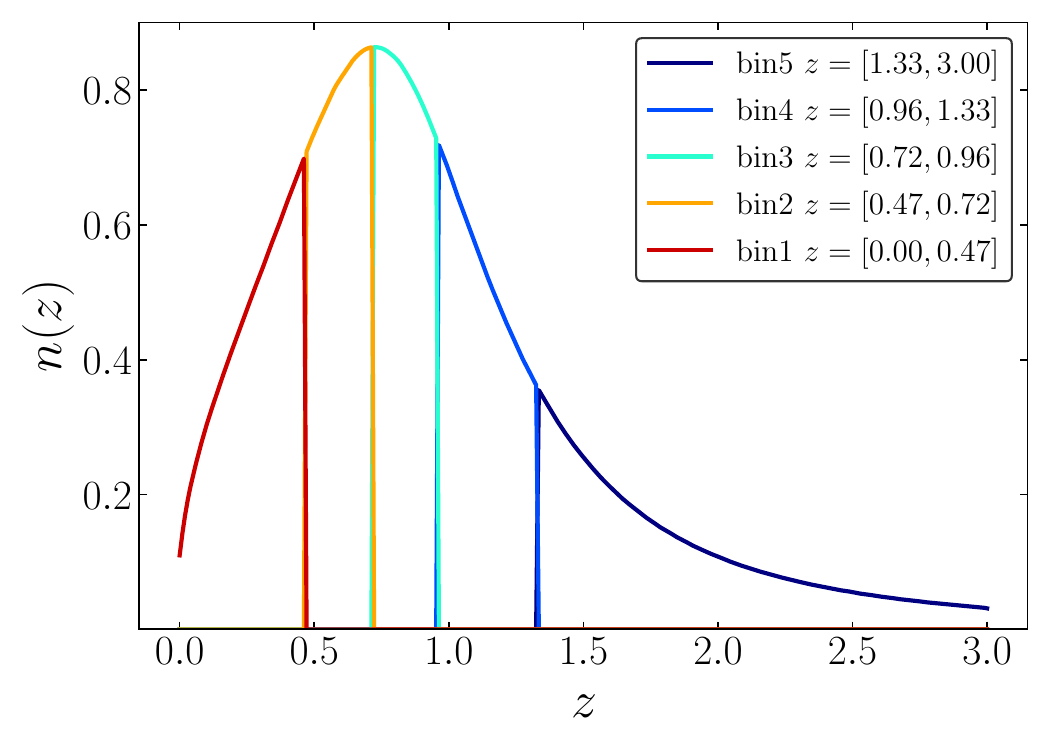}
\includegraphics[width=0.6\textwidth]{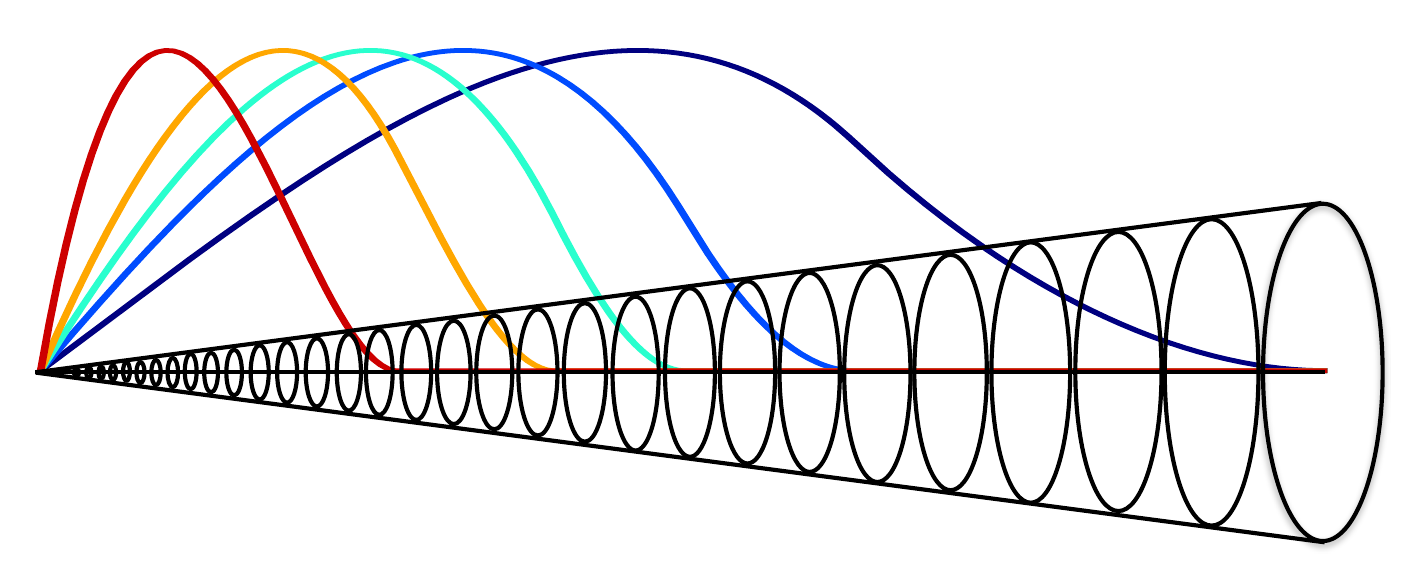}
\caption{(Left) Stage IV-like source redshift distribution probing up to source redshifts of around $z_s\sim 3$ split into 5 equipopulated bins. (Right) Sketch of how the weak lensing convergence field in a fixed aperture is obtained from a superposition of independent layers of the 3D matter density along the line of sight with a weighting according to the weak lensing kernel~\eqref{eq:lensing_kernel}. [Reproduced from Figure~1 in \cite{Castiblanco2024} (CCBY-4.0)]}
\label{fig:WLkernel}
\end{figure}

Both the projected matter density contrast $\delta_{m,\rm proj}$ and the weak lensing convergence $\kappa$ can be considered as line-of-sight projections of the 3-dimensional matter density contrast $\delta_{m,3\mathrm{D}}$. Identifying points on the sky by unit vectors $\mathbf{\hat n}$ on the sphere and assuming a spatially flat Universe this can be written as  \cite{BartelmannSchneider2001,WLreviewPratBacon}
\begin{align}
    \label{eq:def-lens-deltam}
    \delta_{m,\rm proj}(\mathbf{\hat n}) =&\ \int \dd \chi\ w_m(\chi)\ \delta_{m,3\mathrm{D}}(\chi\cdot \mathbf{\hat n},\eta_0 - \chi)\,,\\
    \label{eq:def-convergence}
    \kappa(\mathbf{\hat n}) =&\ \int \dd \chi\ w_l(\chi)\ \delta_{3\mathrm{D}}(\chi\cdot \mathbf{\hat n}, \eta_0 - \chi)\,,
\end{align}
where $\chi$ is comoving distance,  
the line-of-sight projection kernel for the matter $w_m(\chi)$ is given in terms of the redshift distribution $n_g(z)$ of the tracer (lens) galaxies as
\begin{equation}
    w_m(\chi) = n_g(z(\chi))\  \frac{\dd z}{\dd \chi}(\chi)\,,
\end{equation}
and the weak lensing kernel  $w_l(\chi)$ is given by the source galaxy distribution $n_s(z_s)$ and the lensing kernel for a fixed source plane
\begin{align}
\label{eq:lensing_kernel}
w_l(\chi) &= \int dz_s\,n_s(z_s) w_{l,z_s}(z(\chi))\,,\ 
w_{l,z_s}(z) = \frac{3H_{0}^{2}\Omega_{m}}{2c^2} \frac{\chi\big[\chi(z_s)-\chi\big]}{\chi(z_s)}(1+z) \Theta(z_s-z)\,,
\end{align}
where $\Omega_m$ is today's total matter density in units of the critical density, $H_0$ is today's Hubble expansion rate, $c$ is the speed of light and $\Theta$ is the Heaviside function. As illustrated in Figure~\ref{fig:WLkernel}, the weak lensing  kernel is nonzero for all redshifts below the sources, with a weighting that prefers redshifts roughly half way between the sources and the observer. 

As projected quantities, the photometric matter (and hence also galaxy) density and weak lensing convergence have some in-built averaging that tend to gaussianise the fields at large angular scales. This in illustrated in Figure~\ref{fig:density_slice_thick} by comparing a thin and thick slice through a 3-dimensional $N$-body simulation. As weak lensing is an extreme projection, it is beneficial to divide the source distribution into different redshift bins (illustrated in the different colours in the left panel of Figure~\ref{fig:WLkernel}) and do tomography -- jointly considering the resulting weak lensing convergence statistics in different tomographic bins. Similarly, it is beneficial to divide the foreground tracers into several tomographic bins, and do a combined analysis between weak lensing and photometric clustering, which is probing similar slices of the underlying matter density. This analysis based on two-point statistics is called 3x2-point statistics, as it consists of correlation functions for galaxy clustering and weak lensing separately, and their cross-correlation. 

We assume a flat universe and adopt the Limber approximation.  The two-point angular auto power spectra for galaxy clustering can be formulated as
\begin{equation}
    C_{g^i,g^i}(\ell) = b_i^2 \int_0^{\chi_\infty} d\chi \frac{dN_l^i}{dz}^2 P_{g^i,g^i}\left(\frac{\ell}{\chi},\chi\right)\,
\end{equation} 
where $b_i$ denotes the linear galaxy bias, $\frac{dN^i}{dz}$ denotes the photometric redshift distribution in photo-z bin $i$, and $P_{g^i,g^i}$ is the non-linear matter power spectrum for galaxy-galaxy auto-clustering for photo-z bin $i$. The galaxy-galaxy lensing angular power spectra are given by
\begin{equation}
    C_{\kappa^i,g^j}(\ell) = b_j \int_0^{\chi_\infty} d\chi \frac{W_i(\chi)}{\chi} \frac{dN_l^j}{d\chi}P_{m,g^i}\left(\frac{\ell}{\chi},\chi\right)\,
\end{equation}
where the lensing efficiency $W_i(\chi)$ of the tomographic redshift bin i is given by
\begin{equation}
    W_i(\chi) = \frac{3 H_0^2 \Omega_M}{2c^2} \frac{\chi}{a(\chi)}\int_\chi^{\chi_\infty}\frac{\chi'-\chi}{\chi'} \frac{dN_s^i}{d\chi'}d\chi' \,.
\end{equation}
The lensing angular power spectrum is obtained as
\begin{equation}
        C_{\kappa^i, \kappa^j}(\ell) = \int_0^{\chi_\infty} \frac{W_i(\chi) W_j(\chi)}{\chi^2}P_{m,m}\left(\frac{\ell}{\chi},\chi\right).
\end{equation}

\section{Conclusion}
\label{sec:conclusion}
The cosmic large-scale structure is a powerful probe of cosmology and fundamental physics, tracking the growth of structure over time. Major galaxy surveys probe the large-scale distribution of matter through galaxy clustering and weak gravitational lensing. In order to extract information from those probe, we need to model the underlying nonlinear dark matter dynamics and predict informative clustering statistics. On large scales, dark matter can be approximated as an effective fluid and its evolution can be solved perturbatively. Some properties of the clustering such as the mass distribution of bound dark matter structures and the one-point distribution of matter densities can also be predicted with the spherical collapse model. Since galaxies trace the underlying dark matter, we can lift predictions for dark matter to observables with parametrised models of galaxy bias and stochasticity. As surveys observe redshift, we need to model the impact of peculiar velocities on the observed redshift and thus the clustering properties, a phenomenon called redshift space distortions. Weak lensing and photometric clustering provides an additional avenue to simultaneously probe the overall matter distribution and the clustering of tracers, allowing the break degeneracies between galaxy bias and cosmological parameters. 

\section*{Acknowledgements}

I warmly thank the organisers of the Les Houches summer school on `The Dark Universe' and the Heidelberg Graduate School for Physics for the opportunity to give this lecture series, and the staff for providing such a stimulating environment. I also want to thank all of the students for actively participating, catching mistakes, asking questions and engaging in discussions, both scientific and beyond. A special thanks goes to Francisco Villaescusa-Navarra for making the Quijote simulations  \cite{QuijoteSims}  and software package \href{https://github.com/franciscovillaescusa/Pylians3}{Pylians} publicly available through \href{https://quijote-simulations.readthedocs.io/en/latest/access.html}{Globus and Binder}, which was used for producing most of the Figures shown.


\paragraph{Funding information} CU is supported by the European Union (ERC StG, LSS\_BeyondAverage, 101075919). 


\begin{appendix}
\numberwithin{equation}{section}


\end{appendix}





\bibliography{LesHouchesLSS_SciPost_BiBTeX.bib}


\end{document}